\documentclass[a4paper,12pt]{article}
\usepackage{hyperref}
\usepackage[a4paper,left=2.5cm,right=2.5cm,top=3cm,bottom=2.5cm]{geometry}
\usepackage{authblk}
\usepackage{amsmath,bm}
\usepackage{graphicx}
\usepackage{mathrsfs}
\usepackage{amssymb}
\usepackage{physics}
\usepackage{tensor}
\usepackage{xcolor}
\usepackage{tikz}
\usepackage{pgfplots}
\usepackage{amsthm}
\usepackage{booktabs,dcolumn,caption}
\usepackage{multirow}
\usepackage[english]{babel}
\usepackage{soul}

\theoremstyle{remark}
\newtheorem*{remark}{\textbf{Remark}}

\title{Geometrically exact isogeometric Bernoulli-Euler beam based on the Frenet-Serret frame}

\author[1,2]{A. Borkovi\'c}
\author[1]{M. H. Gfrerer}
\author[1]{B. Marussig}

\affil[1]{Institute of Applied Mechanics, Graz University of Technology, Technikerstraße 4/II, 8010 Graz, Austria, aleksandar.borkovic@aggf.unibl.org, aborkovic@tugraz.at}
\affil[2]{University of Banja Luka, Faculty of Architecture, Civil Engineering and Geodesy, Department of Mechanics and Theory of Structures, 78000 Banja Luka, Bosnia and Herzegovina}
\date{}                     
\begin{document}
	\newcommand{\red}[1]{\textcolor{red}{#1}}

	\newcommand{\ssub}[2]{{#1}_{#2}} 
	\newcommand{\vsub}[2]{\textbf{#1}_{#2}} 
	\newcommand{\ssup}[2]{{#1}^{#2}} 
	\newcommand{\vsup}[2]{\textbf{#1}^{#2}} 
	\newcommand{\ssupsub}[3]{{#1}^{#2}_{#3}} 
	\newcommand{\vsupsub}[3]{\textbf{#1}^{#2}_{#3}} 
	
	\newcommand{\veq}[1]{\bar{\textbf{#1}}} 
	\newcommand{\seq}[1]{\bar{#1}} 
	\newcommand{\ve}[1]{\textbf{#1}} 
	\newcommand{\vepre}[1]{\textbf{#1}^\sharp} %
	\newcommand{\sdef}[1]{#1^*} 
	\newcommand{\vdef}[1]{{\textbf{#1}}^*} 
	\newcommand{\vdefeq}[1]{{\bar{\textbf{#1}}}^*} 
	\newcommand{\trans}[1]{\textbf{#1}^\mathsf{T}} 
	\newcommand{\transmd}[1]{\dot{\textbf{#1}}^\mathsf{T}} 
	\newcommand{\mdvdef}[1]{\dot{\textbf{#1}}^*} 
	\newcommand{\mdsdef}[1]{\dot{#1}^*} 
	\newcommand{\mdv}[1]{\dot{\bm{#1}}} 
	\newcommand{\mdvni}[1]{\dot{\textbf{#1}}} 
	\newcommand{\mds}[1]{\dot{#1}} 
	
	\newcommand{\loc}[1]{\hat{#1}} 
	\newcommand{\md}[1]{\dot{#1}} 

	\newcommand{\ii}[3]{{#1}^{#2}_{#3}} 
	\newcommand{\iv}[3]{\textbf{#1}^{#2}_{#3}} 
	\newcommand{\idef}[3]{{#1}^{* #2}_{#3}} 
	\newcommand{\ivdef}[3]{\textbf{#1}^{* #2}_{#3}} 
	\newcommand{\ipre}[3]{{#1}^{\sharp #2}_{#3}} 
	\newcommand{\ivpre}[3]{\textbf{#1}^{\sharp #2}_{#3}} 
	\newcommand{\iloc}[3]{\hat{#1}^{#2}_{#3}} 
	\newcommand{\ieq}[3]{\bar{#1}^{#2}_{#3}} 
	\newcommand{\ic}[3]{\tilde{#1}^{#2}_{#3}} 
	\newcommand{\icdef}[3]{\tilde{#1}^{* #2}_{#3}} 
	\newcommand{\icpre}[3]{\tilde{#1}^{\sharp #2}_{#3}} 
	\newcommand{\iveq}[3]{\bar{\textbf{#1}}^{#2}_{#3}} 
	\newcommand{\ieqdef}[3]{\bar{#1}^{* #2}_{#3}} 
	\newcommand{\iveqdef}[3]{\bar{\textbf{#1}}^{* #2}_{#3}} 
	\newcommand{\ieqmddef}[3]{\dot{\bar{#1}}^{* #2}_{#3}} 
	\newcommand{\icmddef}[3]{\dot{\tilde{#1}}^{* #2}_{#3}} 
	\newcommand{\iveqmddef}[3]{\dot{\bar{\textbf{#1}}}^{* #2}_{#3}} 
	
	\newcommand{\ieqpre}[3]{\bar{#1}^{\sharp #2}_{#3}} 
	\newcommand{\iveqpre}[3]{\bar{\textbf{#1}}^{\sharp #2}_{#3}} 
	\newcommand{\ieqmdpre}[3]{\dot{\bar{#1}}^{\sharp #2}_{#3}} 
	\newcommand{\icmdpre}[3]{\dot{\tilde{#1}}^{\sharp #2}_{#3}} 
	\newcommand{\iveqmdpre}[3]{\dot{\bar{\textbf{#1}}}^{\sharp #2}_{#3}} 
	
	\newcommand{\ieqmd}[3]{\dot{\bar{#1}}^{#2}_{#3}} 
	\newcommand{\icmd}[3]{\dot{\tilde{#1}}^{#2}_{#3}} 
	\newcommand{\iveqmd}[3]{\dot{\bar{\textbf{#1}}}^{#2}_{#3}} 
	
	\newcommand{\imddef}[3]{\dot{#1}^{* #2}_{#3}} 
	\newcommand{\ivmddef}[3]{\dot{\textbf{#1}}^{* #2}_{#3}} 
	
		\newcommand{\imdpre}[3]{\dot{#1}^{\sharp #2}_{#3}} 
	\newcommand{\ivmdpre}[3]{\dot{\textbf{#1}}^{\sharp #2}_{#3}} 
	
	\newcommand{\imd}[3]{\dot{#1}^{#2}_{#3}} 
	\newcommand{\ivmd}[3]{\dot{\textbf{#1}}^{#2}_{#3}} 
	
	\newcommand{\iii}[5]{^{#2}_{#3}{#1}^{#4}_{#5}} 
	\newcommand{\iiv}[5]{^{#2}_{#3}{\textbf{#1}}^{#4}_{#5}} 
	\newcommand{\iivn}[5]{^{#2}_{#3}{\tilde{\textbf{#1}}}^{#4}_{#5}} 
	\newcommand{\iiieq}[5]{^{#2}_{#3}{\bar{#1}}^{#4}_{#5}} 
	\newcommand{\iiieqt}[5]{^{#2}_{#3}{\tilde{#1}}^{#4}_{#5}} 
	
	\newcommand{\eqqref}[1]{Eq.~\eqref{#1}} 
	\newcommand{\fref}[1]{Fig.~\ref{#1}} 

	\maketitle
	
\section*{Abstract}

A novel geometrically exact model of the spatially curved Bernoulli-Euler beam is developed. The formulation utilizes the Frenet-Serret frame as the reference for updating the orientation of a cross section. The weak form is consistently derived and linearized, including the contributions from kinematic constraints and configuration-dependent load. The nonlinear terms with respect to the cross-sectional coordinates are strictly considered, and the obtained constitutive model is scrutinized. The main features of the formulation are invariance with respect to the rigid-body motion, path-independence, and improved accuracy for strongly curved beams.
A new reduced beam model is conceived as a special case, by omitting the rotational DOF. Although rotation-free, the reduced model includes the part of the torsional stiffness that is related to the torsion of the beam axis. This allows simulation of examples where the angle between material axes and Frenet-Serret frame is small. The applicability of the obtained isogeometric finite element is verified via a set of standard academic benchmark examples. The formulation is able to accurately model strongly curved Bernoulli-Euler beams that have well-defined Frenet-Serret frames.

\textbf{Keywords}: spatial Bernoulli-Euler beam; Frenet-Serret frame; rotation-free beam; strongly curved beam; geometrically exact analysis;

\section{Introduction}

The aim of computational mechanics is to develop accurate and efficient models of various mechanical systems. 
The most successful mechanical model for the simulation of slender bodies is \emph{beam}.
The first consistent beam theories were developed in 18th century, and the search for a formulation with optimal balance between accuracy and efficiency is still ongoing. To reduce the problem domain of slender bodies from 3D to 1D, the standard assumption is that the cross sections are rigid, which results with the Simo-Reissner (SR) beam model. By additional assumption that the cross sections remain perpendicular to the deformed axis, the Bernoulli-Euler (BE), also known as Kirchhoff-Love, beam model follows. The subject of the presented research are large deformations of an arbitrarily curved and twisted BE beam with an anisotropic solid cross section, without warping \cite{2019meier}. 

The nonlinear SR beam model has long been the main focus for researchers, partially because its spatial discretization requires only $C^0$-continuous basis functions, such as the Lagrange polynomials. As the name suggests, the SR theory was founded by Reissner \cite{1981reissnera}, and later generalized by Simo \cite{1985simo}, who conceived the term \emph{geometrically exact} beam theory. The main requirement of a geometrically exact formulation is that the relationship between the configuration and the strain is consistent with the balance laws, regardless of the magnitude of displacements and rotations. The adequate description of large rotations is one of the principal challenges since these are not additive nor commutative and constitute nonlinear manifolds. This issue has been a driving force for the formulation of various algorithms for the parameterization and interpolation of rotation \cite{1986simo, 1988cardona, 1988simo,1988iura, 1995ibrahimbegovica, 1997ibrahimbegovic, 1997crisfield}. A turning point in this development was the finding by Crisfield and Jeleni\'c that the interpolation of a rotation field between two configurations cannot preserve objectivity and path-independence \cite{1999crisfielda, 1999jelenica}. The reason is that incremental material rotation vectors, at different instances, do not belong to the same tangent space of the rotation manifold \cite{2008maekinen}. An orthogonal interpolation scheme that is independent of the vector parameterization of a rotation manifold is suggested in \cite{1999crisfielda, 1999jelenica} and several further strategies followed \cite{2009ghosh, 2004romero, 2013zupan}. 

Although the geometrically exact formulations represent the state-of-the-art in beam modeling, their implementation is not straightforward and several alternatives exist, such as the corotational and the Absolute Nodal Coordinate (ANC) approaches. The main idea of the corotational formulations is to decompose the deformation into two parts. The first part is due to large rotations and the second is the local part, measured with respect to the local co-rotated frame. It resembles the strategy employed in \cite{1999crisfielda, 1999jelenica} and allows accurate simulation of large deformations \cite{1999hsiao, 2014le,2020magisano, 2021magisano}. The ANC method is, in essence, a solid finite element for slender bodies. It is well-suited for the implementation of 3D constitutive models, but has issues with engineering structural analysis, where integration with respect to the cross-sectional area is required \cite{2008romeroa,2013gerstmayr}. 

The first BE beam models that are consistent with the geometrically exact theory are \cite{2002weiss} and \cite{2011boyer}. Meier et al. have discussed the issues of objectivity and path-independence in \cite{2014meier}, and proposed an orthogonal interpolation scheme similar to that of Crisfield an Jeleni\'c \cite{1999crisfielda, 1999jelenica}. Membrane locking, contact, and reduced models are considered in subsequent publications \cite{2015meiera, 2017meier}, followed by a comprehensive review \cite{2019meier}. An efficient BE beam formulation based on the Cartan frame was developed in \cite{2022jamunkumar}, where the position and the local frame are observed independently and subsequently related by the Lagrange multipliers. 

The emergence of the spline-based isogeometric analysis (IGA) \cite{2005hughes} has led to the development of a series of SR beam models \cite{2016marino, 2017marino, 2017weeger, 2018weeger, 2019marino, 2020vob, 2020tasora, 2019choi}. The formulation \cite{2021choi} arguably represents the state-of-the-art since it employs extensible directors and models various couplings. One of the main features of IGA is the smoothness of utilized basis functions, a property that benefits the BE beam due to its $C^1$-continuity requirement. The first IGA BE beam models were introduced by Greco et al. in \cite{2013grecoa, 2014greco, 2016greco,2015grecoa}, while the first nonlinear BE model was developed in \cite{2016bauera}. Due to the reduction in number of DOFs, in comparison with the SR model, multi-patch nonlinear analysis of BE beams has received special attention \cite{2020bauer, 2021vo, 2021grecoa, 2022greco}. Invariance of the geometric stiffness matrix in the frame of buckling analysis is considered in \cite{2021yang}, while the effect of initial curvature on the convergence properties of the solution procedure is considered in \cite{2021herath}. The first truly geometrically exact IGA BE model that preserves objectivity and path-independence was developed in \cite{2022borkovicb}.

As emphasized in this brief literature review, the crucial issues of objectivity and path-independence in the geometrically exact beam theory are related to the nonlinear nature of finite rotations. In order to obtain a generally applicable formulation, the orthogonal interpolation schemes or similar procedures must be applied \cite{2014meier}. An alternative approach is to utilize the Frenet-Serret (FS) triad as the reference frame for the update of rotation. This frame does not depend on previous configurations, and the resulting formulation is expected to be objective and path-independent. 
Although a natural choice, the FS frame is avoided for beam analysis since it is not defined for straight segments of a curve. Furthermore, the FS frame exhibits significant rotation around the curve’s tangent vector at inflection points. Due to these issues, the formulation based on the FS frame fails for arbitrary geometries. Nevertheless, the derivation, implementation and verification of such a computational model is of fundamental importance due to the intrinsic relation between the FS frame, the curve and the beam model. In this paper, we develop a formulation of this kind. Configurations for which the FS frame is not well-defined are not generally considered. An approximation of straight initial configuration will nevertheless be considered by imposing a small curvature. Regarding the inflection points, it can be shown that for a regular analytic space curve, which is not a straight line, a point with a zero-curvature is the point of analyticity of torsion \cite{1972hord} . This means that the torsional angle is defined at inflection points. However, due to the large gradients of this angle, the FS frame exhibits significant twisting and poor convergence is expected \cite{2018radenkovicb}. 

The calculation of torsion of the FS frame involves the third order derivative of the position vector, implying that a $C^2$-continuous discretization is required to obtain the torsion field. IGA allows high interelement continuities, up to $C^{p-1}$, where $p$ is the order of the basis functions. This feature makes IGA ideally suited for the implementation of the beam model based on the FS frame. This fact was utilized in \cite{2018radenkovicb} for the linear analysis of such beam model. 

There are several rotation-free formulations in the literature that model a spatial BE beam, e.g. \cite{2013raknesa} and \cite{2015meiera}. However, since they disregard the torsional stiffness, the area of application is reduced to a few specific cases in which the torsion can be neglected, such as cables \cite{2013raknesa}, or where the torsion is not present due to the specific loading conditions \cite{2015meiera}. Starting from the proposed BE model that is based on the FS frame, it is possible to obtain a specific rotation-free model that is more accurate than the existing ones. The main feature of this model is that it contains the torsion of the FS frame.

When a spatially curved beam exhibits large deformations, axial, torsional and bending actions become coupled due to the nonlinear distribution of strain along the cross section. It is common to disregard these couplings when modeling the BE beam \cite{2016bauera, 2013grecoa}. Recently, axial-bending coupling was considered in the frame of linear \cite{2018radenkovicb, 2020radenkovicb} and nonlinear analysis \cite{2022borkovic, 2022borkovicb}. The \emph{curviness} of a beam, $Kd$, is introduced in \cite{2018borkovicb, 2019borkovicb} as a measure of this coupling. Here, $K$ is the curvature of beam axis and $d$ is the maximum dimension of the cross section in the planes parallel to the osculating plane. The curviness parameter allows classification of beams as small-, medium- or big-curvature beams \cite{2010slivker}. The axial-bending coupling is significant for $Kd>0.1$, and these beams belong to the category of big-curvature, also known as \emph{strongly curved}, beams \cite{2015cazzani}. In order to apply appropriate beam models, the current curviness at each configuration must be observed \cite{2022borkovic, 2022borkovicb}. 

The present research is based on the works \cite{2017radenkovic, 2018radenkovicb, 2020radenkovicb, 2022borkovic, 2022borkovicb} with the aim of tending the linear formulation based on the FS frame to the geometrically exact setting and to improve the existing strongly curved BE model. To summarize, this work makes three main contributions. The first is the derivation of the geometrically exact FS beam formulation. It is geometrically exact in a sense that it can model arbitrary large deformations involving finite, but small, strains. The restriction is that the beam must have a well-defined FS frame during the deformation, meaning that straight segments and inflection points should not occur. It turns out that many academic examples satisfy this requirement. 
The second contribution is the consideration of a special case for the proposed formulation that is obtained by omitting the rotational DOF. In contrast to existing rotation-free BE models, this one includes one part of the torsion. The resulting rotation-free BE beam model can give approximate results for specific deformation cases. The third contribution is a rational constitutive model for strongly curved beams. The nonlinear terms of total strain with respect to the cross-sectional axes are taken into account and simplified models are deduced. This approach improves upon the strongly curved beam models considered in \cite{2022borkovic, 2022borkovicb} where only the linear terms of incremental strain were considered. 

The paper is organized as follows. The next section presents the basic relations of the beam metric and kinematic, while the strain and stress measures are defined in Section 3. The finite element formulation is elaborated in Section 4 and numerical examples are presented in Section 5. The conclusions are delivered in the last section.

\section{Configuration of Bernoulli-Euler beam}

A spatial BE beam at an arbitrary reference configuration $\mathbb{C}$ is defined by the position of its axis and the orientation of cross sections. The position of a spatial curve is a vector, while the orientation of the rigid cross section is, in general, described with the rotation tensor. 
Since we are dealing with the BE beam, the cross sections are perpendicular to the beam axis at each configuration, which leaves us to determine only the rotation in the cross-sectional plane. Therefore, the metric of deformed configuration is defined analogously to the metric of an arbitrary reference configuration. 

Boldface letters are used for the notation of vectors and tensors. An asterisk sign is used to designate a current, unknown, configuration. Greek index letters take the values of 2 and 3, while Latin ones take the values of 1, 2 and 3. Partial and covariant derivatives with respect to the $m^{th}$ coordinate of the convective frame $\left(\xi, \eta, \zeta \right)$ are designated with $(\bullet )_{,m}$ and $(\bullet )_{\vert m}$, respectively. Time derivative is marked as $\imd{(\bullet)}{}{}$. Other specific designations will be introduced as they appear in the text.

The details on the NURBS-based IGA modeling of curves are skipped for the sake of brevity since they are readily found elsewhere \cite{2005hughes, 1995piegla}.

\subsection{Metric of the beam axis}
\label{submetric0}

\begin{figure}[h]
	\includegraphics[width=\linewidth]{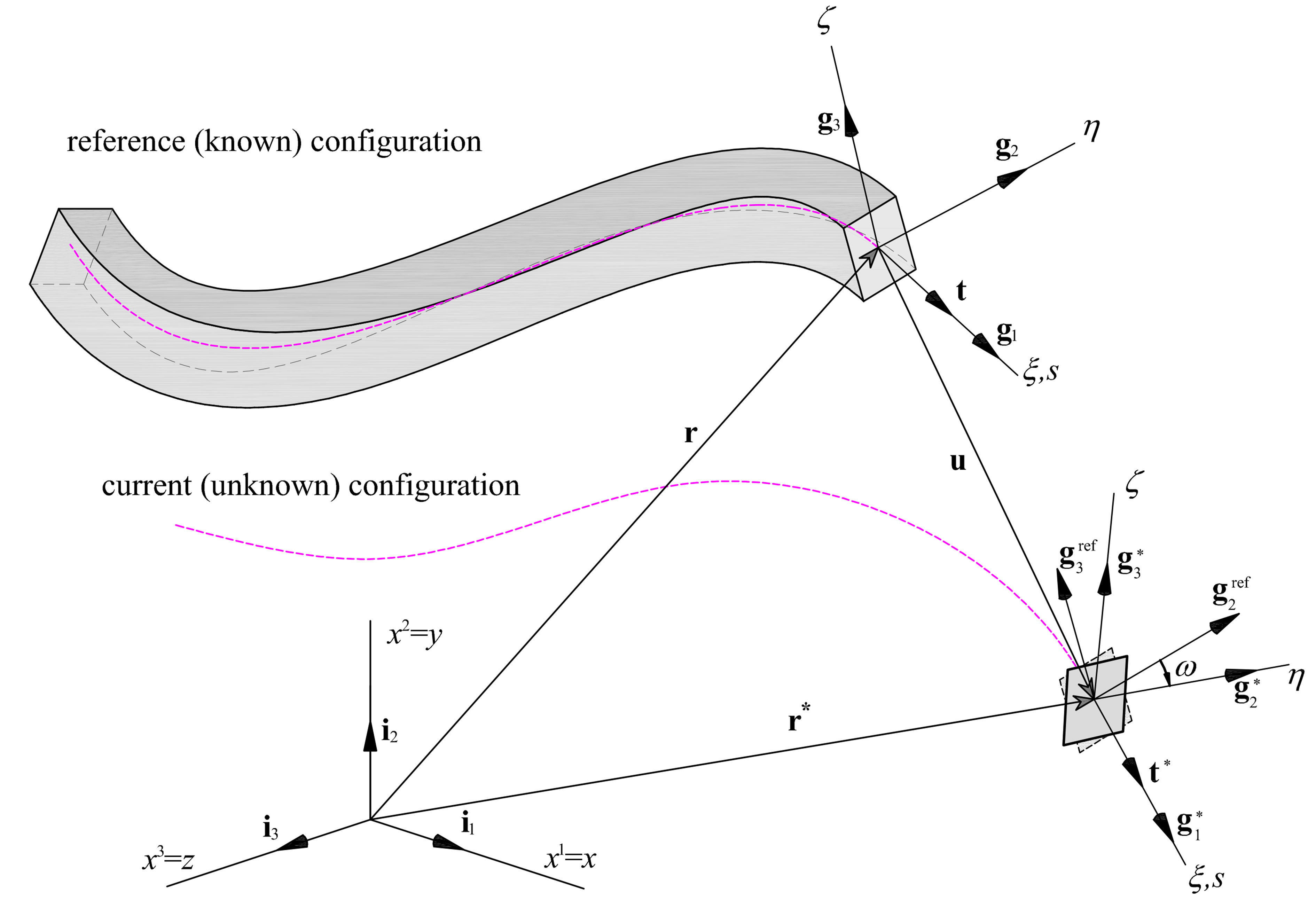}\centering
	\caption{Reference and current configurations of a spatial BE beam. A configuration is defined with the position vector of beam axis and the orientation of cross sections.}
	\label{fig:begin}
\end{figure}

The beam axis is a spatial curve, defined with its position vector:
\begin{equation}
\label{eq: 1}
 \textbf{r} = \iv{r($\xi$)}{}{} = x^m (\xi) \textbf{i}_m = x^m \textbf{i}_m, \quad (x^1=x,\; x^2=y, \;x^3=z),
\end{equation}
where $\ve{i}_m$ are the base vectors of the Cartesian coordinate system, Fig.~\ref{fig:begin}. A curve can be parameterized with either the arc-length coordinate $s \in \left[0,L\right]$ or some arbitrary parametric coordinate $\xi \in \left[0,1\right]$, where $L$ is the length.
For every $C^1$ continuous curve, we can uniquely define a tangent vector $\iv{g}{}{1}$:
\begin{equation}
\label{eq: 2}
\iv{g}{}{1} = \iv{r}{}{,1} = \frac{\dd{\iv{r}{}{}}}{\dd{\xi}} = \ii{x}{m}{,1} \iv{i}{}{m} =\frac{\dd{\iv{r}{}{}}}{\dd{s}} \frac{\dd{s}}{\dd{\xi}}=\frac{\dd{s}}{\dd{\xi}} \iv{t}{}{} = \sqrt{g} \iv{t}{}{}, \quad  \sqrt{g} = \sqrt{\iv{g}{}{1} \cdot \iv{g}{}{1}},
\end{equation}
where $\iv{t}{}{}$ is the unit-length tangent of the beam axis. 

There is an infinite set of local vector bases that can be defined to frame a curve, FS and Bishop frames being the most prominent ones. The unique feature of the FS frame is that one of its base vectors is aligned with the curvature of a line, while the Bishop frame is characterized with zero torsion \cite{1975bishop}. One approach to the BE beam modeling using the ANC formulation and the Bishop frame can be found in \cite{2021ebrahimi}.

Let us now focus on the FS triad that consists of the tangent, normal and binormal $\left(\iv{t}{}{}, \iv{n}{}{}, \iv{b}{}{} \right)$. The normal vector is the unit vector of curvature, while the binormal is perpendicular to the osculating plane and completes the orthonormal FS triad. Due to its intrinsic connection to the curvature, the FS frame cannot be defined for straight lines and has sudden changes near the inflection points. These issues are readily discussed in the context of beam formulations \cite{2014meier, 2018radenkovicb}. The derivatives of FS base vectors are defined with the well-known formulae:
\begin{equation}
\label{eq:18}
\begin{bmatrix}
\iv{t}{}{,s}\\
\iv{n}{}{,s}\\
\iv{b}{}{,s}
\end{bmatrix} = 
\begin{bmatrix}
0 & K & 0 \\
-K & 0 & \tau \\
0 & -\tau & 0
\end{bmatrix}
\begin{bmatrix}
\iv{t}{}{}\\
\iv{n}{}{}\\
\iv{b}{}{}
\end{bmatrix},
\end{equation}
where $K$ is the curvature of a line while $\tau$ is the torsion of the FS frame.

The orientation of the cross section is here defined with two base vectors of unit length, $\iv{g}{}{2}$ and $\iv{g}{}{3}$, that are aligned with the principal axes of inertia of the cross section, Fig.~\ref{fig:begin}. We will refer to the triad $\iv{g}{}{i}$ as the \emph{material vector basis}. Due to the assumption that the cross sections and the beam axis are orthogonal, the orientation of the cross sections is uniquely defined with some reference basis vectors $\iv{g}{ref}{\alpha}$ and an appropriate rotation angle. To obtain the relation between components of curvature with respect to the FS and material vector bases, let us assume that the reference basis vectors $\iv{g}{ref}{\alpha}$ are the normal and binormal. Then, by introducing the angle $\theta^{ref}$, we can define orientation of the cross section:
\begin{equation}
\label{eq:14}
\begin{bmatrix}
\iv{g}{}{2}\\
\iv{g}{}{3}
\end{bmatrix} = 
\begin{bmatrix}
\cos \theta^{ref}  & \sin \theta^{ref} \\
-\sin \theta^{ref} & \cos \theta^{ref}
\end{bmatrix}
\begin{bmatrix}
\iv{n}{}{}\\
\iv{b}{}{}
\end{bmatrix}.
\end{equation}
In this way, the material vector basis is completed, and its metric and reciprocal metric tensors are:
\begin{equation}
\label{eq:4}
\ii{g}{}{ij}=
\begin{bmatrix}
\ii{g}{}{11} & 0 & 0\\
0 & 1 & 0 \\
0 & 0 & 1
\end{bmatrix}, \quad 
\ii{g}{ij}{}=
\begin{bmatrix}
\ii{g}{11}{} & 0 & 0\\
0 & 1 & 0 \\
0 & 0 & 1
\end{bmatrix} ,
\quad \det(\ii{g}{}{ij}) = \ii{g}{}{11} = g=\iv{g}{}{1} \cdot \iv{g}{}{1}, \quad \ii{g}{11}{} = \frac{1}{\ii{g}{}{11}}.
\end{equation}
Furthermore, the derivatives of 
material vector basis are:
\begin{equation}
\label{eq: def: derivatives of base vectors}
\textbf{g}_{i,1} =   \Gamma^{k}_{i1} \textbf{g}_k, \quad
\begin{bmatrix}
\iv{g}{}{1,1}\\
\iv{g}{}{2,1}\\
\iv{g}{}{3,1}
\end{bmatrix}
=
\begin{bmatrix}
\iv{$\Gamma$}{1}{11} & \ic{K}{}{3} & -\ic{K}{}{2}\\
-\ii{K}{}{3} & 0 & \ii{K}{}{1} \\
\ii{K}{}{2} & -\ii{K}{}{1} & 0
\end{bmatrix}
\begin{bmatrix}
\iv{g}{}{1}\\
\iv{g}{}{2}\\
\iv{g}{}{3}
\end{bmatrix},
\end{equation}
where $\Gamma^{k}_{i1}$ are the Christoffel symbols of the second kind, 
while $\ii{K}{}{i}$ are the components of the curvature vector with respect to the reciprocal basis $\iv{g}{i}{}$. In concrete terms, $\ii{K}{}{1}$ is the torsion of the material vector basis, while $\ii{K}{}{\alpha}$ are the appropriate components of curvature. Due to an arbitrary parametrization of the axis, $\ic{K}{}{\alpha} = g \ii{K}{}{\alpha}$ are the components of curvature measured with respect to the parametric coordinate:
\begin{equation}
\label{eq: 25}
\ii{K}{}{1} = \iv{g}{}{2,1}\cdot \iv{g}{}{3} , \quad \ic{K}{}{2} = -\iv{g}{}{1,1} \cdot \iv{g}{}{3}, \quad \ic{K}{}{3} = \iv{g}{}{1,1} \cdot \iv{g}{}{2}.
\end{equation}
By using Eqs.~\eqref{eq:14}, \eqref{eq: def: derivatives of base vectors} and \eqref{eq: 25}, the relation between components of curvature of the two local vector triads are:
\begin{equation}
\label{eq:21}
\begin{aligned}
\ii{K}{1}{} &= \frac{1}{\sqrt{g}} \tau + \frac{1}{g} \ii{\theta}{ref}{,1} \implies \ii{K}{}{1} = \ii{g}{}{11} \ii{K}{1}{}, \\
\ii{K}{2}{} &= \ii{K}{}{2} = K \sin \theta^{ref}, \quad \ii{K}{3}{} = \ii{K}{}{3} = K \cos \theta^{ref}.
\end{aligned}
\end{equation}
Finally, let us note the relation between the normal and binormal, and the tangent vector $\iv{g}{}{1}$:
\begin{equation}
\label{eq:17}
\iv{n}{}{}=\frac{1}{\ic{K}{}{}} \left(\iv{g}{}{1,1} - \ii{\Gamma}{1}{11} \iv{g}{}{1}\right), \quad\iv{b}{}{}=\frac{1}{\sqrt{g}} \left(\iv{g}{}{1} \times \iv{n}{}{}\right),
\end{equation}
where $\ic{K}{}{} = Kg$ is the modulus of curvature with respect to the parametric coordinate.

\subsection{Metric of a generic point in beam continuum}
\label{submetric}

In order to reduce the problem from 3D to 1D, the metric of a beam continuum will be represented via a finite set of reference quantities.  
Let us define an \textit{equidistant line} which is a set of points for which $ \left(\eta, \zeta \right) = const$. Its position and tangent base vectors are:
\begin{equation}
\label{eq:def:r_eq}
\begin{aligned}
\veq{r} \left(\xi,\eta,\zeta \right) &= \veq{r}= \ve{r} + \eta \ve{g}_2 + \zeta \ve{g}_3, \\
\veq{g}_1 \left(\xi,\eta,\zeta \right) &= \veq{g}_1 = \iveq{r}{}{,1} = \ve{g}_1 + \eta \iv{g}{}{2,1} + \zeta \iv{g}{}{3,1},
\end{aligned}
\end{equation}
where $\ieq{(\bullet)}{}{}$ designates quantities at equidistant line. In order to enable a concise derivation, we will abuse the notation slightly by setting: $\xi^2=\xi_2=\zeta$ and $\xi^3=\xi_3=-\eta$. 
The base vectors of an equidistant line are now: 
\begin{equation}
	\label{eq:eq strain3DGG}
	\begin{aligned}
		\iveq{g}{}{1}&=\ii{g}{}{0} \iv{g}{}{1} - \zeta \ii{K}{}{1} \iv{g}{}{2} + \eta \ii{K}{}{1} \iv{g}{}{3} = \ii{g}{}{0} \iv{g}{}{1} - \xi^\alpha K_1 \iv{g}{}{\alpha}, \quad \iveq{g}{}{2} = \iv{g}{}{2}, \quad \iveq{g}{}{3} = \iv{g}{}{3}.
	\end{aligned}
\end{equation}
The quantity $g_0=1+ \xi^\alpha \ii{K}{}{\alpha}$ is sometimes referred to as the \emph{shifter} \cite{2021radenkovicb} or the \emph{curvature correction term} \cite{2003kapaniaa}. The metric tensor at an equidistant line is:
\begin{equation}
	\label{eq:metrictensor}
	\begin{aligned}
		\ieq{g}{}{ij}&=
		\begin{bmatrix}
			\ieq{g}{}{11} & -\zeta K_1 & \eta K_1\\		
			-\zeta K_1 & 1 & 0 \\
			\eta K_1 & 0 & 1
		\end{bmatrix}, \quad \det(\ieq{g}{}{ij}) = \ieq{g}{}{} = \ii{g}{2}{0} \ii{g}{}{11}, \quad
		\ieq{g}{}{11} = 
		\ii{g}{2}{0} \ii{g}{}{11} + \left(\eta^2+\zeta^2 \right) \ii{K}{2}{1},
	\end{aligned}
\end{equation}
while its reciprocal counterpart is:
\begin{equation}
	\label{eq:4GGrec}
	\begin{aligned}
		\ieq{g}{ij}{} &=\frac{1}{\ieq{g}{}{}}
		\begin{bmatrix}
			1 & \zeta K_1 & -\eta K_1\\
			\zeta K_1 & \ieq{g}{}{11} - \eta^2 \ii{K}{2}{1} & -\eta \zeta \ii{K}{2}{1} \\
			-\eta K_1 & -\eta \zeta \ii{K}{2}{1} & \ieq{g}{}{11} - \zeta^2 \ii{K}{2}{1}
		\end{bmatrix}.
	\end{aligned}
\end{equation}
These rigorous relations show that the vector basis at an equidistant line, $\iveq{g}{}{i}$, is not orthogonal, which results with the coupling of axial, bending and torsional actions. 
It is possible to analytically decouple these actions in the frame of linear analysis \cite{2018radenkovicb}, but, in the nonlinear analysis, the coupling is always present. A standard approach is to simplify the beam model by setting $g_0\rightarrow1$ and neglect higher order terms \cite{2016bauera}. Here, we will also decouple axial and torsional actions by simplifying the metric tensors. However, we will not simplify the shifter $g_0$, which will allow us to accurately model axial-bending coupling. In concrete terms, we are neglecting the higher order terms with respect to the torsional curvature and the off-diagonal terms of reciprocal metric tensor:
\begin{equation}
	\label{eq:metricsimpl}
	\ieq{g}{}{11} \approx \ii{g}{2}{0} \ii{g}{}{11}, \quad \ieq{g}{ij}{} \approx
	\begin{bmatrix}
		1\slash \ieq{g}{}{} & 0 & 0\\
		0 & 1  & 0 \\
		0 & 0 & 1
	\end{bmatrix}.
\end{equation}

\subsection{Kinematics of beam axis and cross section}

As discussed at the beginning of this section, it is assumed that a cross section is rigid and remains perpendicular to the beam axis in all configurations. The task is to find the material vector triad $\iv{g}{*}{i}$ at the current (unknown) configuration by the update of some reference configuration, Fig.~\ref{fig:begin}.

Finding the position and tangent of the beam axis at the current configuration are straightforward tasks:
\begin{equation}
\label{eq:def:r equidistant def1}
\begin{aligned}
\idef{\ve{r}}{}{} &=\idef{\ve{r}}{}{} (\xi) = \ve{r} + \iv{u}{}{},\\
\idef{\ve{g}}{}{1} &= \idef{\ve{g}}{}{1} (\xi) = \idef{\ve{r}}{}{,1} = \iv{g}{}{1} + \iv{u}{}{,1},
\end{aligned}
\end{equation}
where $\iv{u}{}{}$ is the displacement vector of the beam axis. The other two basis vectors must be found via rotation:
\begin{equation}
\label{eq:40}
\ivdef{g}{}{\alpha} = \iv{R}{}{} \iv{g}{}{\alpha} , 
\end{equation}
where $\ve{R}$ is the rotation tensor. In general, this tensor is a member of the special orthogonal group SO(3) that is a nonlinear manifold. This fact motivated the development of various strategies for the interpolation and parameterization of the rotation tensor \cite{1995geradin, 1997ibrahimbegovic}. In essence, the rotation tensor must be linearized and this is achieved via its linear tangent space so(3). In formal terms, we can choose to follow the velocity or the variation of the current configuration, and the former option is adopted here. 

Since the velocity of the beam axis is $\iv{v}{}{} = \ivmddef{r}{}{} = \ivmd{u}{}{}$, the velocity gradient with respect to the coordinate $\xi$ follows directly:
\begin{equation}
	\iv{v}{}{,1} = \ivmddef{g}{}{1} = \ivmd{u}{}{,1}.
\end{equation}
Let us adopt that $\ivdef{g}{}{\alpha}=\iv{g}{}{\alpha} + \iv{u}{}{,\alpha}$, where vectors $\iv{u}{}{,\alpha}$ represent increments of the basis vectors $\iv{g}{}{\alpha}$. We will refer to their time derivatives formally as the velocity gradients along the material axes $\eta$ and $\zeta$, and express as a function of the rotation tensor, \eqqref{eq:40}:
\begin{equation}
\label{eq:41}
\iv{v}{}{,\alpha} = \ivmd{u}{}{,\alpha} = \ivmddef{g}{}{\alpha} = \ivmd{R}{}{} \iv{g}{}{\alpha} + \iv{R}{}{} \ivmd{g}{}{\alpha} = \ivmd{R}{}{} \iv{g}{}{\alpha}.
\end{equation}
The members of the so(3) group are skew-symmetric tensors (spinors) which allow an exponential mapping of the elements of the SO(3) group \cite{2003kapaniaa}. In the case of the finite rotation tensor $\ve{R}$, an appropriate spinor $\ve{$\bm{\Phi$}}$ is the antisymmetric part of the displacement gradient, and its elements are infinitesimal rotations. 
Now, the exponential mapping $\iv{R}{}{} = \mathrm{e}^{\bm{\Phi}}$ allows us to find time derivative of the rotation tensor and to calculate the velocity gradients in \eqqref{eq:41}:
\begin{equation}
\label{eq:415}
\ivmd{R}{}{} =   \mdv{\Phi} \iv{R}{}{} \implies \iv{v}{}{,\alpha} = \mdv{\Phi} \ivdef{g}{}{\alpha}.
\end{equation}
The components of the spinor $\mdv{\Phi}$ are angular velocities:
\begin{equation}
	\label{eq:44}
	\imd{\omega}{i}{} = \frac{1}{2} \ii{\epsilon}{ijk}{} \iloc{v}{}{k \vert j} \implies \imd{\omega}{1}{}= \frac{1}{\sqrt{g^*}} \iloc{v}{}{3 \vert 2}, \;\; \imd{\omega}{2}{}=-\frac{1}{\sqrt{g^*}} \iloc{v}{}{3 \vert 1}, \;\; \imd{\omega}{3}{} = \frac{1}{\sqrt{g^*}} \iloc{v}{}{2 \vert 1},
\end{equation}
where $\epsilon^{ijk}$ is the Levi-Civita symbol and $\iloc{v}{}{k}$ are the components of velocity with respect to the local material triad $\iv{v}{}{} = \iloc{v}{}{k} \ivdef{g}{k}{}$. If we represent $\mdv{\Phi}$ via its axial vector 
 $\ivmd{$\pmb{\omega}$}{}{}= \imd{\omega}{i}{} \ivdef{g}{}{i}$, \eqqref{eq:415} reduces to, \cite{1985simo}:
\begin{equation}
\label{eq:43}
\iv{v}{}{,\alpha} = \ivmd{$\pmb{\omega}$}{}{} \cross \ivdef{g}{}{\alpha}.
\end{equation}
Note that $\iloc{v}{}{3 \vert 2} = - \iloc{v}{}{2 \vert 3}$ due to the assumption of the rigid cross sections, while the assumption of orthogonality of the cross section and beam axis gives $\iloc{v}{}{2 \vert 1} = - \iloc{v}{}{1 \vert 2}$ and $\iloc{v}{}{3 \vert 1} = - \iloc{v}{}{1 \vert 3}$.
These relations allow the representation of components $\imd{\omega}{2}{}$ and $\imd{\omega}{3}{}$ via the velocity of beam axis:
\begin{equation}
\label{eq:45}
\begin{aligned}
\imd{\omega}{2}{} &= -\frac{1}{\sqrt{g^*}} ~ \iloc{v}{}{3 \vert 1} = -\frac{1}{\sqrt{g^*}} ~\ivdef{g}{}{3} \cdot \iv{v}{}{,1},\\
\imd{\omega}{3}{} &= \frac{1}{\sqrt{g^*}} ~\iloc{v}{}{2 \vert 1} = \frac{1}{\sqrt{g^*}} ~\ivdef{g}{}{2} \cdot \iv{v}{}{,1}.
\end{aligned}
\end{equation}
Since these two components of angular velocity are not independent quantities, it follows that there is only one independent component of the angular velocity of the BE beam:
\begin{equation}
\label{eq:46}
\imd{\omega}{1}{} = \frac{1}{\sqrt{g^*}} ~ \iloc{v}{}{3 \vert 2} = \frac{1}{\sqrt{g^*}} ~ \md{\omega}.
\end{equation}
This quantity represents the angular velocity of a cross section with respect to the tangent of the beam axis, and it is often referred to as the \emph{twist velocity}. For simplicity, we will designate its physical counterpart with $\md{\omega}$. In this way, generalized coordinates of the BE beam are the components of the velocity of the beam axis and the twist component of angular velocity of the cross section. In contrast to the SR beam model, the rotation of a cross section of the BE beam belongs to the SO(2) group of in-plane rotations \cite{2014meier}. For another mathematically sound discussion on the decomposition of the BE beam rotation, reference \cite{2022greco} is recommended.

Once the current triad $\ivdef{g}{}{i}$ is found, we can define the complete metric of the current configuration by employing the expressions \eqref{eq:4}, \eqref{eq:eq strain3DGG}, \eqref{eq:metrictensor} and \eqref{eq:4GGrec}. To obtain the relationship between the reference and current configurations, we must represent quantities $\iv{v}{}{,\alpha}$ as functions of the generalized coordinates \cite{2022borkovicb}. By using Eqs. \eqref{eq:43} and \eqref{eq:45}, we find:
\begin{equation}
\label{eq:461}
\begin{aligned}
\iv{v}{}{,2} &= \mdv{\Phi} \ivdef{g}{}{2} = \ivmd{$\pmb{\omega}$}{}{} \times \ivdef{g}{}{2} = - \frac{1}{g^*} \iloc{v}{}{2 \vert 1} \ivdef{g}{}{1} + \iloc{v}{}{3 \vert 2} \ivdef{g}{}{3} = - \frac{1}{g^*} \left(\ivdef{g}{}{2} \cdot \iv{v}{}{,1} \right) \ivdef{g}{}{1} + \md{\omega} \ivdef{g}{}{3}, \\
\iv{v}{}{,3} &= \mdv{\Phi} \ivdef{g}{}{3} = \ivmd{$\pmb{\omega}$}{}{} \times \ivdef{g}{}{3} = - \frac{1}{g^*} \iloc{v}{}{3 \vert 1} \ivdef{g}{}{1} - \iloc{v}{}{3 \vert 2} \ivdef{g}{}{2} = - \frac{1}{g^*} \left(\ivdef{g}{}{3} \cdot \iv{v}{}{,1} \right) \ivdef{g}{}{1} - \md{\omega} \ivdef{g}{}{2},
\end{aligned}
\end{equation}
while their derivatives along the beam axis are:
\begin{equation}
\label{eq: gradients v21 v31}
\begin{aligned}
\iv{v}{}{,21} &= \left[ -\frac{1}{g^*} \left( \ivdef{g}{}{2} \cdot \iv{v}{}{,1} \right) \ivdef{g}{}{1} + \md{\omega}  \ivdef{g}{}{3}  \right]_{,1}\\
&= -\frac{1}{g^*} \bigg\{ \left( \ivdef{g}{}{2} \cdot \iv{v}{}{,1} \right) \ivdef{g}{}{1,1}   + \left[ \left( \idef{K}{}{1} \ivdef{g}{}{3} -\idef{K}{}{3} \ivdef{g}{}{1} \right) \cdot \iv{v}{}{,1} \right] \ivdef{g}{}{1} \\
& \;\;\;\; + \left( \ivdef{g}{}{2} \cdot \iv{v}{}{,11} \right) \ivdef{g}{}{1} -2 ~ \idef{\Gamma}{1}{11} \left( \ivdef{g}{}{2} \cdot \iv{v}{}{,1} \right) \ivdef{g}{}{1}  \bigg\} + \md{\omega} \left( \idef{K}{}{2} \ivdef{g}{}{1} - \idef{K}{}{1} \ivdef{g}{}{2}\right)  + \md{\omega}_{,1}  \ivdef{g}{}{3} ,\\
\iv{v}{}{,31} &= \left[ -\frac{1}{g^*} \left( \ivdef{g}{}{3} \cdot \iv{v}{}{,1} \right) \ivdef{g}{}{1} -  \omega \ivdef{g}{}{2} \right]_{,1} \\
&= -\frac{1}{g^*} \bigg\{   \left( \ivdef{g}{}{3} \cdot \iv{v}{}{,1} \right) \ivdef{g}{}{1,1} + \left[ \left( \idef{K}{}{2} \ivdef{g}{}{1} - \idef{K}{}{1} \ivdef{g}{}{2}\right) \cdot \iv{v}{}{,1} \right] \ivdef{g}{}{1} \\
& \;\;\;\; + \left( \ivdef{g}{}{3} \cdot \iv{v}{}{,11} \right) \ivdef{g}{}{1} - 2~ \idef{\Gamma}{1}{11} \left( \ivdef{g}{}{3} \cdot \iv{v}{}{,1} \right)\ivdef{g}{}{1} \bigg\} + \md{\omega} \left( \idef{K}{}{3} \ivdef{g}{}{1} - \idef{K}{}{1} \ivdef{g}{}{3}\right)  - \md{\omega }_{,1}  \ivdef{g}{}{2} .
\end{aligned}
\end{equation}

\subsection{Update of the local vector basis}
\label{subrot}

\begin{figure}[h]
	\includegraphics[width=9 cm]{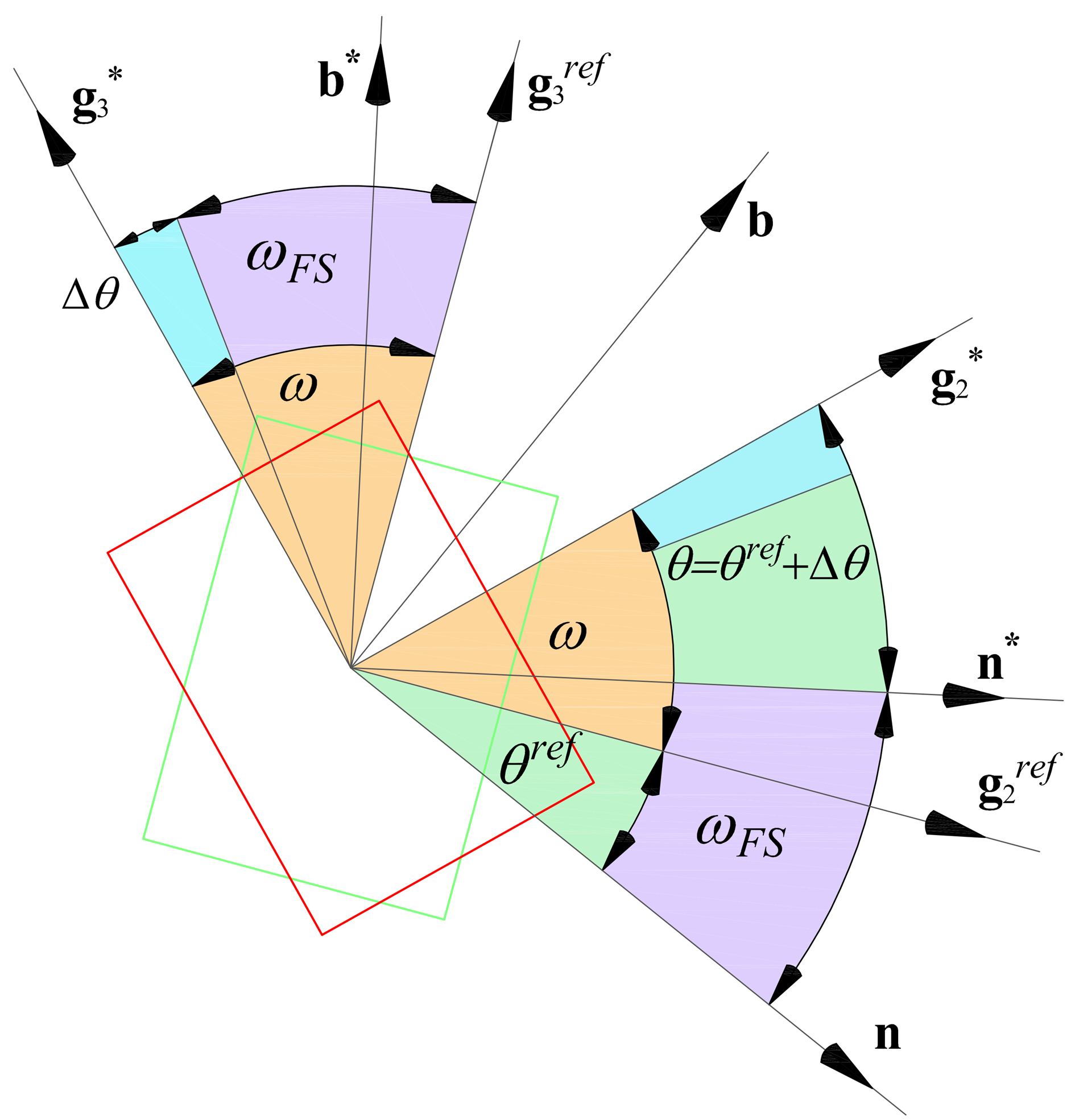}\centering
	\caption{In-plane rotation of a rectangular cross section (reference configuration - green, current configuration - red). The material basis vectors $\iv{g}{*}{\alpha}$ can be updated with respect to the reference material basis vectors $\iv{g}{ref}{\alpha}$ using the total twist angle $\omega$, or with respect to the current FS basis $\left(\iv{n}{*}{},\iv{b}{*}{}\right)$, using the independent twist angle $\theta$. Additive decomposition of the total twist angle $\omega=\omega_{FS} + \Delta \theta$ is evident. }
	\label{fig:rotations}
\end{figure}

In order to define the orientation of a cross section, the base vectors $\ivdef{g}{}{\alpha}$ of the BE beam are found through the rotation of the reference basis vectors $\iv{g}{ref}{\alpha}$ in the current cross-sectional plane, analogous to \eqqref{eq:14}, see Fig.~\ref{fig:rotations}. 
The definition of these reference basis vectors is not unique and three procedures are discussed here: the Smallest Rotation (SR), the Nodal Smallest Rotation Smallest Rotation Interpolation (NSRISR), and the Frenet-Serret Rotation (FSR).

The SR mapping defines a reference vector basis by the rotation of the triad from the reference configuration such that the tangents from both configurations align.
The distinguishing property of the SR algorithm is that this rotation angle is minimized, which gives the procedure's name \cite{1997crisfield}.
The SR procedure is readily used due to its simplicity and satisfactory accuracy \cite{2021vo}. 
However, it is solely based on the rotation between the current and reference configurations, and the interpolation of such rotations, in general, includes rigid-body motion \cite{1999crisfielda}. This error is often ignored since it mitigates with $h$-refinement \cite{2020erdelj}.

In order to overcome the deficiency of the interpolation of the rotations between the current and reference configurations, a linear interpolation of the relative rotation between the element nodes is suggested in \cite{1999crisfielda}. Since this relative rotation is free from any rigid-body motion, the objectivity of the discretized strain measures is preserved. One such algorithm is the NSRISR that is based on a specific double implementation of the SR procedure \cite{2014meier}. The first step is similar to that of the standard SR algorithm, but new triads are defined only at the start and at the end of the finite element ($\iv{g}{SR}{i, start}$, $\iv{g}{SR}{i, end}$). The reference triad $\iv{g}{ref}{i}$ is then obtained by another SR mapping, but this time via the mapping of the triad at the start of element $\iv{g}{SR}{i, start}$ along the length of finite element. The reference triad that is free from the rigid-body motion is obtained by this means. In order to compensate for the definition of the new reference frame, the twist angle DOF must be modified. The required correction angle equals the angle between the $\iv{g}{SR}{i, start}$ and $\iv{g}{ref}{i}$ triads, and it is obtained through the linear interpolation between the start and the end of the element. A weak point of the NSRISR algorithm is the requirement that the rotation field must be $C^0$ continuous \cite{2014meier, 2022borkovicb}.

Both SR and NSRISR procedures utilize material basis vectors at some reference configuration and the total twist angle $\omega$. In order to apply the update, the basis vectors $\iv{g}{}{\alpha}$ and ($\iv{n}{}{}$, $\ve{b}$) are first rotated to the current cross-sectional plane without twist, using the SR algorithm. Let us observe one such current cross-sectional plane, where the material and FS vectors in both configurations are designated, \fref{fig:rotations}. Evidently, the total twist angle can be the additively decomposed as:
\begin{equation}
	\label{eq:dec}
\omega=\omega_{FS} + \Delta \theta,
\end{equation}
where $\ii{\omega}{}{FS}$ is the twist of FS frame, while $\Delta \theta$ is the incremental change of the twist angle between the material and FS vector bases, Fig.~\ref{fig:rotations}. Note that the twist of the FS frame can be found in several ways, e.g. from $\cos\ii{\omega}{}{FS}= \ivdef{n}{}{} \cdot \ve{n}$. The decomposition \eqref{eq:dec} suggests that only the $\Delta \theta$ part of the total twist represents the independent rotation field. Following this observation, it is evident that the independent twist angle $\theta$ (or its increment $\Delta \theta$) can be introduced as the rotational DOF. The FSR formulation for the update of the local vector basis is introduced by this means. The formulation utilizes the FS basis at the current configuration as the reference vector basis. This is a straightforward approach that is frequently mentioned in the literature \cite{2014meier, 2020yang}, but, to the best of our knowledge, was never implemented in the context of the geometrically exact beam theory. The linear IGA was developed in \cite{2018radenkovicb}, 
while the nonlinear case is considered here.

\section{Stress and strain}
In this section, strain at an equidistant line is defined as a function of strains of the beam axis. Then, the strain rates are introduced and the relations between the strain rates and generalized coordinates are set. Finally, the constitutive relation is derived and section forces are defined. For convenience, we will assume that the reference configuration is the initial stress-free configuration.
 
\subsection{Strain measure}

Components of the Green-Lagrange and Almansi strain tensors are the same. 
The axial strain component at an equidistant line follows from Eqs.~\eqref{eq:metrictensor} and \eqref{eq:metricsimpl}:
\begin{equation}
	\label{eq:e11eq1}
	\begin{aligned}
\ieq{\epsilon}{}{11}&= \frac{1}{2} \left( \ieqdef{g}{}{11} - \ieq{g}{}{11} \right)= \frac{1}{2} \left( \idef{g}{2}{0}\idef{g}{}{11}  - \ii{g}{2}{0}\ii{g}{}{11}  \right) =  \frac{1}{2} \left[ \left(1+\xi^\alpha \idef{K}{}{\alpha} \right)^2 \idef{g}{}{11} - \left(1+\xi^\alpha \ii{K}{}{\alpha} \right)^2 \ii{g}{}{11} \right]\\
&= \frac{1}{2} \left[ \left(\idef{g}{}{11} - \ii{g}{}{11} \right) + 2 \xi^\alpha \left( \idef{K}{}{\alpha} \idef{g}{}{11} - \ii{K}{}{\alpha} \ii{g}{}{11} \right) + \xi^\alpha \xi^\beta \left(\idef{K}{}{\alpha} \idef{K}{}{\beta} \idef{g}{}{11} - \ii{K}{}{\alpha} \ii{K}{}{\beta} \ii{g}{}{11}\right) \right].		
	\end{aligned}
\end{equation}
Recalling that for the definition of equidistant quantities, we abuse the notation by setting: $\xi^2=\xi_2=\zeta$ and $\xi^3=\xi_3=-\eta$. Let us introduce the axial strain of the beam axis:
\begin{equation}
	\label{eq:e11}
	\begin{aligned}
		\ii{\epsilon}{}{11}=  \frac{1}{2} \left(\idef{g}{}{11} - \ii{g}{}{11} \right),
	\end{aligned}
\end{equation}
and the changes of bending curvatures of beam axis with respect to the parametric convective coordinate:
\begin{equation}
	\label{eq:kapa}
	\begin{aligned}
		\ii{\kappa}{}{\alpha}&=  \idef{K}{}{\alpha} \idef{g}{}{11} - \ii{K}{}{\alpha} \ii{g}{}{11} =\icdef{K}{}{\alpha} - \ic{K}{}{\alpha}. 
	\end{aligned}
\end{equation}
Furthermore, we need to define changes of curvature with respect to convective arc-length coordinate \cite{2020radenkovicb}:
\begin{equation}
	\label{eq:hi}
	\ii{\chi}{}{\alpha}=\idef{K}{}{\alpha} - \ii{K}{}{\alpha},
\end{equation}
where these two measures of curvature change of the beam axis relate as: 
\begin{equation}
	\label{eq:kapahi}
	\ii{\kappa}{}{\alpha} = \ii{\chi}{}{\alpha} \idef{g}{}{11}+ 2 \ii{\epsilon}{}{11} K_\alpha.
\end{equation}
Evidently, $\kappa_{\alpha}$ and $\chi_\alpha$ differ due to the parameterization, the change of length of the beam axis and the initial curvature. Let us now rewrite the last term in parentheses of \eqqref{eq:e11eq1} as:

\begin{equation}
	\label{eq:eq strain plane 23D}
	\begin{aligned}
\idef{K}{}{\alpha} \idef{K}{}{\beta} \idef{g}{}{11} - \ii{K}{}{\alpha}\ii{K}{}{\beta} \ii{g}{}{11} &= \left(\ii{K}{}{\alpha} +\ii{\chi}{}{\alpha} \right) \idef{K}{}{\beta} \idef{g}{}{11} - \ii{K}{}{\alpha} \ii{K}{}{\beta} \ii{g}{}{11} = \ii{K}{}{\alpha} \ii{\kappa}{}{\beta} + \ii{\chi}{}{\alpha} \idef{K}{}{\beta} \idef{g}{}{11}\\
		&= \ii{K}{}{\alpha} \ii{\kappa}{}{\beta} + \left(\ii{\kappa}{}{\alpha}-2 \ii{K}{}{\alpha} \ii{\epsilon}{}{11}\right) \left(\ii{K}{}{\beta} + \ii{\chi}{}{\beta}\right) \\
		&= 2 \ii{K}{}{\alpha} \left( \ii{\kappa}{}{\beta} -\ii{K}{}{\beta} \ii{\epsilon}{}{11} -\ii{\chi}{}{\beta} \ii{\epsilon}{}{11} \right) + \ii{\kappa}{}{\alpha}\ii{\chi}{}{\beta}.
	\end{aligned}
\end{equation}
By inserting Eqs.~\eqref{eq:eq strain plane 23D}, \eqref{eq:kapa}, and \eqref{eq:e11} into \eqqref{eq:e11eq1}, we obtain:
\begin{equation}
	\label{eq:e11eq}
	\begin{aligned}
		\ieq{\epsilon}{}{11}&= \ii{\epsilon}{}{11} + \xi^\alpha \kappa_\alpha +\xi^\alpha \xi^\beta \left(K_\alpha \kappa_\beta + \frac{1}{2} \chi_\beta \kappa_\alpha - \ii{\epsilon}{}{11} K_\alpha K_\beta- \ii{\epsilon}{}{11} \chi_\beta K_\alpha\right) \\
		&=g_0 \left[\left(1-\xi^\alpha K_\alpha \right) \ii{\epsilon}{}{11}+ \xi^\alpha \kappa_\alpha\right] + \xi^\alpha \xi^\beta \chi_\beta \left(\frac{1}{2} \kappa_\alpha - K_\alpha \ii{\epsilon}{}{11} \right).	
	\end{aligned}
\end{equation}
The first term of this expression corresponds to the linear analysis \cite{2018radenkovicb}. The second term is nonlinear with respect to strains and it consists of two parts. The first is usually dominant as a product of curvature changes. The second part consists of the product of curvature change and axial strain, and it only exist for initially curved beams. The nonlinear relation between the equidistant axial strain and the changes of curvatures results with a strong coupling between bending and axial actions, even for an initially straight beam:
\begin{equation}
	\label{eq:eq strain plane 73D13}
\begin{aligned}
	\ieq{\epsilon}{}{11}	&=\ii{\epsilon}{}{11}+ \xi^\alpha \kappa_\alpha + \frac{1}{2} \xi^\alpha \xi^\beta \chi_\beta \kappa_\alpha.
\end{aligned}
\end{equation}

Relations for the equidistant shear strains due to the torsion are much simpler:
\begin{equation}
	\label{eq:eq shear strain}
	\ieq{\gamma}{}{1 \alpha} = 2 \ieq{\epsilon}{}{1\alpha}	=\ieqdef{g}{}{1 \alpha} - \ieq{g}{}{1 \alpha}  = -\xi^\alpha \kappa_1
\end{equation}
where $\kappa_1 = \idef{K}{}{1} - \ii{K}{}{1}$ is the change of the beam's torsional curvature.
With Eqs.~\eqref{eq:e11eq} and \eqref{eq:eq shear strain}, the strain field of the BE beam continuum is defined as a function of strains of the beam axis: $\ii{\epsilon}{}{11}, \kappa_1, \kappa_2$, and $\kappa_3$. We will refer to these quantities as the \emph{reference strains} of the BE beam. 

\subsection{Strain rate}

The strain rates are required, since we will be using the equation of virtual power. They follow as the time derivatives of strain components $\ieqmd{\epsilon}{}{ij}=\ieq{d}{}{ij}$. The rate of axial strain at an equidistant line is:
\begin{equation}
	\label{eq:eq strain rate1}
	\begin{aligned}
		\ieq{d}{}{11}&= \frac{1}{2} \ieqmddef{g}{}{11}  = \idef{g}{}{0} \left(\imddef{g}{}{0} \idef{g}{}{11} + \idef{g}{}{0} d_{11} \right) =\idef{g}{}{0} \left[\xi^\alpha \imddef{K}{}{\alpha} \idef{g}{}{11} + \left(1+\xi^\alpha \idef{K}{}{\alpha}\right) d_{11} \right] \\
		&=\idef{g}{}{0} \left[\xi^\alpha \left(\imd{\kappa}{}{\alpha} - 2 \xi^\alpha \idef{K}{}{\alpha}\right) + \left(1+\xi^\alpha \idef{K}{}{\alpha}\right) d_{11} \right] = \idef{g}{}{0} \left[ \left(1-\xi^\alpha \idef{K}{}{\alpha}\right) d_{11} + \xi^\alpha \imd{\kappa}{}{\alpha}\right],
	\end{aligned}
\end{equation}
where $\ii{d}{}{11}$ and $\imd{\kappa}{}{\alpha}$ are the respective rates of axial strain and curvature changes of beam axis. The obtained equidistant strain rate, $\ieq{d}{}{11}$, is analogous to the linear part of the strain in \eqqref{eq:e11eq}, but expressed with respect to the current configuration. Additionally, the rates of equidistant shear strains are:
\begin{equation}
	\label{eq:shearrate}
	\ieq{d}{}{1\alpha}=\ieqmd{\gamma}{}{1 \alpha} = 2 \ieqmd{\epsilon}{}{1\alpha}	=\ieqmddef{g}{}{1 \alpha}  = -\xi^\alpha \imd{\kappa}{}{1},
\end{equation}
where $\imd{\kappa}{}{1}$ is the rate of change of torsional curvature.

\subsection{Reference strain rates and generalized coordinates}
\label{subsrefstrrates}

The rates of reference strains allow us to find the strain rates at every point of the beam continuum. The relation between the rate of reference  axial strain and generalized coordinates is simple:
\begin{equation}
	\label{eq: 99}
	\ii{d}{}{11} = \frac{1}{2} \imddef{g}{}{11} = \ivdef{g}{}{1} \cdot \iv{v}{}{,1},
\end{equation}
while the rates of the curvature components are more involved. They follow from \eqqref{eq: 25}:
\begin{equation}
	\label{eq: 991}
	\begin{aligned}
		\imd{\kappa}{}{1} &= \imddef{K}{}{1} = \iv{v}{}{,21} \cdot \ivdef{g}{}{3} + \ivdef{g}{}{2,1} \cdot \iv{v}{}{,3}, \\
		\imd{\kappa}{}{2} &= \icmddef{K}{}{2} = -\iv{v}{}{,11} \cdot \ivdef{g}{}{3} - \ivdef{g}{}{1,1} \cdot \iv{v}{}{,3},  \\
		\imd{\kappa}{}{3} &= \icmddef{K}{}{3} = \iv{v}{}{,11} \cdot \ivdef{g}{}{2} + \ivdef{g}{}{1,1} \cdot \iv{v}{}{,2},
	\end{aligned}
\end{equation}
and by inserting Eqs.~\eqref{eq:461} and \eqref{eq: gradients v21 v31} into \eqqref{eq: 991}, the relations between the rates of curvatures and generalized coordinates become:
\begin{equation}
	\label{eq: rs1}
	\begin{aligned}
		\imd{\kappa}{}{1} &= \idef{K}{}{2} \left( \ivdef{g}{}{2} \cdot \iv{v}{}{,1} \right) + \idef{K}{}{3} \left( \ivdef{g}{}{3} \cdot \iv{v}{}{,1} \right) + \ii{\md{\omega}}{}{,1}, \\
		\imd{\kappa}{}{2} &=  -\ivdef{g}{}{3} \cdot \left( \iv{v}{}{,11} - \idef{\Gamma}{1}{11} \iv{v}{}{,1}\right) + \icdef{K}{}{3} \md{\omega},  \\
		\imd{\kappa}{}{3} &= \ivdef{g}{}{2} \cdot \left( \iv{v}{}{,11} - \idef{\Gamma}{1}{11} \iv{v}{}{,1}\right) - \icdef{K}{}{2} \md{\omega}. 
	\end{aligned}
\end{equation}
These expressions employ velocity of the beam axis $\ve{v}$ and the total twist velocity of the cross section $\md{\omega}$ as DOFs, which is a standard approach for the BE beam \cite{2022borkovicb}. 

In order to derive equations of the FSR formulation, we must additively decompose the total twist angular velocity $\md{\omega}$, analogously to \eqqref{eq:dec}:
\begin{equation}
	\label{eq: rs21}
	\md{\omega}=\md{\omega}_{FS} + \md{\theta},
\end{equation}
where $\md{\omega}_{FS}$ is the twist angular velocity of the normal plane that is a function of velocity of the beam axis. This quantity is derived by the linearization of the normal and binormal in \cite{2018radenkovicb}:
\begin{equation}
	\label{eq: rs2}
	\md{\omega}_{FS} = \iloc{v}{}{\bar{3} \vert \bar{2}} = \ivdef{b}{}{} \cdot \iv{v}{}{,\bar{2}} = - \iloc{v}{}{\bar{2} \vert \bar{3}} = - \ivdef{n}{}{} \cdot \iv{v}{}{,\bar{3}} = \frac{1}{\icdef{K}{}{}} \ \ivdef{b}{}{} \cdot \left(\iv{v}{}{,11} - \idef{\Gamma}{1}{11} \iv{v}{}{,1}\right).
\end{equation}
Here, we have used indices with overbars $(\bullet )_{\bar{\alpha},\bar{\beta}}$ to distinguish the components and derivatives with respect to the axes of the FS frame from those with respect to the axes of material basis $\iv{g}{}{\alpha}$. Furthermore, $\md{\theta}$ in \eqqref{eq: rs21} represents the part of total angular velocity that is independent of the velocity of the beam axis, and we will refer to it as the \emph{independent twist angular velocity}.  

By introducing Eqs.~\eqref{eq: rs21} and \eqref{eq: rs2} into \eqqref{eq: rs1}, we obtain:
\begin{equation}
	\label{eq: new rs}
	\begin{aligned}
		\imd{\kappa}{}{1} &= \ivdef{T}{}{1}\cdot \iv{v}{}{,1} - \ivdef{T}{}{2}\cdot \iv{v}{}{,11} +\ivdef{T}{}{3}\cdot \iv{v}{}{,111} + \imd{\theta}{}{,1}, \\
		\imd{\kappa}{}{2} &= \ivdef{n}{}{} \cdot \left( \iv{v}{}{,11} - \idef{\Gamma}{1}{11} \iv{v}{}{,1}\right) \sin \idef{\theta}{}{} + \icdef{K}{}{3} \imd{\theta}{}{}, \\
		\imd{\kappa}{}{3} &= \ivdef{n}{}{} \cdot \left( \ivdef{v}{}{,11} - \idef{\Gamma}{1}{11} \iv{v}{}{,1}\right) \cos \idef{\theta}{}{} - \icdef{K}{}{2} \imd{\theta}{}{}.
	\end{aligned}
\end{equation}
where:
\begin{equation}
	\label{eq: new rs1}
	\begin{aligned}
		\iv{T}{}{1} &= \frac{1}{\ic{K}{}{}} \left[ \ii{\Gamma}{1}{11} \ieq{\tau}{}{} \iv{n}{}{} + \left( \ii{\Gamma}{1}{11} \frac{\ic{K}{}{,1}}{\ic{K}{}{}} - \ii{\Gamma}{1}{11,1} + \ii{K}{}{} \ic{K}{}{} \right) \iv{b}{}{} \right], \\
		\iv{T}{}{2} &= \frac{1}{\ic{K}{}{}} \left[ \ieq{\tau}{}{} \iv{n}{}{} + \left( \ii{\Gamma}{1}{11} + \frac{\ic{K}{}{,1}}{\ic{K}{}{}} \right) \iv{b}{}{} \right], \\
		\iv{T}{}{3} &= \frac{1}{\ic{K}{}{}} \iv{b}{}{}.
	\end{aligned}
\end{equation}
With the decomposition of the total angular velocity, the reference strain rates are represented as a function of the velocity of the axis $\ve{v}$ and the independent twist velocity $\md{\theta}$. This approach comes at the cost of having to introduce the third order derivatives of the velocity of the beam axis, which will require at least $C^2$-continuous spatial discretization. We should note that \eqqref{eq: new rs} can be derived directly from \eqqref{eq:21}, as in \cite{2018radenkovicb}. A derivation of this kind does not require the introduction and decomposition of the total twist velocity, and $\imd{\theta}{}{}$ naturally follows as a generalized coordinate.

Note that we have omitted the configuration designation (asterisk sign) in definition \eqref{eq: new rs1}. Such notation entails that a vector $\iv{P}{t}{}$, in a configuration $\mathbb{C}^t$, is expressed as a function of the geometric quantities (i.e.~$\ii{K}{t}{}$, $\ii{\tau}{t}{}$, $\iv{g}{t}{1}$, $\iv{b}{t}{}$) in the corresponding configuration $\mathbb{C}^t$. The metric of an arbitrary configuration in Subsections \ref{submetric0} and \ref{submetric} is already defined in that manner. This notation will allow us to simplify the writing, especially after we introduce a previously calculated configuration in Section \ref{secfem}.

Next, let us define the variations of equidistant strain rates:
\begin{equation}
	\label{eq:strainratevar}
	\begin{aligned}
		\delta \ieq{d}{}{11}& = \idef{g}{}{0} \left[ \left(1-\xi^\alpha \idef{K}{}{\alpha}\right) \delta d_{11} + \xi^\alpha \delta \imd{\kappa}{}{\alpha}\right] , \quad \delta \ieq{d}{}{1\alpha}= -\xi^\alpha \delta \imd{\kappa}{}{1},
	\end{aligned}
\end{equation}
and the variations of reference strain rates:
\begin{equation}
	\label{eq: varrefstrrates}
	\begin{aligned}
      \delta \ii{d}{}{11} &= \ivdef{g}{}{1} \cdot \delta \iv{v}{}{,1},\\
      \delta	\imd{\kappa}{}{1} &= \idef{K}{}{2} \left( \ivdef{g}{}{2} \cdot  \delta\iv{v}{}{,1} \right) + \idef{K}{}{3} \left( \ivdef{g}{}{3} \cdot  \delta\iv{v}{}{,1} \right) +  \delta\ii{\md{\omega}}{}{,1}=\ivdef{T}{}{1}\cdot \delta \iv{v}{}{,1} - \ivdef{T}{}{2}\cdot \delta \iv{v}{}{,11} +\ivdef{T}{}{3}\cdot \delta \iv{v}{}{,111} + \delta \imd{\theta}{}{,1}, \\
      \delta \imd{\kappa}{}{2} &=  -\ivdef{g}{}{3} \cdot \left(  \delta\iv{v}{}{,11} - \idef{\Gamma}{1}{11}  \delta\iv{v}{}{,1}\right) + \icdef{K}{}{3}  \delta\md{\omega}=\ivdef{n}{}{} \cdot \left( \delta \iv{v}{}{,11} - \idef{\Gamma}{1}{11} \delta \iv{v}{}{,1}\right) \sin \idef{\theta}{}{} + \icdef{K}{}{3} \delta \imd{\theta}{}{},\\
      \delta \imd{\kappa}{}{3} &= \ivdef{g}{}{2} \cdot \left(  \delta\iv{v}{}{,11} - \idef{\Gamma}{1}{11} \delta\iv{v}{}{,1}\right) - \icdef{K}{}{2}  \delta\md{\omega}= \ivdef{n}{}{} \cdot \left( \delta\ivdef{v}{}{,11} - \idef{\Gamma}{1}{11} \delta\iv{v}{}{,1}\right) \cos \idef{\theta}{}{} - \icdef{K}{}{2} \delta\imd{\theta}{}{}.
	\end{aligned}
\end{equation}

\subsection{Constitutive relation}

We are considering hyperelastic St.~Venant - Kirchhoff material:
\begin{equation}
	\label{eq:stress strain}
	\ieq{S}{ij}{} = 2 \mu \ieq{\epsilon}{ij}{} + \lambda \ii{\delta}{ij}{} \ii{\epsilon}{mm}{},
\end{equation}
where $\mu$ and $\lambda$ are Lamé material parameters while $\ieq{S}{ij}{}$ are contravariant components of the second Piola-Kirchhoff stress tensor. From the conditions $\ieq{S}{22}{} = \ieq{S}{33}{} = 0$, we obtain:
\begin{equation}
	\label{eq:stress strain2}
	\ieq{S}{ij}{} = 2 \mu \left( \ieq{\epsilon}{ij}{} + \nu \ii{\delta}{ij}{} \ieq{\epsilon}{11}{}\right) = 2 \mu \left( \ieq{g}{ik}{} \ieq{g}{jl}{} \ieq{\epsilon}{}{kl} + \nu \ii{\delta}{ij}{} \ieq{g}{1k}{} \ieq{g}{1l}{}\ieq{\epsilon}{}{kl}\right),
\end{equation}
where $\nu$ is the Poisson's ratio. By employing the simplification of the reciprocal metric tensor in \eqqref{eq:metricsimpl}, three non-zero stress components of the BE beam are:
\begin{equation}
	\label{eq:stress strain4}
	\ieq{S}{11}{} = E \ieq{g}{11}{} \ieq{g}{11}{} \ieq{\epsilon}{}{11}  \quad \text{and} \quad \ieq{S}{1\alpha}{} = 2 \mu \ieq{g}{11}{} \ieq{g}{\alpha \alpha}{} \ieq{\epsilon}{}{1\alpha} = \mu \ieq{g}{11}{} \ieq{g}{\alpha\alpha}{} \ieq{\gamma}{}{1\alpha} \: \left(\text{no summation over }  \alpha \right), 
\end{equation}
where $E$ is the Young's modulus of elasticity.

Let us note that, for the BE beam, the relation between the components of the Cauchy stress $\ieq{\sigma}{ij}{}$ and the second Piola-Kirchhoff stress $\ieq{S}{ij}{}$ are:
\begin{equation}
	\label{eq:stress strain5}
	\ieq{\sigma}{ij}{} = \frac{\sqrt{\ieq{g}{}{}}}{\sqrt{\ieqdef{g}{}{}}} \ieq{S}{ij}{}.
\end{equation}
This relation follows from the fact that the area of the cross section does not change, and the change of the volume element is only due to the length change along the direction of the beam axis.

\subsection{Stress resultant and stress couples} 
\label{subs stress res}

The stress resultant is defined as the integral of the tractions at current configuration. For the BE beam, the stress resultant has direction of the tangent of the beam axis:
\begin{equation}
	\label{eq:stress resultants}
	\begin{aligned}
		\ve{N} &= N \ivdef{t}{}{} = \int_{A^*}^{} \iveq{$\bm{\sigma}$}{}{} \iveqdef{t}{}{} \dd{A^*} = \int_{A}^{} \ieq{\sigma}{11}{} \left(\iveqdef{g}{}{1} \otimes \iveqdef{g}{}{1} \right)\ivdef{t}{}{} \dd{A} = \ivdef{t}{}{} \int_{A}^{} \ieq{S}{11}{} \ii{g}{*}{0} \ii{g}{}{0} \sqrt{g} \sqrt{g^*}\dd{A}, \\
		N&=	\int_{A}^{} \ieq{S}{11}{} \sqrt{\ieqdef{g}{}{}} \sqrt{\ieq{g}{}{}}\dd{A} =	\int_{A}^{} \ieq{S}{11}{} \ii{g}{*}{0} \ii{g}{}{0} \sqrt{g} \sqrt{g^*}\dd{A},
	\end{aligned}
\end{equation}
where we note that $\idef{\dd{A}}{}{}=\dd{A}$ and $\iveqdef{t}{}{} = \ivdef{t}{}{}$. Here, $\iveq{$\bm{\sigma}$}{}{}$ is the Cauchy stress tensor while $N$ is the physical normal force.
Next, let us define the stress couples:
\begin{equation}
	\label{eq:stress couples}
	\begin{aligned}
		\ve{M} &=  \int_{A^*}^{}  \iv{$\pmb{\rho}$}{}{} \cross \iveq{$\bm{\sigma}$}{}{} \iveqdef{t}{}{} \dd{A^*} = \int_{A}^{} \left(\eta \ivdef{g}{}{2} + \zeta \ivdef{g}{}{3}\right) \cross  \ieq{\sigma}{k1}{} \left(\iveqdef{g}{}{k} \otimes \iveqdef{g}{}{1} \right)\ivdef{t}{}{} \dd{A} = \ii{M}{1}{} \ivdef{t}{}{} + \ii{M}{\alpha}{} \ivdef{g}{}{\alpha}, \\
		\ii{M}{1}{}&= -\int_{A}^{} \xi_\alpha \ieq{S}{\alpha 1}{} \sqrt{\ieq{g}{}{}} \dd{A}, \quad \ii{M}{\alpha}{}= \int_{A}^{} \xi^\alpha \ieq{S}{11}{} \sqrt{\ieqdef{g}{}{}} \sqrt{\ieq{g}{}{}} \dd{A},
	\end{aligned}
\end{equation}
where $M^1$ is the physical torsional moment, while $M^{\alpha}$ are the physical bending moments.

By the insertion of Eqs.~\eqref{eq:e11eq}, \eqref{eq:eq shear strain} and \eqref{eq:stress strain4} into Eqs.~\eqref{eq:stress resultants} and \eqref{eq:stress couples}, the stress resultant and stress couples can be expressed as functions of the reference strains. However, due to the presence of shifters in both initial and current configurations, $g_0$ and $g_0^*$, the exact expressions are cumbersome. Our aim is to present a simplified relation between the stress resultants and couples, and the reference strains. For this, the exact constitutive relation is approximated using the Taylor series with respect to $\xi^\alpha$ coordinates, and the higher order terms of strains and initial curvatures are neglected. In this way, the following constitutive relation is obtained:
\begin{equation}
	\label{eq:section forces}
	\begin{aligned}
		\trans{S} &= \begin{bmatrix}
			\ii{N}{}{} & \ii{M}{1}{} &\ii{M}{2}{} &\ii{M}{3}{}
		\end{bmatrix}, 
		\quad \trans{$\pmb{\epsilon}$} = \begin{bmatrix}
			\epsilon_{11} & \ii{\kappa}{}{1} & \ii{\kappa}{}{2} &\ii{\kappa}{}{3}
		\end{bmatrix}, \quad \iv{S}{}{} = \iv{D}{SF}{} \pmb{\epsilon},\\
		\iv{D}{SF}{} &= \frac{E}{g} \sqrt{\frac{g^*}{g}}\begin{bmatrix}
			A & 0 &\ii{I}{}{\zeta \zeta} c_{13} &\ii{I}{}{\eta \eta} c_{14} \\
			0 & \mu g \ii{I}{}{t}/(E\sqrt{g^*}) &0 &0\\
			\ii{I}{}{\zeta \zeta} c_{31} & 0 &\ii{I}{}{\zeta \zeta} &0\\
			\ii{I}{}{\eta \eta} c_{41} & 0 &0 &\ii{I}{}{\eta \eta}\\
		\end{bmatrix}, \\
		c_{13} &= 1.5 \chi_2 - K_2 , \quad c_{14} = 1.5 \chi_3 - K_3, \quad c_{31} = \chi_2 - 2 K_2 , \quad c_{41} = \chi_3 - 2 K_3, \\
		A &=\int_{A}^{} \dd{\eta} \dd{\zeta} , \quad \ii{I}{}{\zeta \zeta} = \int_{A}^{}  \zeta ^2 \dd{\eta} \dd{\zeta}, \quad \ii{I}{}{\eta \eta} = \int_{A}^{}  \eta ^2 \dd{\eta} \dd{\zeta}, \quad \ii{I}{}{t} = \int_{A}^{}  \left(\eta ^2 + \zeta ^2 \right) \dd{\eta} \dd{\zeta}.
	\end{aligned}
\end{equation}
Axial and bending actions are evidently coupled, which mostly affects the stress resultant $N$. For circular cross section, the term $I_{t}$ reduces to the polar moment of area. However, for all the other cross-section shapes, the term $I_{t}$ equals so-called \emph{torsional constant} which must be calculated approximately \cite{2021vo}.

\section{Finite element formulation}
\label{secfem}

In this Section, the equation of motion of isogeometric spatial BE element based on the FS frame is derived, spatially discretized, and linearized. 

\subsection{Principle of virtual power}

The principle of virtual power represents the weak form of the equilibrium. It states that at any instance of time, the total virtual power of the external, internal and inertial forces is zero for any admissible virtual state of motion. If the inertial effects are neglected and the loads are applied with respect to the beam axis, the equation of motion is: 
\begin{equation}
\label{eq:virtual power}
\delta \ii{P}{}{} =  \delta \ii{P}{}{int}+\delta \ii{P}{}{ext}=\int_{V^*}^{} \iveq{\bm{$\sigma$}}{*}{} : \delta \iveq{d}{}{} \dd{\ieq{V}{*}{}} - \int_{\xi}^{} \left( \iv{p}{*}{} \cdot \delta \ve{v} + \iv{m}{*}{} \cdot \delta \imd{\bm{\omega}}{}{} \right) \sqrt{g^*} \dd{\xi}=0,
\end{equation}
where $\ve{d}$ is the strain rate tensor, while $\ve{p}$ and $\ve{m}$ are the vectors of external distributed line forces and moments, respectively. All these quantities are defined at the current, unknown, configuration $\mathbb{C}^*$. 

Equation \eqref{eq:virtual power} is nonlinear, and we will solve it by Newton's method which requires the linearization of the equation. Assuming that the external load is configuration-independent, only the internal virtual power must be linearized:
\begin{equation}
\label{eq:linearization of stress}
\begin{aligned}
\bm{L} \; \iveq{\bm{$\sigma$}}{*}{} &= \iveq{\bm{$\sigma$}}{\sharp}{} +\ieqmd{\bm{\sigma}}{}{} \Delta \ii{t}{}{}, \\ 
\bm{L} \; \delta \iveq{d}{}{} &= \delta \iveq{d}{\sharp}{} +\Delta \delta \iveq{d}{}{}, \\
\bm{L} \; \left( \iveq{\bm{$\sigma$}}{*}{} \delta \iveq{d}{}{} \right) &= \iveq{\bm{$\sigma$}}{\sharp}{} \delta \iveq{d}{\sharp}{}  + \ieqmd{\bm{\sigma}}{}{} \delta \iveq{d}{\sharp}{} \Delta \ii{t}{}{} + \iveq{\bm{$\sigma$}}{\sharp}{} \Delta \delta \iveq{d}{}{}, 
\end{aligned}
\end{equation}
where $\bm{L}$ marks the linearization, while $\left(\bullet \right)^\sharp$ designates sharp quantities, i.e., values from the previously calculated configuration $\mathbb{C}^\sharp$, which is generally not in equilibrium. $\ieqmd{\bm{\sigma}}{}{}$ is the stress rate tensor which is calculated as the Lie derivative of current stress. Since the components of the stress rate tensor are equal to the material time derivatives of the components of the stress tensor, \cite{2021radenkovicb}, the linearized form of the internal virtual power is:
\begin{equation}
	\label{eq:linearized virtual power}
	\begin{aligned}
	\bm{L} \; \delta P_{int} = \int_{V^*}^{} \ieq{\sigma}{\sharp 1k}{} \delta \ieq{d}{\: \sharp }{1k} 
	\dd{\ieqdef{V}{}{}}  + \int_{V^*}^{} \ieqmd{\sigma}{1k}{} \delta \ieq{d}{\: \sharp}{1k} 
	\dd{\ieqdef{V}{}{}} \Delta t + \int_{\idef{V}{}{}}^{} \ieq{\sigma}{\sharp 1k}{} \Delta \delta \ieq{d}{}{1k} \dd{\ieqdef{V}{}{}}.
\end{aligned}
\end{equation}
Using relations \eqref{eq:stress strain5} and $\dd \ieq{V}{*}{} = \sqrt{\ieqdef{g}{}{} \slash \ieq{g}{}{}} \; \dd \ieq{V}{}{}$, we can switch to the components of the second Piola-Kirchhoff stress and integrate with respect to the initial volume:
\begin{equation}
	\begin{aligned}
	\label{eq:linearized virtual power PK2}
		\bm{L} \; \delta P_{int} = \int_{V}^{} \ieq{S}{\sharp 1k}{} \delta \ieq{d}{\: \sharp }{1k} 
	\dd{\ieq{V}{}{}}  + \int_{V}^{} \ieqmd{S}{1k}{} \delta \ieq{d}{\: \sharp}{1k} 
	\dd{\ieq{V}{}{}} \Delta t + \int_{\ii{V}{}{}}^{} \ieq{S}{\sharp 1k}{} \Delta \delta \ieq{d}{}{1k} \dd{\ieq{V}{}{}}.
\end{aligned}
\end{equation}
By integrating \eqqref{eq:linearized virtual power PK2} with respect to the area of the cross section, the integrals over the 3D volume reduce to integrals along the beam axis:
\begin{equation}
	\begin{aligned}
\label{eq:from 3D to 2D}
\bm{L} \; \delta P_{int} &= \int_{\xi}^{} \left( \ic{N}{\sharp}{}  \delta \ii{d}{\: \sharp}{11} + \ic{M}{\sharp i}{} \delta \imd{\kappa}{\sharp}{i} \right) \sqrt{g}  \dd{\xi} + \int_{\xi}^{} \left( \icmd{N}{}{} \delta \ii{d}{\: \sharp}{11} + \icmd{M}{i}{} \delta \imd{\kappa}{\sharp}{i} \right) \sqrt{g} \dd{\xi} \Delta t \\
&\quad + \int_{\xi}^{} \left( \ic{N}{\sharp}{} \Delta \delta \ii{d}{}{11} + \ic{M}{\sharp i}{} \Delta \delta \imd{\kappa}{}{i}  
\right) \sqrt{g}  \dd{\xi},
\end{aligned}
\end{equation}
where $\ic{N}{}{}$ and  $\ic{M}{i}{}$ are stress resultant and stress couples 
that are energetically conjugated with the reference strain rates of the beam axis, $\ii{d}{}{11}$ and $\imd{\kappa}{}{j}$, while $\icmd{N}{}{}$ and $\icmd{M}{i}{}$ are their respective rates.
If we introduce the vectors: 
\begin{equation}
\label{eq:vectors of section forces and stran rates}
\begin{aligned}
	\trans{$\iv{f}{}{}$} &= 
	\begin{bmatrix}
		\ic{N}{}{} & \ic{M}{1}{} & \ic{M}{2}{} & \ic{M}{3}{}
	\end{bmatrix}, \quad
\trans{$\iv{e}{}{}$} = 
\begin{bmatrix}
	\ii{d}{}{11} & \imd{\kappa}{}{1} & \imd{\kappa}{}{2} & \imd{\kappa}{}{3} 
\end{bmatrix}, \\
\trans{p} &= 
\begin{bmatrix}
\ii{p}{}{1} & \ii{p}{}{2} & \ii{p}{}{3} 
\end{bmatrix}, \quad
\trans{m} = 
\begin{bmatrix}
\iloc{m}{}{1} & \iloc{m}{}{2} & \iloc{m}{}{3} 
\end{bmatrix},
\end{aligned}
\end{equation}	
linearized virtual power can be expressed in compact matrix form as:
\begin{equation}
\label{matrix form of linearized virtual power}
\int_{\xi}^{} \trans{$\iv{f}{\: \sharp}{}$} \delta \iv{e}{\sharp}{} \sqrt{g} \dd{\xi}  + \int_{\xi}^{} \transmd{f} \delta \iv{e}{\sharp}{} \sqrt{g} \dd{\xi} \Delta t + \int_{\xi}^{} \trans{$\iv{f}{\: \sharp}{}$} \Delta \delta \ve{e} \sqrt{g} \dd{\xi} = \int_{\xi}^{} \left( \trans{$\ivdef{p}{}{}$}  \delta \ve{v} + \trans{$\ivdef{m}{}{}$} \delta \imd{\bm{\omega}}{}{} \right) \sqrt{g^*} \dd{\xi}. 
\end{equation}

\subsection{Calculation of the stress rate resultant and couples}

Let us find the relation between the stress rate resultant and stress rate couples, and the reference strain rates. This relation is required for the integration of internal virtual power over the cross-sectional area, see \eqqref{eq:from 3D to 2D}:
\begin{equation}
	\label{integration of stress rate}
	\int_{A}^{} \ieqmd{S}{1k}{} \delta \ieq{d}{\: \sharp}{1k} g_0
	\dd{A}= \transmd{f} \delta \iv{e}{\sharp}{} = \icmd{N}{}{} \delta \ii{d}{\: \sharp}{11} + \icmd{M}{i}{} \delta \imd{\kappa}{\sharp}{i}.
\end{equation}
Variations of equidistant strain rates are given with \eqref{eq:strainratevar}, while the stress rates follow from \eqref{eq:stress strain4}. Similar to the stress resultant and stress couples in Subsection \ref{subs stress res}, the exact expressions are cumbersome, and an analogous approximation is made. The resulting simplified constitutive relation is symmetric, as required:
\begin{equation}
	\label{eq:constTAN}
	\begin{aligned}
		\imd{\ve{f}}{}{}  &= \begin{bmatrix}
			\icmd{N}{}{} & \icmd{M}{1}{} &\icmd{M}{2}{} &\icmd{M}{3}{}
		\end{bmatrix}^\mathsf{T}, 
		\quad \imd{\ve{f}}{}{} = \ic{\ve{D}}{\sharp M}{} \iv{e}{\sharp}{},\\
		\ic{\ve{D}}{M}{}  &= \frac{E}{g^2} \begin{bmatrix}
			A & 0 &\ii{I}{}{\zeta \zeta} \ic{a}{}{13} &\ii{I}{}{\eta \eta} \ic{a}{}{14} \\
			0 & \mu g \ii{I}{}{t}/E &0 &0\\
			\ii{I}{}{\zeta \zeta} \ic{a}{}{31} & 0 &\ii{I}{}{\zeta \zeta} &0\\
			\ii{I}{}{\eta \eta} \ic{a}{}{41} & 0 &0 &\ii{I}{}{\eta \eta}\\
		\end{bmatrix}, \\
		\ic{a}{}{13} &= \ic{a}{}{31} =  \chi_2 - 2 K_2 , \quad \ic{a}{}{14} = \ic{a}{}{41} = \chi_3 - 2 K_3,
	\end{aligned}
\end{equation}

\subsection{Calculation of internal forces}

The correct calculation of the internal forces is crucial for accurate simulations. The internal forces follow from Eqs.~\eqref{eq:stress strain4}, \eqref{eq:e11eq}, \eqref{eq:eq shear strain}, and \eqref{eq:linearized virtual power PK2}:
\begin{equation}
	\label{eq:int forces3Dx}
	\begin{aligned}
\int_{A}^{} \ieq{S}{\sharp 1k}{} \delta \ieq{d}{\: \sharp}{1k} \ii{g}{}{0}\dd{A} 
= \ic{N}{\sharp}{}\delta \ii{d}{\: \sharp}{11} +\ic{M}{\sharp i}{} \delta \imd{\kappa}{\sharp}{i}, 
	\end{aligned}
\end{equation}
and they allow the reduction given by \eqqref{eq:from 3D to 2D}. Again, after the approximation with Taylor series and by neglecting higher order terms with respect to strains and initial curvatures, the relation between internal forces and the reference strains is:
\begin{equation}
	\label{eq:constif}
		\iv{f}{\: \sharp}{} = \ic{\ve{D}}{\sharp}{} \iv{$\pmb{\epsilon}$}{\sharp}{}, \quad
		\ic{\ve{D}}{}{} = \frac{E}{g^2} \begin{bmatrix}
			A & 0 &\ii{I}{}{\zeta \zeta} a_{13} &\ii{I}{}{\eta \eta} a_{14} \\
			0 & \mu g \ii{I}{}{t}/E &0 &0\\
			\ii{I}{}{\zeta \zeta} a_{31} & 0 &\ii{I}{}{\zeta \zeta} &0\\
			\ii{I}{}{\eta \eta} a_{41} & 0 &0 &\ii{I}{}{\eta \eta}\\
		\end{bmatrix},
\end{equation}
where the coupling coefficients $a$ are:
\begin{equation}
	\label{eq:ac}	
	a_{13} = 0.5 \chi_2 - 2 K_2 , \quad a_{14} = 0.5 \chi_3 - 2 K_3, \quad a_{31} = \chi_2 - 2 K_2 , \quad a_{41} = \chi_3 - 2 K_3.
\end{equation}
In contrast to the matrix $\ic{\ve{D}}{M}{}$, the constitutive matrix $\ic{\ve{D}}{}{}$ is not symmetric, since it relates total values of stress resultant and couples, and reference strains. With the coupling coefficients of \eqqref{eq:ac}, the effect of strong curvature is correctly captured. 

Regarding the definition of the curviness parameter, it is evident from \eqqref{eq:ac} that the initial curvature has more influence on the axial-bending coupling than the change of curvature. This suggest that the current curviness parameter $\idef{K}{}{}d=\left(\chi + K \right)d$, as defined in the introduction, is not comprehensive and it can be refined. For simplicity, we will keep the standard definition in this paper.

One of our aims is to examine the influence that strong curvature has on the response of the beam. The correct constitutive relation for the calculation of internal forces is given by Eqs.~\eqref{eq:constif} and \eqref{eq:ac}, and we will designate that model with $D^{C}$. Furthermore, let us introduce two reduced constitutive models, as in \cite{2022borkovicb}. The first one is the standard decoupled model that ignores off-diagonal terms in the constitutive relation \eqref{eq:constif} - $D^0$ model. The second reduced model is obtained by setting shifters to one  $\left( g_0 \rightarrow 1, \: \ii{g}{*}{0} \rightarrow 1 \right)$ and by neglecting the nonlinear terms in \eqqref{eq:e11eq}. This approximation returns the constitutive model that restricts the change of the length of the axis due to bending. We will refer to this constitutive model as the \emph{small-curvature model} and designate it with $D^1$. The coupling coefficients in \eqqref{eq:constif} for the $D^1$ model are:
\begin{equation}
	\label{eq:D1}
	a_{13} = -(\chi_2 + K_2)=-K^*_2 , \quad a_{14} = -(\chi_3 + K_3) =-K^*_3, \quad a_{31} = - K_2 , \quad a_{41} = - K_3.
\end{equation}

\subsection{Spatial discretization}

Using IGA, both geometry and kinematics are here discretized with the same univariate NURBS functions $R_I$:
\begin{equation}
	\label{eq:def:u}
	\begin{aligned}
		\ve{r} &= \sum\limits_{I =1}^{N} R_{I} (\xi) \ve{r}_{I}, \quad \theta = \sum\limits_{I =1}^{N} R_{I} (\xi) \theta_{I},\\
		\quad \ve{v} &= \sum\limits_{I =1}^{N} R_{I} (\xi) \ve{v}_{I}, \quad  \imd{\theta}{}{} = \sum\limits_{I =1}^{N} \ii{R}{}{I} (\xi) \imd{\theta}{}{I},
	\end{aligned}
\end{equation}
where $\left(\bullet \right)_{I}$ stands for the value of quantity at the $I^{th}$ control point. 
If we introduce a vector of generalized coordinates, $\iv{v}{\mathsf{T}}{\theta}= \left[\trans{v} \; \md{\theta}\right]$, and the matrix of basis functions $ \ve{N} $, the kinematic field of the beam can be represented as:
\begin{equation}
	\iv{v}{}{\theta} = \ve{N} \ivmd{q}{}{},
\end{equation}
where:
\setcounter{MaxMatrixCols}{20}
\begin{equation}
	\label{eq:def:u via matrices2}
	\begin{aligned}
		\trans{$\ivmd{q}{}{}$} &=
		\begin{bmatrix}
			\ivmd{q}{}{1} & \ivmd{q}{}{2} & ... & \ivmd{q}{}{I} & ... & \ivmd{q}{}{N}
		\end{bmatrix}, \quad
		\ivmd{q}{}{I}=
		\begin{bmatrix}
			\ii{v}{1}{I} & \ii{v}{2}{I} & \ii{v}{3}{I} & \imd{\theta}{}{I} 
		\end{bmatrix} 
		\\
		\ve{N} &= 
		\begin{bmatrix}
			\iv{N}{}{1} & \iv{N}{}{2} & ... & \iv{N}{}{I} & ... & \iv{N}{}{N} \\
			\iv{N}{\theta}{1} & \iv{N}{\theta}{2} & ... & \iv{N}{\theta}{I} & ... & \iv{N}{\theta}{N} 			
		\end{bmatrix}, \quad
		\iv{N}{}{I} = 
		\begin{bmatrix}
			\ii{R}{}{I} & 0 & 0 & 0 \\
			0 & \ii{R}{}{I} & 0 & 0\\
			0 & 0 & \ii{R}{}{I} & 0
		\end{bmatrix}, \quad
		\trans{$\iv{N}{\theta}{I}$} = 
		\begin{bmatrix}
			\textbf{0}_{3\times1} \\ \ii{R}{}{I} 
		\end{bmatrix}.
	\end{aligned}
\end{equation}
Here, $\textbf{0}_{m\times n}$ designates the $m\times n$ zero matrix.

\subsection{Discrete equations of motion}

To define spatially discretized equations of motion, we must relate the reference strain rates with the generalized coordinates at the control points by using Eqs.~\eqref{eq: 99}, \eqref{eq: new rs1} and \eqref{eq:def:u via matrices2}:
\begin{equation}
	\label{eq: vector of reference strains matrix form}
	\ve{e} = \iv{B}{*}{L} \ivmd{q}{}{}= \ve{H}^* \ve{B} \ivmd{q}{}{},
\end{equation}
where:
\setcounter{MaxMatrixCols}{20}
\begin{equation}
	\label{eq: vector of reference strains matrix form2}
	\begin{aligned}
		\ve{B} &= 
		\begin{bmatrix}
			\iv{B}{}{1} & \iv{B}{}{2} & ... & \iv{B}{}{I} & ... & \iv{B}{}{N} 
		\end{bmatrix}, \quad
		\trans{$\iv{B}{}{I} $}=
		\begin{bmatrix}
			\trans{$\iv{N}{}{I,1}$}  &
			\trans{$\iv{N}{}{I,11}$}  &
			\trans{$\iv{N}{}{I,111}$}  &
			\trans{$\iv{N}{\theta}{I}$}  &
			\trans{$\iv{N}{\theta}{I,1}$}
		\end{bmatrix}, \\
		\ve{H} &=
\begin{bmatrix}
	\trans{$\iv{g}{}{1}$} & \textbf{0}_{1\times3} & \textbf{0}_{1\times3} & 0 & 0  \\
	\trans{$\iv{T}{}{1}$} & -\trans{$\iv{T}{}{2}$} &  \trans{$\iv{T}{}{3}$} & 0 & 1\\
	-\ii{\Gamma}{1}{11} \sin \ii{\theta}{}{} \:\trans{$\iv{n}{}{}$} &  \sin \ii{\theta}{}{} \: \trans{$\iv{n}{}{}$} & \textbf{0}_{1\times3}  &  \ic{K}{}{3} & 0\\
	-\ii{\Gamma}{1}{11} \cos \ii{\theta}{}{} \: \trans{$\iv{n}{}{}$}  &  \cos \ii{\theta}{}{} \:\trans{$\iv{n}{}{}$} & \textbf{0}_{1\times3}  & -\ic{K}{}{2} & 0
\end{bmatrix}.
	\end{aligned}
\end{equation}
The variation of the vector of reference strain rates is, cf. Eqs.~\eqref{eq: varrefstrrates} and \eqref{eq: vector of reference strains matrix form}:
\begin{equation}
	\label{eq: var e}
	\delta \ve{e}^\sharp = \ve{H}^\sharp \ve{B} \delta \ivmd{q}{}{},
\end{equation}
while its linearized increment is:
\begin{equation}
	\label{eq: 1 e=Hw, BL=HB}
	\Delta \delta \ve{e} = \Delta \left( \ve{H}^* \ve{B} \delta \ivmd{q}{}{} \right) = \Delta \ve{H}^* \ve{B} \delta \ivmd{q}{}{},  
\end{equation}
where the increment of the operator $\ve{H}^*$ is quite involved, and it is given in detail in Appendix A. To continue the derivation, let us refer to the matrix of basis functions $\ve{B}_G$ and the matrix of generalized section forces $\ve{G}$ that are defined in Appendix A by Eqs.~\eqref{eq: apnestoG matrix def} and \eqref{eq: appB}. With these expressions, and Eqs.~\eqref{eq:constTAN}, \eqref{eq: var e}, and \eqref{eq: 1 e=Hw, BL=HB}, we can rewrite the integrands on the left-hand side of \eqqref{matrix form of linearized virtual power} in a spatially discretized form:
\begin{equation}
	\begin{aligned}
	\label{eq: part of vp generated by known stress and variation of strain rate}
\trans{$\iv{f}{\: \sharp}{}$} \delta \iv{e}{\sharp}{}  &= \trans{$\iv{f}{\: \sharp}{}$} \ve{H}^\sharp \ve{B} \delta \ivmd{q}{}{} =\trans{$\iv{f}{\: \sharp}{}$} \iv{B}{\sharp}{L} \delta \ivmd{q}{}{}, \\
\transmd{f} \delta \iv{e}{\sharp}{} &= \trans{$\iv{e}{\sharp}{}$} \ic{\ve{D}}{\sharp M}{} \delta \iv{e}{\sharp}{} = 
 \transmd{q} \trans{$\iv{B}{\sharp}{L}$} \ic{\ve{D}}{\sharp M}{} \iv{B}{\sharp}{L} \delta \ivmd{q}{}{} , \\
\trans{$\iv{f}{\: \sharp}{}$} \Delta \delta \ve{e}   &= \trans{$\iv{f}{\: \sharp}{}$} \Delta \ve{H}^* \ve{B} \delta \ivmd{q}{}{}  = \transmd{q} \trans{$\iv{B}{}{G}$} \iv{G}{\sharp}{} \iv{B}{}{G}  \delta \ivmd{q}{}{} \Delta t. \\
\end{aligned}
\end{equation}
Let us note that there is a virtual power that stems from the imposition of boundary conditions. These contributions are discussed in Subsection \ref{sec lag mul} and Appendix B.

Regarding the external virtual power, it can be spatially discretized via the vector of the external load, $\ve{Q}$:
\begin{equation}
	\label{eq: virtual equilibrium0}
	\int_{\xi}^{} \left( \trans{$\ivdef{p}{}{}$}  \delta \ve{v} + \trans{$\ivdef{m}{}{}$} \delta \md{\bm{\omega}} \right) \sqrt{g^*} \dd{\xi} = \trans{$\iv{Q}{*}{}$} \delta \ivmd{q}{}{},
\end{equation}
which is the same as in the linear analysis \cite{2018radenkovicb}. However, if the vector of external load depends on the configuration, it must be linearized as well. For this case, a contribution to the geometric stiffness is derived in Appendix C. 

With all these ingredients, the linearized equation of equilibrium is:
\begin{equation}
	\label{eq: virtual equilibrium}
	\transmd{q} \int_{\xi}^{} \left( \trans{$\iv{B}{\sharp}{L}$} \ic{\ve{D}}{\sharp M}{} \iv{B}{\sharp}{L} + \trans{$\iv{B}{}{G}$} \ve{G}^\sharp \iv{B}{}{G} \right) \sqrt{g} \dd{\xi} \delta \ivmd{q}{}{} \Delta t =\trans{$\ivdef{Q}{}{}$} \delta \ivmd{q}{}{} - \int_{\xi}^{} \trans{$\iv{f}{\: \sharp}{}$} \iv{B}{\sharp}{L} \sqrt{g} \dd{\xi} \delta \ivmd{q}{}{},
\end{equation}
which can be cast into the standard form:
\begin{equation}
	\label{eq:standard form of equlibrium}
	\iv{K}{\sharp}{T} \Delta \ve{q} = \iv{Q}{*}{} - \iv{F}{\sharp}{}, \quad \left( \Delta \ve{q} = \ivmd{q}{}{} \Delta t\right),
\end{equation}
where:
\begin{equation}
	\label{eq: Kt}
	\iv{K}{\sharp}{T} = \iv{K}{\sharp}{M}+\iv{K}{\sharp}{G}=\int_{\xi}^{} \trans{$\iv{B}{\sharp}{L}$} \ic{\ve{D}}{\sharp M}{} \iv{B}{\sharp}{L} \sqrt{g} \dd{\xi} + \int_{\xi}^{}  \trans{$\iv{B}{}{G}$} \ve{G}^\sharp \iv{B}{}{G}\sqrt{g} \dd{\xi}, 
\end{equation}
is the tangent stiffness matrix that consists of material and geometric parts, while:
\begin{equation}
	\label{eq: Q and F}
	\ivpre{F}{}{} = \int_{\xi}^{} \trans{$\iv{B}{\sharp}{L}$} \iv{f}{\: \sharp}{} \sqrt{g} \dd{\xi},
\end{equation}
is the vector of internal forces. The vector $\Delta \ve{q}$ in \eqqref{eq:standard form of equlibrium} contains increments of displacement and independent twist angle at control points with respect to the previous configuration. This vector allows us to update the configuration to check if the equilibrium is satisfied, that is, if the residual $\iv{$\pmb{\Psi}$}{\sharp}{}=\iv{Q}{*}{} - \iv{F}{\sharp}{}$ is less than the prescribed error tolerance. In addition to the standard Newton-Raphson method, the Arc-length method is also employed here in order to simulate responses that include load limit points.

\subsection{Imposition of boundary conditions}
\label{sec lag mul}

Kinematic boundary conditions with respect to the displacements can be implemented in a straightforward manner. Regarding the rotations, the situation is more involved. The rotation components $ \ii{\omega}{2}{}$ and $\ii{\omega}{3}{}$ are not independent quantities and their values can be imposed at section $\xi=\xi_c$ in a standard manner by using the parent-child approach for constraining the DOFs. In essence, these conditions require that the tangent at $\xi_c$ does not rotate. For $\xi_c=0$ and $\xi_c=1$, the tangent is aligned with the control polygon, and only two control points influence the rotation. The resulting constraint conditions are linear and straightforward to implement. 

On the other hand, the twist angle $\omega^1=\omega$ consists of two parts. One is the rotation of the normal plane $ \ii{\omega}{}{FS}$ which depends on the displacement of the axis, while the other is the independent twist angle $\Delta \theta$. For simplicity, let us consider the procedure required to impose homogeneous boundary condition at current configuration for $\xi=0$ , $\omega^* (0)=0$. The constraint equation is:
\begin{equation}
	\label{eq: constraint}
	c^* = \omega^*= \ii{\omega}{*}{FS}  + \Delta \ii{\theta}{}{}= 0.
\end{equation}
Since $\ii{\omega}{*}{FS} = \arccos(\ivdef{n}{}{} \cdot \ve{n})$, this constraint is nonlinear and the parent-child approach is not suitable. Therefore, we will implement the constraint \eqref{eq: constraint} via Lagrange multiplier $\lambda^*$, by requiring that:
\begin{equation}
	\label{eq:cons}
\lambda^* c^* = 0.
\end{equation}

\begin{remark}
When the current and initial FS frames are known, and the cross section is fixed, the calculation of the angle $\ii{\omega}{*}{FS}$ is trivial. However, the issue occurs if the cross section rotates arbitrarily. To find the twist of the FS frame in this case, we must rotate the FS frame from the reference to the current configuration with the SR algorithm, see Subsection \ref{subrot}. The obtained vectors are designated with ($\iv{n}{}{SR}$, $\iv{b}{}{SR}$), and the angle $\ii{\omega}{*}{FS}$ follows as $\ii{\omega}{*}{FS} = \text{sgn} \left(\ivdef{n}{}{} \cdot \iv{b}{}{SR}\right) \arccos(\ivdef{n}{}{} \cdot \ve{n}_{SR})$. 
\end{remark}

The virtual rate of the condition \eqref{eq:cons} is added to the virtual power \eqref{eq:virtual power}:
\begin{equation}
	\label{eq: vp constraint}
	\delta P + c^* \delta \md{\lambda}  + \lambda^* \delta \md{c}  =0.
\end{equation}
Since the current configuration $\mathbb{C}^*$ is unknown, the next step is to linearize the constraint with respect to the previous configuration $\mathbb{C}^\sharp$. 
Let us represent $\md{c}$ as a function of generalized coordinates, see \eqqref{eq: rs2}:
\begin{equation}
	\label{eq: vp constraint amtrix}
	\md{c} =\frac{1}{\icdef{K}{}{}} \ \ivdef{b}{}{} \cdot \left(\iv{v}{}{,11} - \idef{\Gamma}{1}{11} \iv{v}{}{,1}\right) + \md{\theta}=\ivdef{H}{}{\lambda} \iv{B}{}{\lambda} \ivmd{q}{}{\lambda} = \ivdef{K}{}{\lambda} \ivmd{q}{}{\lambda} ,
\end{equation}
where
\setcounter{MaxMatrixCols}{20}
\begin{equation}
	\label{eq:constr via matrices2}
	\begin{aligned}
		\ve{B}_\lambda &= 
			\begin{bmatrix}
				\iv{N}{}{\lambda 1,1}  & \iv{0}{}{3 \times 1} & \iv{N}{}{\lambda 2,1}  & \iv{0}{}{3 \times 1} & ... & \iv{N}{}{\lambda I,1}  &\iv{0}{}{3 \times 1}&... & \iv{N}{}{\lambda N,1} &\iv{0}{}{3 \times 1} &0\\
				\iv{N}{}{\lambda 1,11}  & \iv{0}{}{3 \times 1} & \iv{N}{}{\lambda 2,11}  & \iv{0}{}{3 \times 1} & ... & \iv{N}{}{\lambda I,11}  &\iv{0}{}{3 \times 1}&... & \iv{N}{}{\lambda N,11} &\iv{0}{}{3 \times 1} & 0\\
				\iv{0}{}{1 \times 3} & 1 &\iv{0}{}{1 \times 3} & 0 &... &\iv{0}{}{1 \times 3} & 0&...&\iv{0}{}{1 \times 3} & 0& 0
				\end{bmatrix}, \\			
			\iv{N}{}{\lambda I} &= 
				\begin{bmatrix}
					\ii{R}{}{I} & 0 & 0  \\
					0 & \ii{R}{}{I} & 0 \\
					0 & 0 & \ii{R}{}{I}
				\end{bmatrix}, \quad 
			\ve{H}_\lambda =\frac{\trans{$\ve{b}$}}{\ic{K}{}{}}
			\begin{bmatrix}
				-\ii{\Gamma}{1}{11} & 1 & 0 
			\end{bmatrix}, \quad \trans{$\ivmd{q}{}{\lambda}$} = \begin{bmatrix} \trans{$\ivmd{q}{}{}$} \; \md{\lambda} \end{bmatrix}.
	\end{aligned}
\end{equation}
Linearization of the virtual power due to the constraint gives:
\begin{equation}
	\label{eq: linearization LM}
	\begin{aligned}
		\bm{L}  \left(c^* \delta \md{\lambda}  + \lambda^* \delta \md{c} \right)= \ipre{c}{}{} \delta \imdpre{\lambda}{}{} + \ipre{c}{}{} \Delta \delta \imd{\lambda}{}{} + \delta \imdpre{\lambda}{}{} \Delta c + \ipre{\lambda}{}{} \delta \imdpre{c}{}{} + \ipre{\lambda}{}{} \Delta \delta \imd{c}{}{} + \delta \imdpre{c}{}{} \Delta \lambda,
	\end{aligned}
\end{equation}
where $\Delta \lambda = \md{\lambda} \Delta t$ and  $\Delta c = \md{c} \Delta t$. The terms of linearized virtual power in \eqqref{eq: linearization LM} can be represented in a spatially discretized form as:
\begin{equation}
	\label{eq: constraint matrixx}
	\begin{aligned}
		\ipre{c}{}{} \delta \imdpre{\lambda}{}{} + \ipre{\lambda}{}{} \delta \imdpre{c}{}{} &= \delta \imdpre{\lambda}{}{} \; \ivpre{K}{}{\lambda} \ivmd{q}{}{\lambda} +	\delta \trans{$\ivmd{q}{}{\lambda}$} \trans{$\iv{K}{\sharp}{\lambda}$} \ipre{\lambda}{}{} = \delta \trans{$\ivmd{q}{}{\lambda}$} \ivpre{F}{}{\lambda}, \\
		\ipre{c}{}{} \Delta \delta \imd{\lambda}{}{} + \ipre{\lambda}{}{} \Delta \delta \imd{c}{}{} &= 0+ 
		  \delta \trans{$\ivmd{q}{}{\lambda}$} \iv{K}{\sharp}{G\lambda} \ivmd{q}{}{\lambda} \Delta t, \\
		\delta \imdpre{\lambda}{}{} \Delta c + \delta \imdpre{c}{}{} \Delta \lambda &= \delta \imdpre{\lambda}{}{} \; \iv{K}{\sharp}{\lambda} \ivmd{q}{}{\lambda} \Delta t + \delta \trans{$\ivmd{q}{}{\lambda}$} \trans{$\iv{K}{\sharp}{\lambda}$} \md{\lambda} \Delta t,
	\end{aligned}
\end{equation}
where $\iv{K}{\sharp}{\lambda}$ is the vector that must be added to the material part of the tangent stiffness matrix as its row and column, at the position corresponding to the $\md{\lambda}$ DOF. Vector $\trans{$\iv{F}{\sharp}{\lambda}$} = \begin{bmatrix} \lambda^\sharp \iv{K}{\sharp}{\lambda} \; c^\sharp \end{bmatrix}$ represents the contribution to the internal force vector that comes from the constraint, while $\iv{K}{}{G \lambda}$ is the contribution to the geometric stiffness. Detail derivation of the matrix $\iv{K}{}{G \lambda}$ is given in Appendix B.

\section{Numerical examples}

The aim of the following numerical studies is to verify and benchmark the derived formulation. The Dirichlet boundary conditions are imposed in a well-known manner where the rotations are treated with special care, cf. Subsection \ref{sec lag mul}. The global components of external moments are applied as force couples which must be updated at each iteration \cite{2022borkovic}. Standard Gauss quadrature with $p+1$ integration points per element are applied. All the results are presented with respect to the load proportionality factor (LPF), rather than to the load intensity itself. 

By removing the independent twist angular velocity $\imd{\theta}{}{}$ from the vector of unknowns, a novel rotation-free formulation of the spatial BE beam is obtained. In contrast to existing reduced models \cite{2015meiera, 2013raknesa}, which completely neglect the twist DOF, this formulation incorporates one part of the twist velocity - $\imd{\omega}{}{FS}$. It should return better results than existing reduced models because it does not completely neglect torsional stiffness. The formulation is designated as the Frenet-Serret Rotation Twist-Free (FSR TF). Therefore, four element formulations are considered: FSR, FSR TF, SR and NSRISR. For the interpolation of kinematic quantities, the highest available interelement continuity is applied exclusively. The exception is the NSRISR formulation which employs $C^0$ continuity for the approximation of twist. 

In some examples, the error of vector $\iv{a}{}{}$ is calculated using the following relative $L^2$-error norm:
\begin{equation}
\label{eq: l2 definition}
\ii{\norm{e}}{2}{} =\frac{1}{a_{max}} \sqrt{\frac{1}{l} \int_{0}^{l} \norm{\iv{a}{}{h}-\iv{a}{}{ref}}^2 \dd{s}},
\end{equation}
where $l$ is the length of the beam and $a_{max}$ is the maximum component of the observed vector. $\iv{a}{}{h}$ represents the approximate solution, while $\iv{a}{}{ref}$ is the appropriate reference solution.

The convergence criteria for the nonlinear solvers is set with respect to the values of both displacement and force error norms, as in \cite{2022borkovicb}. 
The tolerance for these error norms is $10^{-4}$ in all examples.

\subsection{Pre-twisted circular beam}

\subsubsection{Path-independence}

\begin{figure}
	\includegraphics[width=\linewidth]{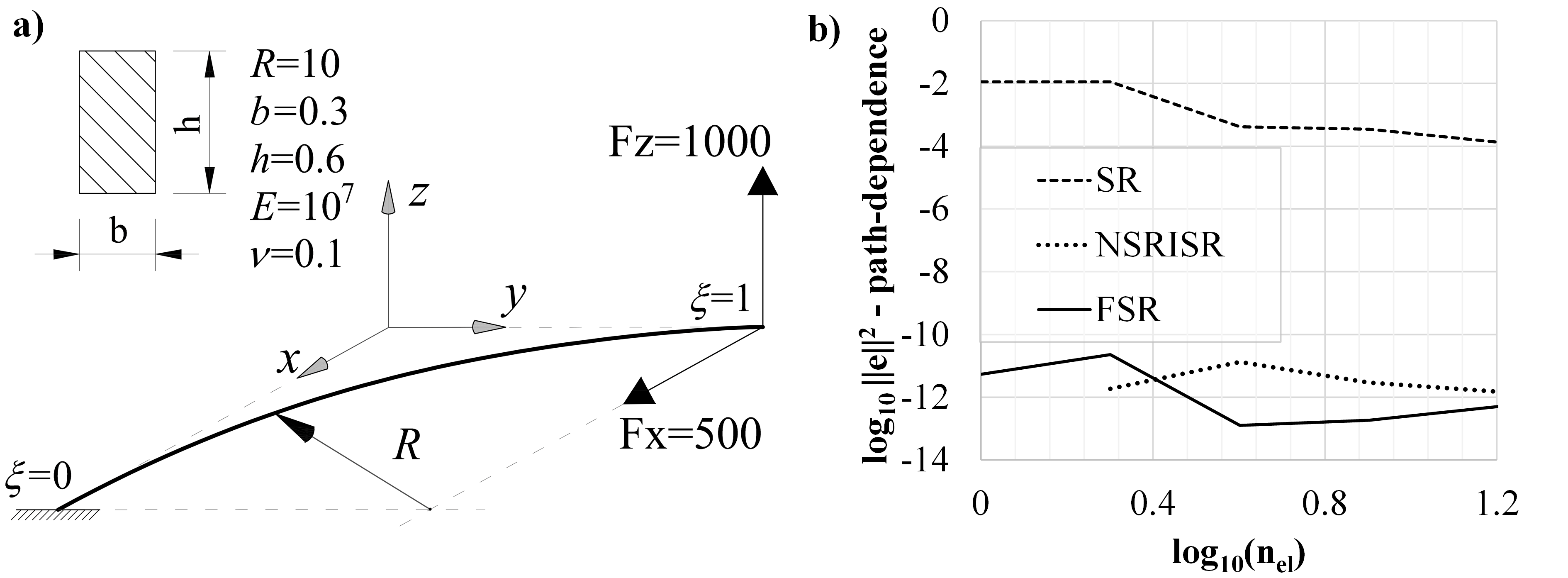}\centering
	\caption{Pre-twisted circular beam. a) Geometry and load. b) Path-dependence test for three formulations and cubic splines. Difference between SUCXZ and SIM load cases for LPF=1 vs. the number of elements.}
	\label{fig:path disposition}
\end{figure}

Path-independence of a computational formulation can be analyzed in various ways. Some authors simply apply different sizes of load increments \cite{1999jelenica}, while the others change the order of the applied load \cite{2014meier, 2022borkovicb}. Here, we employ the latter approach and analyze a quarter-circle cantilever beam loaded with two forces at the free end, as shown in Fig.~\ref{fig:path disposition}a. 
A special feature of this example is that the beam is pre-twisted with an angle of $\theta_{PT}=\pi\xi/2 $. Three cases of the application of load are considered. First, both $\iv{F}{}{X}$ and $\iv{F}{}{Z}$ are applied simultaneously - SIM case. For the other two cases, the forces are applied successively, one for 0$<$LPF$<$0.5 and the other for 0.5$<$LPF$<$1. The case when the $\iv{F}{}{X}$ is applied first is designated with SUCXZ while the other case is marked with SUCZX. For all cases, the load is applied in 20 increments.

For a path-independent solution, the final configurations must be the same, that is invariant to the load order. The difference of position between the SIM and SUCZX loadings are calculated for LPF=1 using \eqqref{eq: l2 definition}. The local vector basis is updated incrementally, with respect to the previously converged configuration. The results for different cubic NURBS meshes are shown in Fig.~\ref{fig:path disposition}b for all three formulations. The results indicate that the presented FSR formulation is indeed path-independent since the observed difference is practically zero. This is expected since there is no interpolation hidden in the history of evolution of the independent twist angle \cite{1999crisfielda}. Additionally, it is confirmed that the NSRISR formulation is path-independent, while the SR is not \cite{2022borkovicb}.

\begin{figure}
	\includegraphics[width=\linewidth]{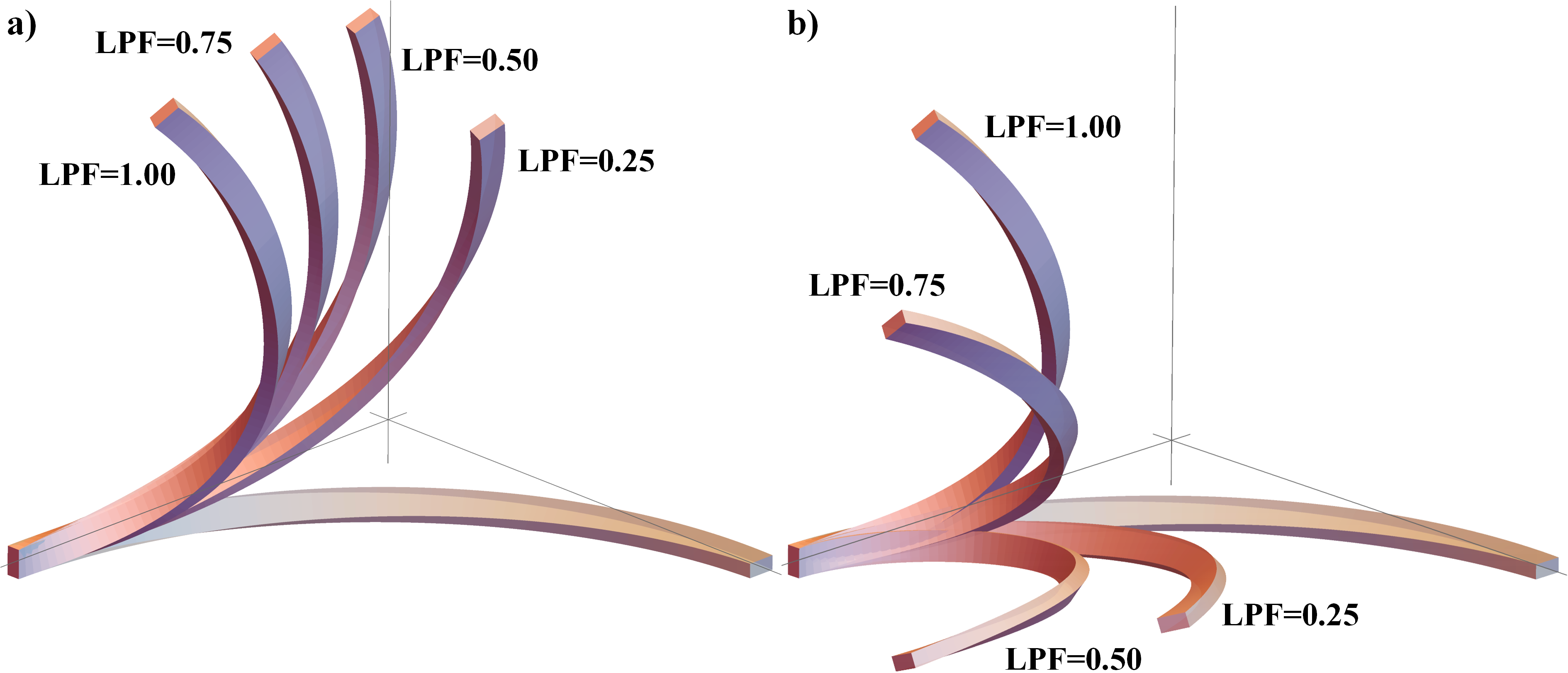}\centering
	\caption{Path-independence of a pre-twisted beam. Deformed configurations of the beam for two different loading orders and four values of LPFs: a) SUCZX, b) SUCXZ }
	\label{fig:path ind config}
\end{figure}

For visualization purposes, the deformed beam configurations for SUCXZ and SUCZX, and the four characteristic LPFs are shown in Fig.~\ref{fig:path ind config}. 
Apparently, both load cases yield similar final configurations in visual terms, but each with a different deformation history.

\subsubsection{Objectivity}

The invariance of a computational formulation with respect to the rigid-body motions is designated as objectivity. This means the structure subjected to a rigid-body motion should not be strained. Invariance with respect to the translations of beams is readily satisfied while the invariance with respect to the rotation requires special attention \cite{2012armeroa}. 

The present example is based on \cite{2014meier, 2022borkovicb} where a quarter-circular cantilever beam is rotated ten times around its clamped end with respect to the $x$-direction, see Fig.~\ref{fig:path disposition}a. For a deformation case of this kind, an objective formulation should not produce any internal strain energy. In contrast to the previous studies \cite{2014meier, 2022borkovicb}, the beam is pre-twisted here.  
 
\begin{figure}
	\includegraphics[width=\linewidth]{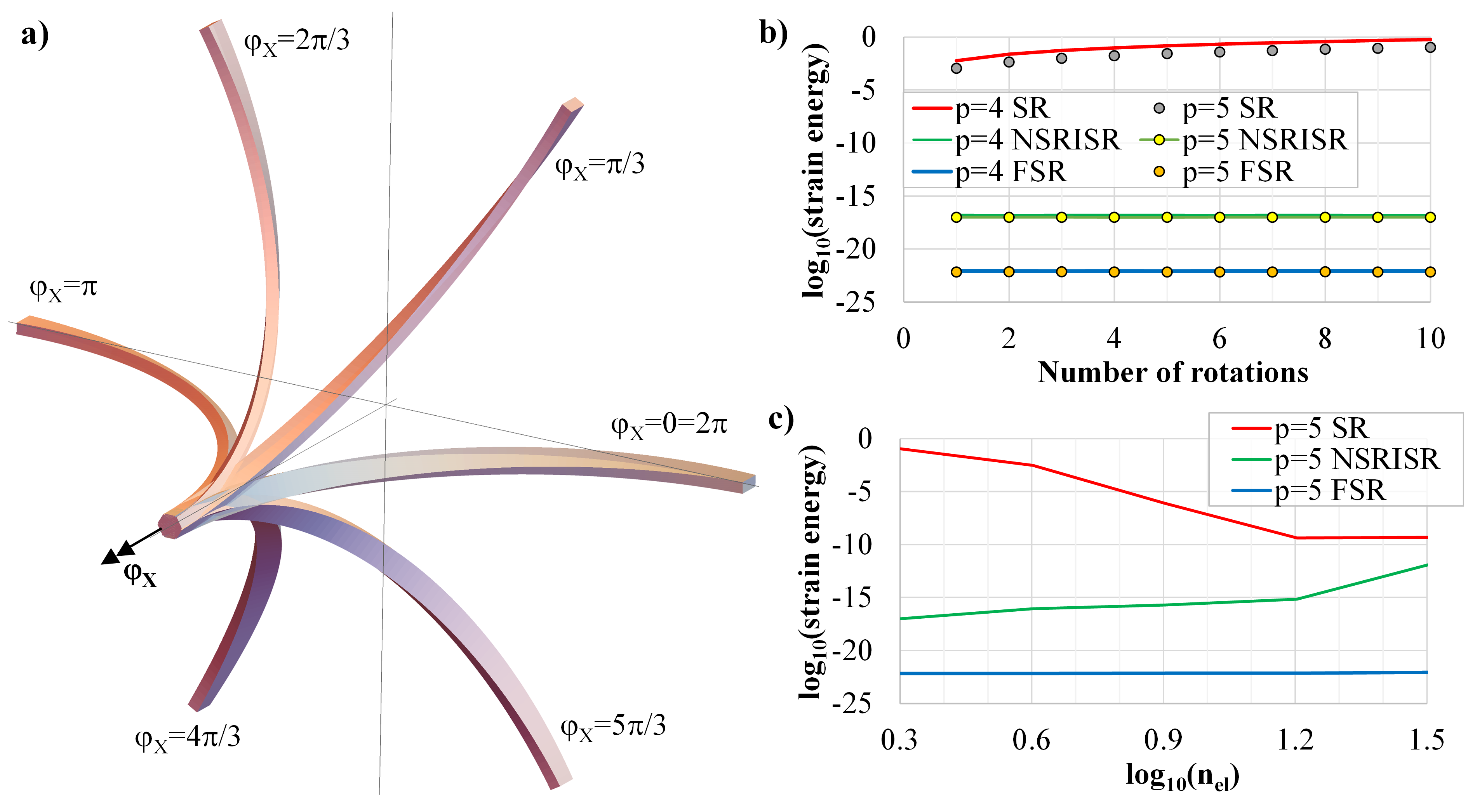}\centering
	\caption{Objectivity of a pre-twisted beam. a) Visualization of deformed configurations. b) Evolution of internal strain energy with respect to the number of rotations. c) Internal strain energy at final configuration for different formulations using quintic elements. }
	\label{fig:objectivity}
\end{figure}

The beam is discretized with two elements, and two different NURBS orders are considered, $p=4$ and $p=5$. The non-homogeneous boundary condition, $\varphi_x=20\pi$, is applied in 100 increments. Characteristic configurations during the first cycle of rotation are visualized in Fig.~\ref{fig:objectivity}a. 
The internal strain energy in the final configuration is plotted in Fig.~\ref{fig:objectivity}b with respect to the number of rotations. These results suggest that the FSR formulation is indeed objective since the internal strain energy is practically equal to zero. Additionally, it is confirmed that the NSRISR formulation is objective while the SR is not. These observations are invariant with respect to the NURBS order. The same energy is observed as a function of mesh density in Fig.~\ref{fig:objectivity}c. The results indicate that the problem with the representation of rigid-body motion mitigates for the SR formulation when the number of elements is increased. This is a well-known fact which sometimes justifies the application of non-objective formulations in quasi-static analyses \cite{2020erdelj}. Furthermore, our implementation of the NSRISR method shows an increase of the strain energy, similar to that in \cite{2022borkovicb}.  Nevertheless, if the scaling factor is included here, as in \cite{2014meier, 2022borkovicb}, the normalized internal energy would equal zero up to the machine precision. Regarding the FSR formulation, the results are completely invariant with respect to the number of rotations and are equal zero.

Let us note in passing that the rotation-free FSR TF formulation can describe rigid-body motion of this beam by default.

\subsubsection{Convergence}
\begin{figure}[h]
	\includegraphics[width=\linewidth]{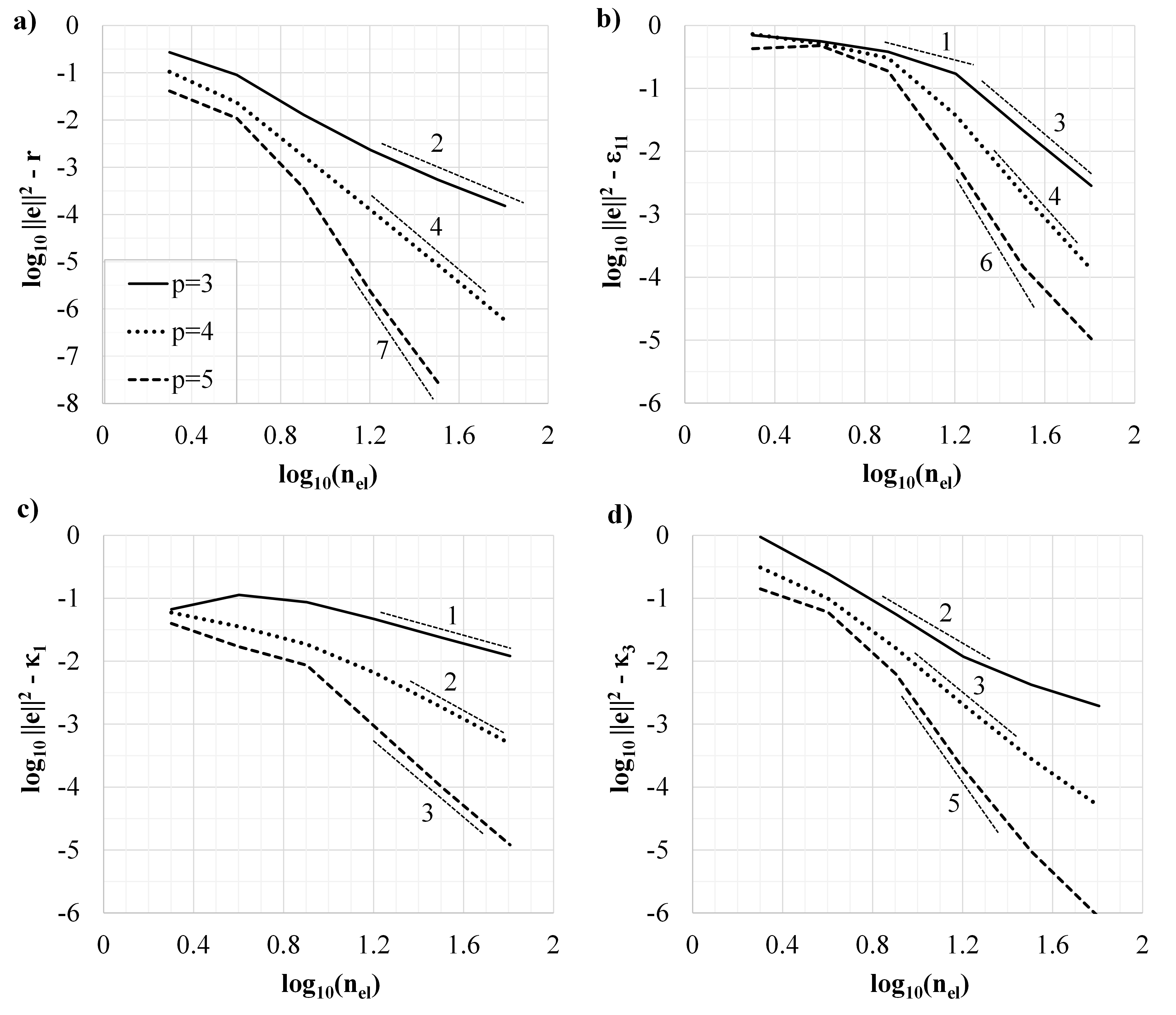}\centering
	\caption{Convergence test of a pre-twisted beam. Number of elements vs. relative $L^2$-error of: a) position, b) axial strain, c) torsional strain, d) bending strain. }
	\label{fig:path converg}
\end{figure}

Next, we examine the convergence behavior of the FSR formulation, again applied to the pre-twisted beam example. The reference solution for the SIM load case is obtained using a quintic NURBS mesh with 128 elements. The position of the axis and three reference strains are observed at the final configuration, Fig.~\ref{fig:path converg}. 
The theoretical convergence rates are $min\left[p+1,\;2\left(p-m+1\right)\right]$, where $m$ is the highest derivative appearing in the weak form \cite{2014meier}. Since $m=3$ for the FSR formulation, the expected convergence rates for the position using the cubic, quartic, and quintic NURBS are 2, 4, and 6, respectively. The obtained rates in Fig.~\ref{fig:path converg} are generally in-line with these predictions, while the quintic mesh slightly exceeds theoretical expectations. 

Next, we investigate the performance of our nonlinear solvers. The number of required increments and iterations for the convergence of five quintic meshes are shown in Table~\ref{tab:pretwisted00}. 
\begin{table}[h!]
	\begin{center}
		\caption{Number of required increments/iterations for the convergence using quintic elements. }
		\label{tab:pretwisted00}
		
		\begin{tabular}{c c c c c c c}

		&	& $n_{el}=5$ & $n_{el}=10$ & $n_{el}=20$ & $n_{el}=40$ &$n_{el}=80$\\
			\hline
 		\multirow{3}{*}{Newton-Raphson} & NSRISR	 & 29/158  & 29/158 & 28/152 & 28/152 &28/152    \\
	                           &  FSR	& 27/146  & 27/146 & 27/146 & 36/202 & 59/345     \\
	                           &  FSR TF	& 24/130  & 24/127 & 24/130 & 30/165 & 44/254     \\
	              	\hline
			
		\multirow{3}{*}{Arc-length} &NSRISR	 & 21/102  & 20/98 & 19/94 & 18/90 & 18/92   \\
									&	FSR	& 23/106  & 23/106 & 23/106 & 29/146 & 38/204  \\
									&	FSR	TF & 21/94  & 21/94 & 21/94 & 24/117 &  31/162 \\
			\hline
			
		\end{tabular}
	\end{center}
\end{table}
The first increment is applied as LPF=0.01.  
An automatic incrementation algorithm is used and the desired number of iterations per increment is set as $n_d=6$. The new increment for the Newton-Raphson solver is calculated by scaling the previous one with $n_d \slash n_c$, where $n_c$ is the number of iterations required for the convergence of previous increment. For the Arc-length procedure, the arc-length of the predictor is calculated by scaling the previous one with $\sqrt{n_d \slash n_c}$. The results suggest that our implementation of the Arc-length has superior convergence over the Newton-Raphson implementation. This is due both to the flexibility of the Arc-length method, where the increment size is varied during one load step, and also to the specific automatic incrementation setup. 

\begin{figure}[h]
	\includegraphics[width=8cm]{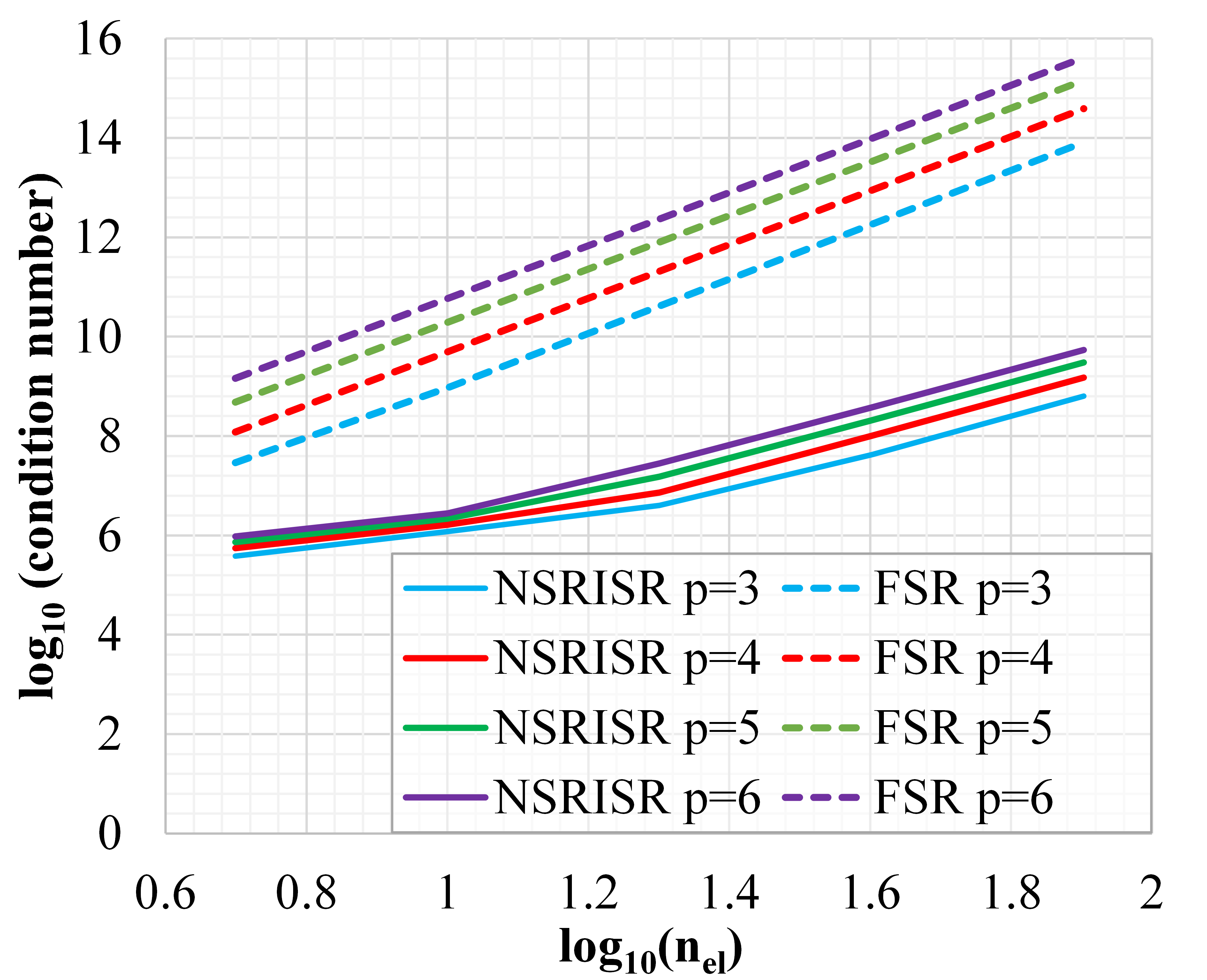}\centering
	\caption{Pre-twisted beam. The condition number of the linear stiffness matrix vs.~the number of elements. }
	\label{fig:connum}
\end{figure}

The NSRISR formulation returns the most consistent results and the convergence properties do not change significantly with the mesh density. The opposite holds for the FSR and FSR TF formulations. For $n_{el} > 20$, the required increments and iterations increase with a decrease in element size. To gain more insight, \fref{fig:connum} illustrates the condition number of the linear stiffness matrix for different meshes and NURBS orders.
It is evident that the condition number and its increase rate are significantly larger for FSR in comparison with NSRISR. These facts provide a rationale for the increase of the required increments and iterations for FSR. We attribute this behavior to the presence of the third order derivatives of basis functions in the FSR formulation. The results for FSR TF are indistinguishable from those of FSR, and they are thus omitted.

\subsubsection{Twist-free model}

\begin{figure}
	\includegraphics[width=\linewidth]{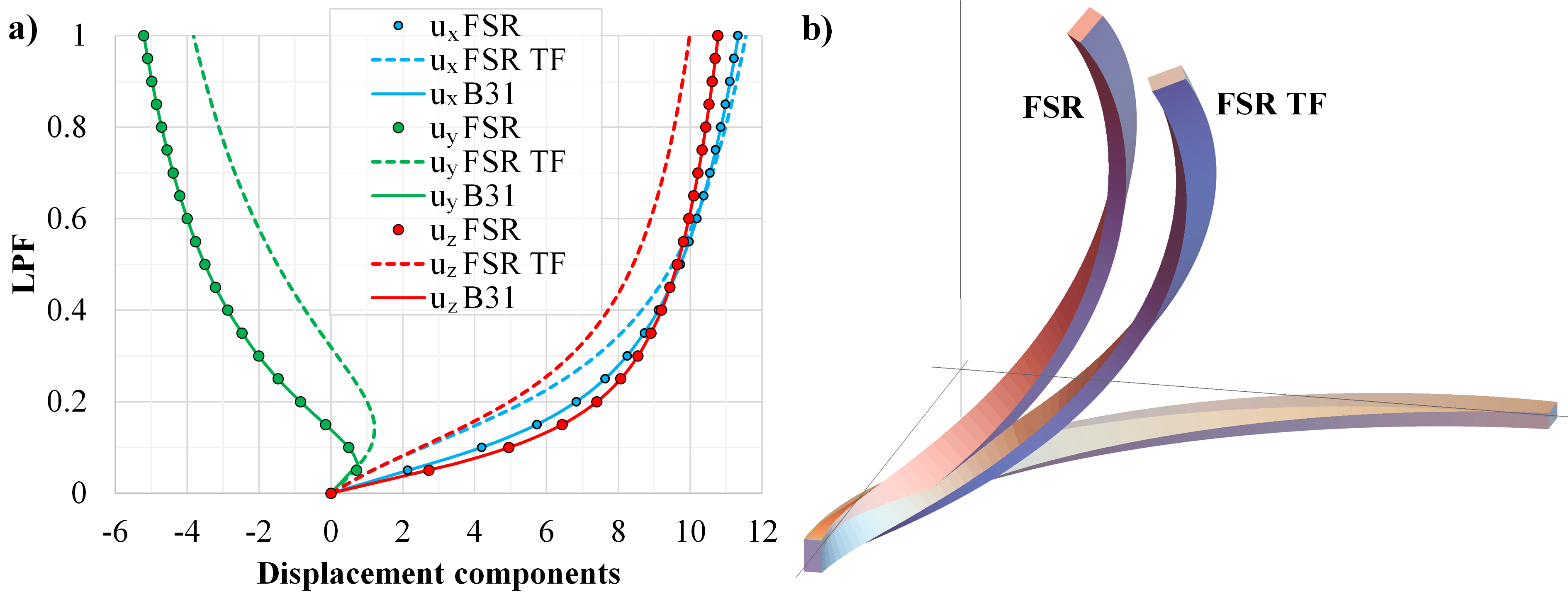}\centering
	\caption{A pre-twisted beam. a) Comparison of displacement components vs. LPF using the FSR, FSR TF, and Abaqus. b) Comparison of deformed configurations for LPF=1, using the FSR and FSR TF.}
	\label{fig: FSR vs FSR TF}
\end{figure}

In order to assess the influence of the independent twist angle, the pre-twisted beam is analyzed with the FSR TF model and the results are compared in Fig.~\ref{fig: FSR vs FSR TF}a, while the deformed configurations are visualized in Fig.~\ref{fig: FSR vs FSR TF}b.
Additionally, the FSR is verified by a comparison with the results obtained using Abaqus and a mesh of 314 B31 elements. Although the B31 element models shear deformable beams \cite{2009smith}, the results are practically indistinguishable from those of the FSR model. Interestingly, the cubic element B33 that models BE beams could not converge for such large deformations.

It is evident that the FSR TF formulation returns erroneous results in this case. The error depends on the boundary conditions and load, and the usage of the FSR TF cannot be generally recommended. For large load values, the error decreases as the axial strain of the beam axis increases. 

\subsection{Pure bending of a cantilever beam}
\label{pureplane}

\begin{figure}
	\includegraphics[width=8cm]{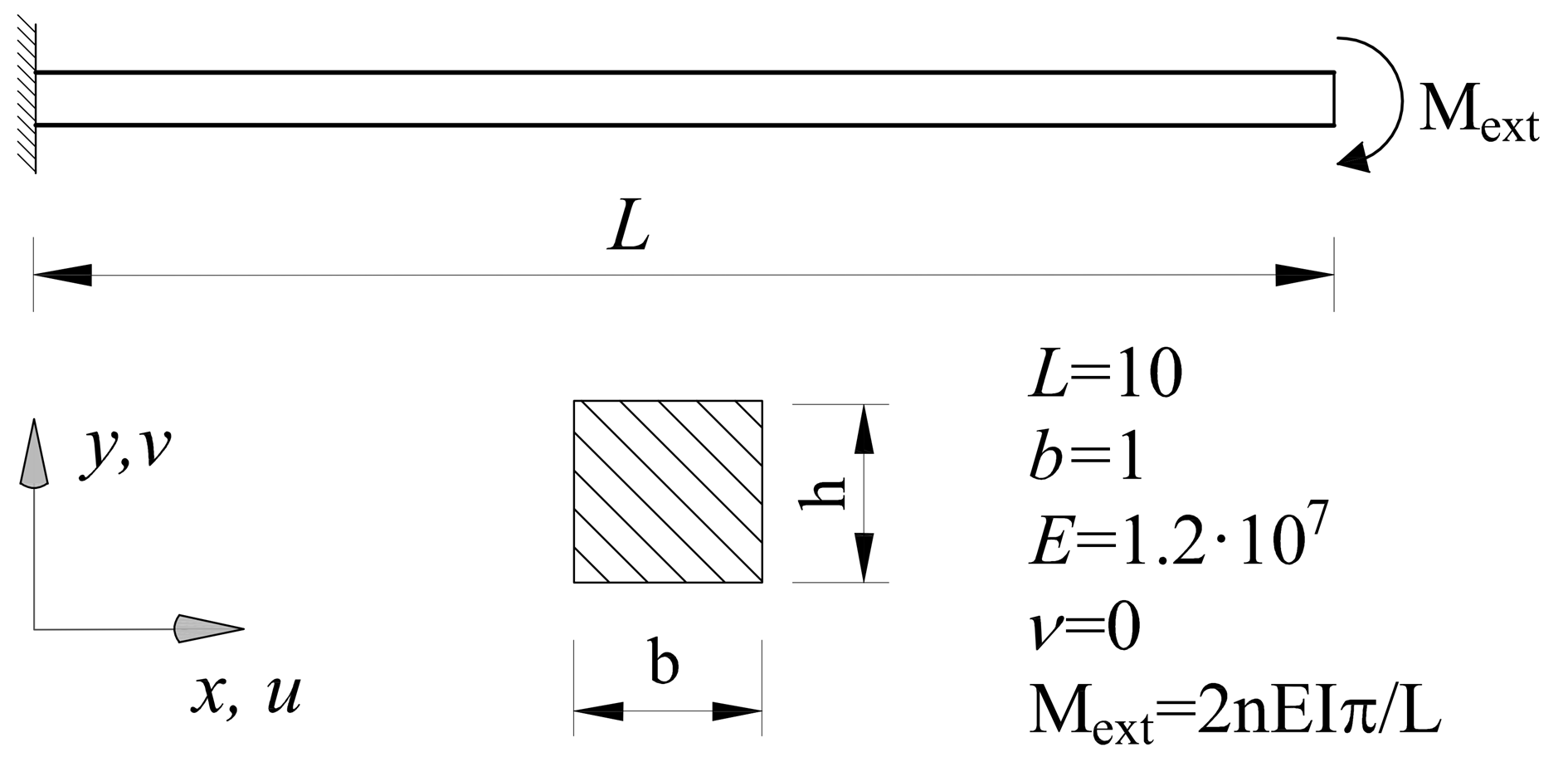}\centering
	\caption{Pure bending of a cantilever beam. Geometry and applied load.}
	\label{fig: Main1}
\end{figure}

\begin{figure}
	\includegraphics[width=\linewidth]{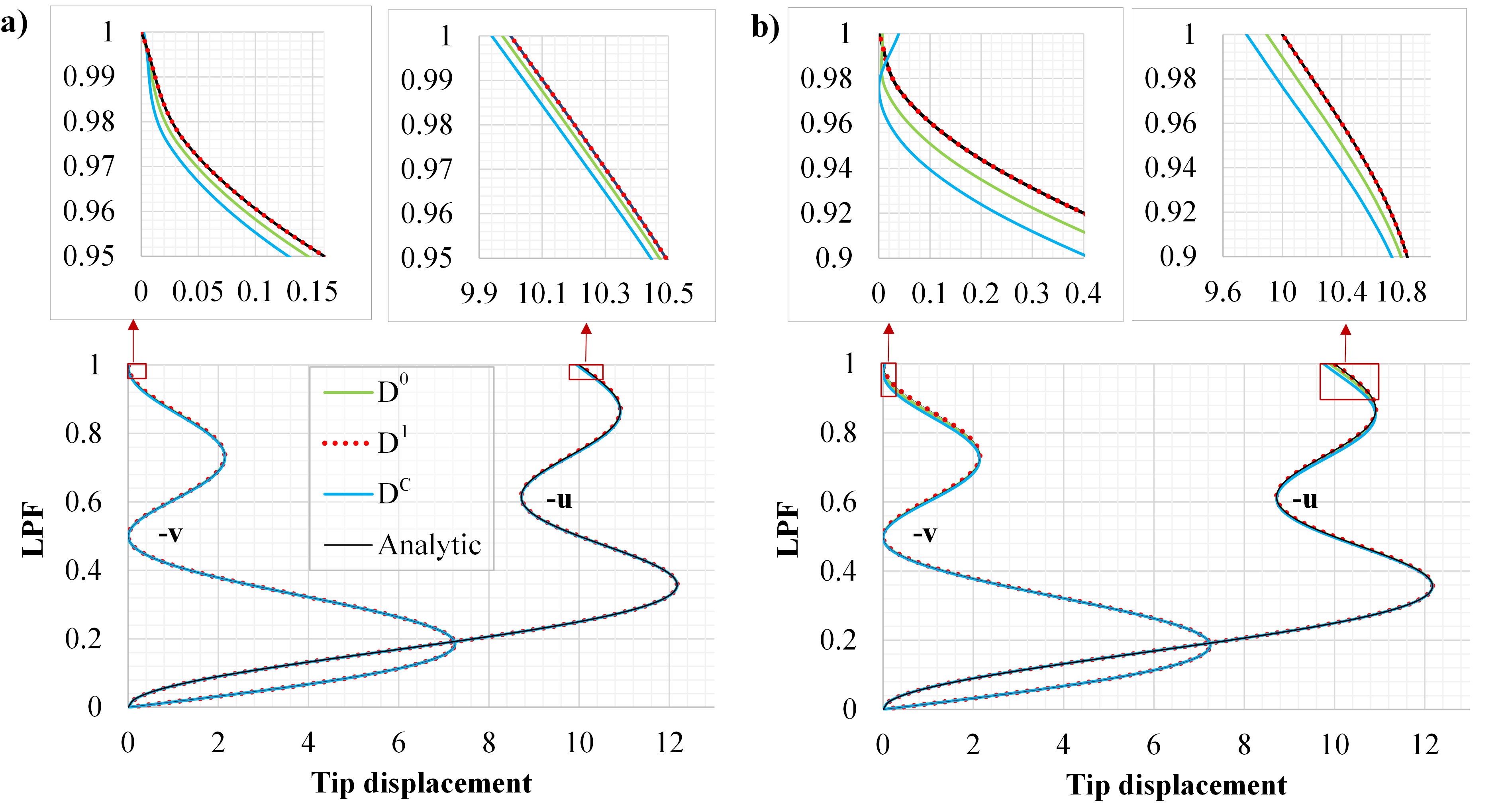}\centering
	\caption{Pure bending of a cantilever beam. Displacement of the tip for $n=2$: a) $h=0.1$, b) $h=0.2$.}
	\label{fig: Main2}
\end{figure}

This is a standard benchmark example for in-plane nonlinear beam formulations. A cantilever is loaded with a tip moment, causing the state of pure bending, Fig.~\ref{fig: Main1}. 
Since the beam deforms in a plane, there is no twisting, and we can apply the rotation-free plane IGA model, see \cite{2022borkovic}. The purpose of the example is to investigate the influence of large curviness on the beam response. If the condition of inextensibilty of the beam axis is enforced, the beam deforms into a circle with curvature $\chi=2n\pi\slash L \left(n=1,2,...\right)$. For this case, the analytical solution is straightforward \cite{2021choi}. However, if the nonlinear distribution of axial strain along the cross section is considered, coupling between the bending and axial actions occurs. The results for the displacement components of the tip for $n=2$, and cross-sectional heights $h=0.1$ and $h=0.2$ are compared in Fig.~\ref{fig: Main2}.
The values of curviness at the final configuration are 0.126 and 0.251, for cases $h=0.1$ and $h=0.2$, respectively. Evidently, all constitutive models are in agreement with analytical predictions for small load values. As the load and curviness increase, the differences in displacement components become apparent, as emphasized in the zoomed parts of the graphs. The constitutive model $D^1$ is fully aligned with the analytical solution, which suggests that \eqqref{eq:D1} indeed results with near-zero axial strain of the beam axis.

\begin{figure}
	\includegraphics[width=\linewidth]{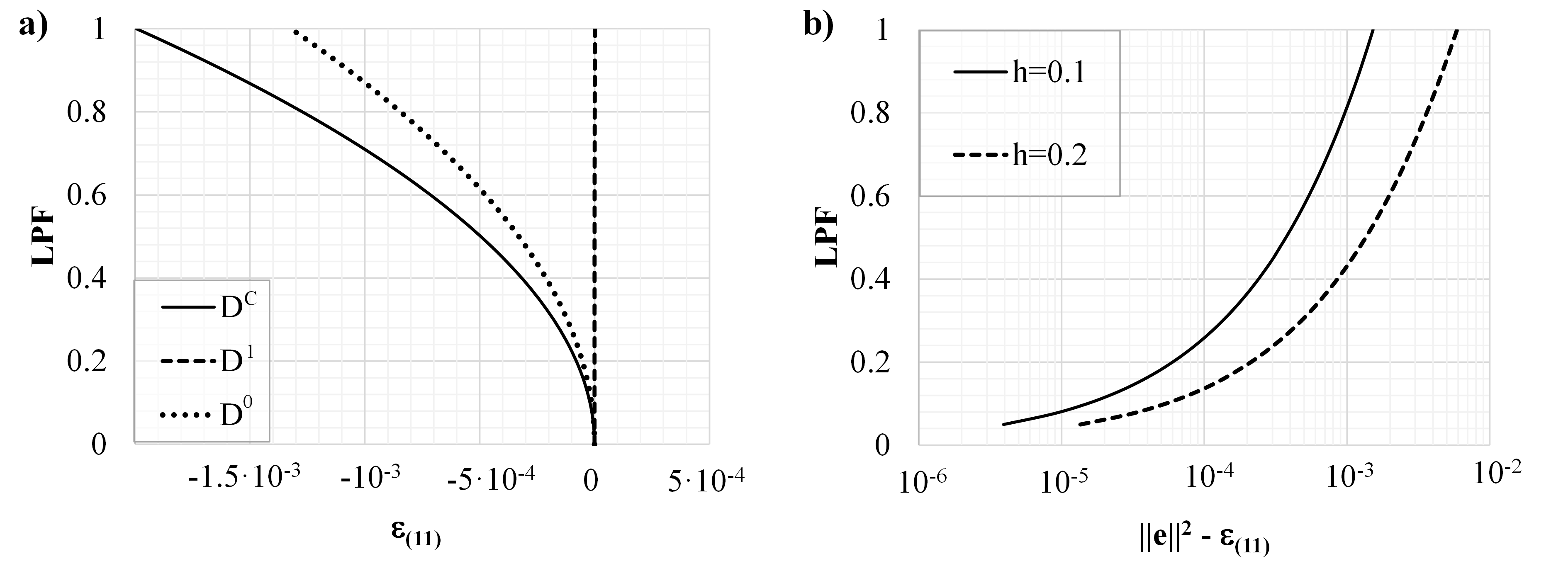}\centering
	\caption{Pure bending of a cantilever beam. a) Axial strain of the beam axis for three constitutive models ($h=0.1$, $n=2$). b) $L^2$ norm of a difference between the axial strain of beam axis calculated by solving Eqs.~\eqref{eq:sfplane} and by the $D^C$ model.}
	\label{fig: Main3}
\end{figure}

Furthermore, due to the pure bending conditions, this example is ideally suited for validating the strongly curved beam model. The expressions for physical normal force and bending moment are given by \eqqref{eq:section forces}. For an in-plane beam they reduce to:
\begin{equation}
	\label{eq:sfplane}
	N=\frac{E}{g} \sqrt{\frac{g^*}{g}}\left(A \ii{\epsilon}{}{11} + 1.5 I \chi \kappa \right) \quad \text{and} \quad M=\frac{EI}{g} \sqrt{\frac{g^*}{g}} \left(\chi \ii{\epsilon}{}{11} + \kappa \right).
\end{equation}
It is reasonable to assume that both, the axial strain and the change of curvature, are constant along the beam axis. By imposing the conditions $N=0$ and $M=M_{ext}$, and having the relation $\kappa=\chi \left(2\ii{\epsilon}{}{11} + g\right)$ in mind, Eqs.~\eqref{eq:sfplane} reduce to a system of two nonlinear algebraic equations with two unknowns. This allows us to calculate the reference strains of the beam axis and to test our computational isogeometric model. 
For this, a dense mesh of 64 quartic elements is used and the equilibrium paths of the reference axial strain are shown in Fig.~\ref{fig: Main3}a for the three constitutive models and $h=0.1$. 
This evolution is nonlinear for the $D^C$ and $D^0$ models, while the $D^1$ model returns effectively zero axial strain. Using the $D^C$ model, we have obtained the reference axial strain of $\idef{\epsilon}{}{(11)}=-0.0004948$ for n=1, which is in agreement with the values presented in \cite{2021choi}. It is interesting that the axis compresses, but the ends of the beam overlap. This is due to the fact that the curvature at the final configuration is not exactly $2 n \pi \slash L$, but is slightly larger. In concrete terms we have obtained $\idef{K}{}{} = \chi^* = -0.62956$, for n=1 and h=0.1, while $2 n \pi \slash L=0.2\pi=0.62832$. At first, this behavior is counter-intuitive, and it can result in a misinterpretation of the sign for axial strain as in \cite{2022borkovic}. 

Additionally, we have obtained the internal normal force using the equations of the $D^C$ model \eqref{eq:constif} and \eqref{eq:ac}. Its tensorial value is $\ic{N}{}{}g=-395.7$, which is again in agreement with the value reported in \cite{2021choi} where this quantity is named \emph{effective axial stress resultant}. 

Finally, we have compared the values of reference axial strain for $n=2$, and $h=0.1$ and $h=0.2$. Relative differences of results calculated by the IGA $D^C$ model and Eqs.~\eqref{eq:sfplane} are displayed in Fig.~\ref{fig: Main3}b. The differences are small, but increase with both the load and curviness. Further test, which we have left out of the manuscript, show that this difference reduces with $h$-refinement.

\subsection{Circular ring subjected to twisting}

\begin{figure}
	\includegraphics[width=\linewidth]{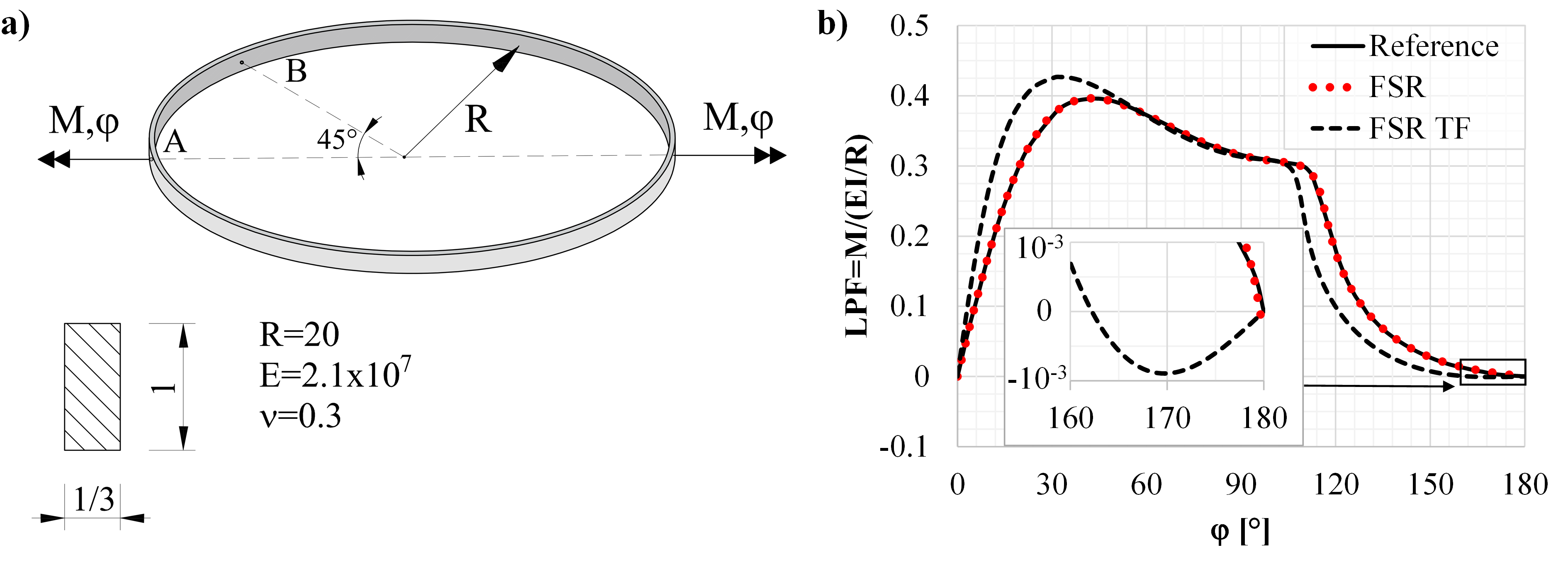}\centering
	\caption{Circular ring subjected to twisting. a) Load and geometry. b) Comparison of LPF vs. rotation at the point of the application of load.  }
	\label{fig:Ring1}
\end{figure}

A circular ring that is subjected to symmetrical twisting is a well-known test for the verification of formulations involving large rotations and small strains of spatial beams \cite{2014meier, 2020magisano,2022borkovicb}. The geometry and load are displayed in Fig.~\ref{fig:Ring1}a.
Here, the external twist is applied through a pair of concentrated moments $M=EI/R$. Due to the symmetry of the load and the geometry, only a quarter of the ring is modeled \cite{1992yoshiaki,2022borkovicb}. The equilibrium path of the external angle of twist is commonly observed for the verification of computational models. The results obtained with a quintic mesh with 16 elements are compared with the reference solution from \cite{1996pai} in Fig.~\ref{fig:Ring1}b. The FSR and reference results are in full agreement. As expected, the response obtained with the FSR TF formulation differs. However, for approximately $\varphi \in \left(50^\circ, 100^\circ \right)$, the equilibrium path is well-aligned with the exact one. The zoomed part in Fig.~\ref{fig:Ring1}b shows that the FSR TF passes LPF=0 twice, near $162^\circ$ and $180^\circ$.  
While the FSR TF formulation disagrees with standard formulations, it is astonishing that this simplified rotation-free model can approximate such complex behavior.

Our tests show that there is no noticeable difference in displacements between the different constitutive models since the beam has small curviness \cite{2022borkovicb}. The reference axial strain and normal force, however, are affected by the constitutive relation as will be discussed later. 

\begin{figure}
	\includegraphics[width=\linewidth]{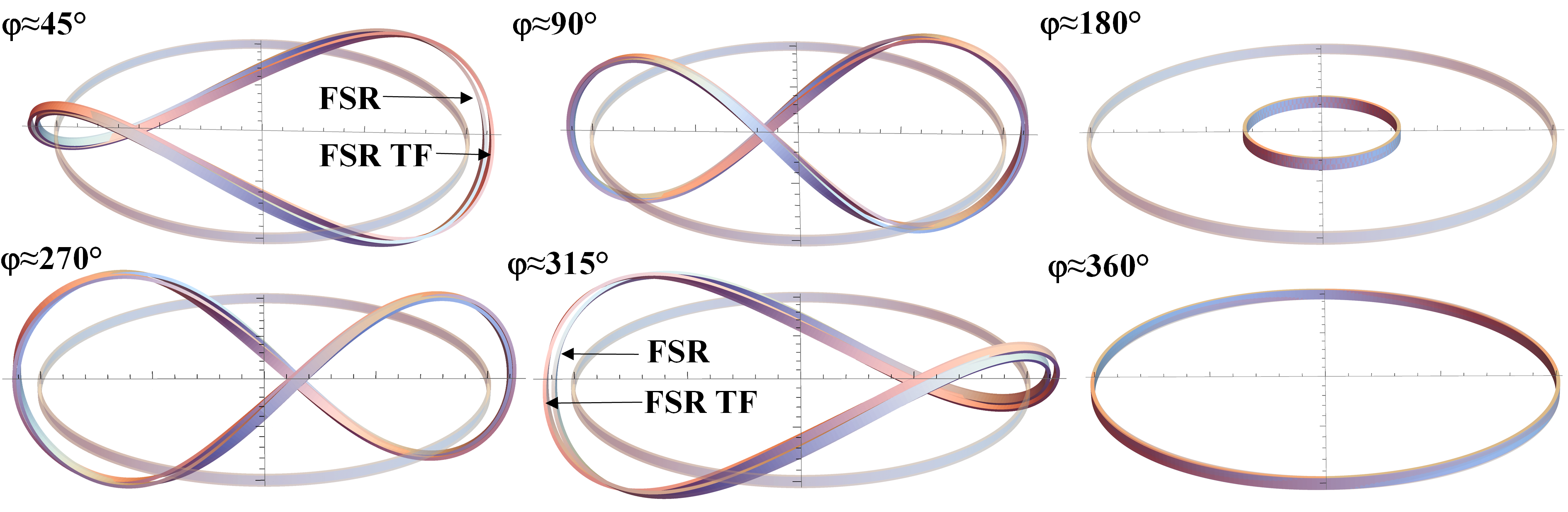}\centering
	\caption{Circular ring subjected to twisting.  Deformed configurations calculated with the FSR and FSR TF formulations. 
	}
	\label{fig:Ring2}
\end{figure}

A characteristic feature of this example is that after the external twisting of $\varphi=180^{\circ}$, the ring deforms into a smaller ring, with a diameter reduced by a factor of three. Additional application of the external twisting returns the ring into its original configuration for $\varphi=360^{\circ}$. A graphical representation is given in Fig.~\ref{fig:Ring2} for both FSR and FSR TF formulations. 
This visualization confirms that the FSR TF formulation fairly approximate the exact behavior of this beam for some values of the external load.

\begin{figure}
	\includegraphics[width=\linewidth]{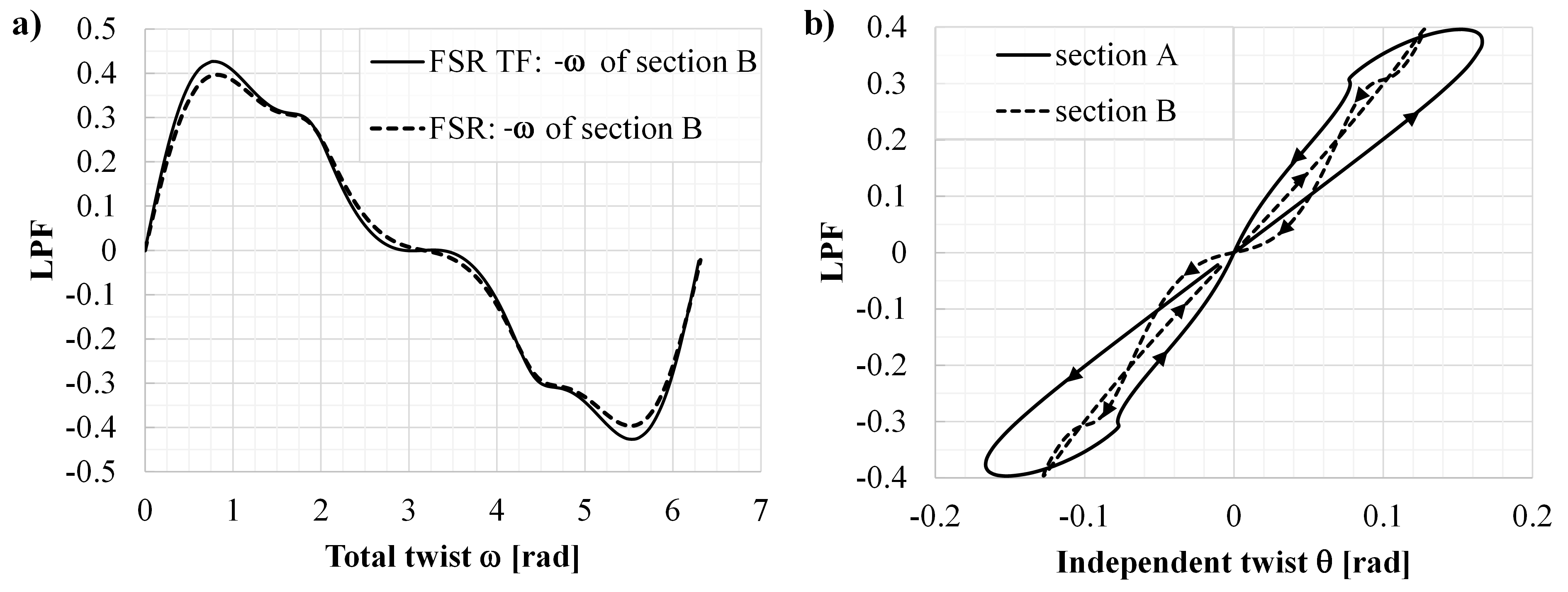}\centering
	\caption{Circular ring subjected to twisting.  Comparison of twist angles. a) Total twist $\omega$ of section B for the FSR and FSR TF formulations. b) Independent twist angle $\theta$ at sections A and B using the FSR formulation. }
	\label{fig:Ring3}
\end{figure}

In order to closely examine this phenomena, the twist angles of cross sections A and B (marked in Fig.~\ref{fig:Ring1}a) are observed for $\varphi=360^{\circ}$. The total twist angle $\omega$ of the cross section B is displayed in Fig.~\ref{fig:Ring3}a where similar equilibrium paths for both formulations are observed. 
A cause for the partial agreement of the two formulations is revealed in Fig.~\ref{fig:Ring3}b, where the equilibrium paths of the independent twist angle $\theta$ at sections A and B is given. This angle is relatively small in this case, which allows approximate modeling with the FSR TF formulation. 

\begin{figure}
	\includegraphics[width=\linewidth]{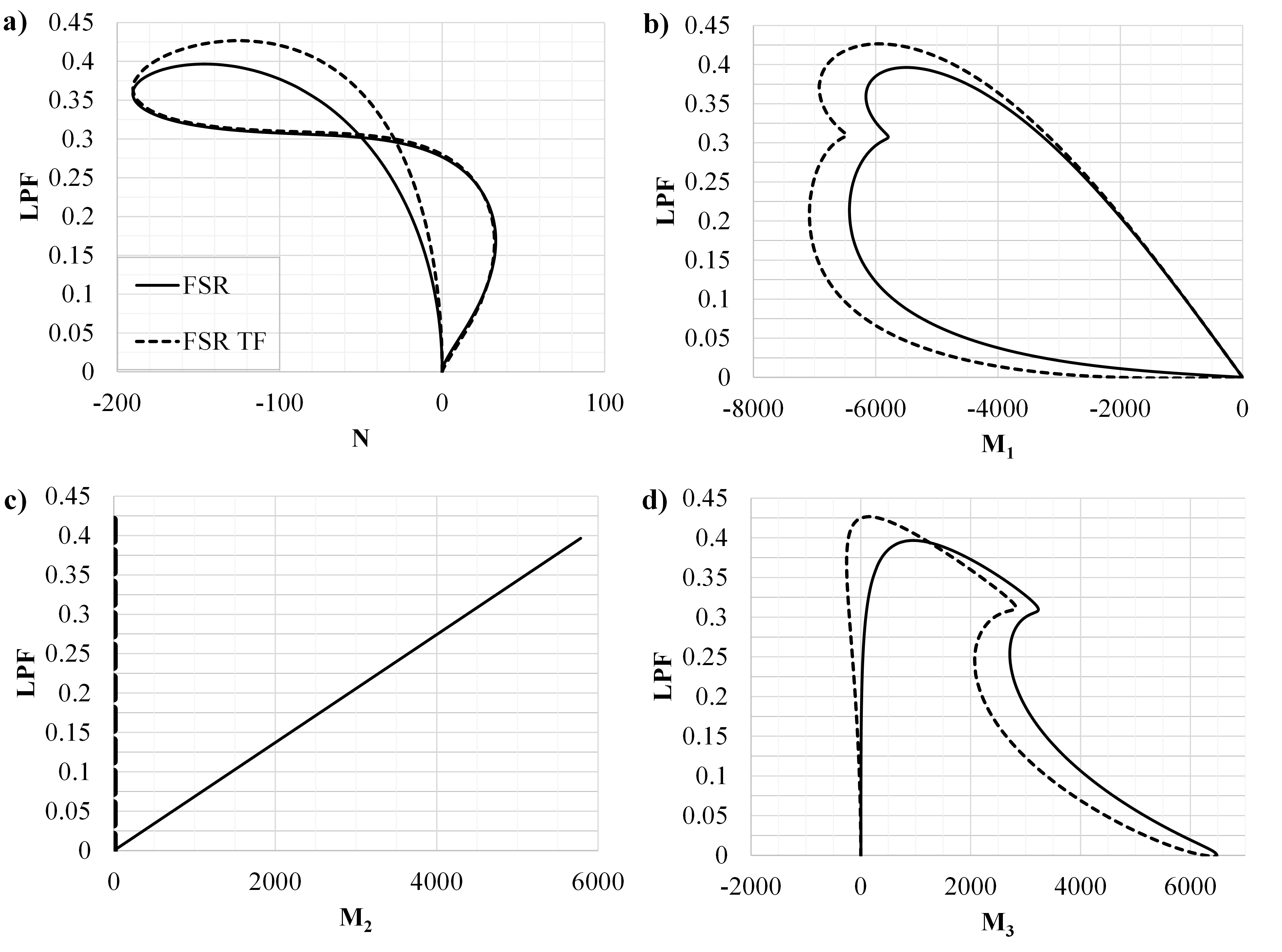}\centering
	\caption{Circular ring subjected to twisting.  Comparison of the FSR and FSR TF formulations. Stress resultant and stress couples at section A: a) normal force, b) torsional moment, c) bending moment $M_2$, d) bending moment $M_3$. }
	\label{fig:Ring5x}
\end{figure}

Further investigation is made by comparing the values of section forces using the $D^C$ model and a dense mesh of 32 quintic elements, see Fig.~\ref{fig:Ring5x}. 
These results suggest that the FSR TF vaguely follows the true equilibrium paths but gives erroneous values. The error is particularly pronounced for the moment $M_2$, due to the fact that the material basis vectors of the FSR TF model are aligned with the normal and binormal. 
The effect of this error on the structural response is partly compensated with the opposite sign of $M_3$ for the FSR and FSR TF before the load limit point, and by the linear response of $M_2$. 
The value of $M_2$ is large because $I_{\zeta \zeta}=9 I_{\eta \eta}$, but the changes of curvature $\chi_2$ and $\chi_3$ are both around 0.01 near the load limit point, after which $\chi_2$ returns to zero, while $\chi_3$ shows snap-through behavior for both formulations. 
All in all, this example shows that the FSR TF formulation can approximate some parts of the equilibrium path for specific deformation cases.

\begin{figure}
	\includegraphics[width=\linewidth]{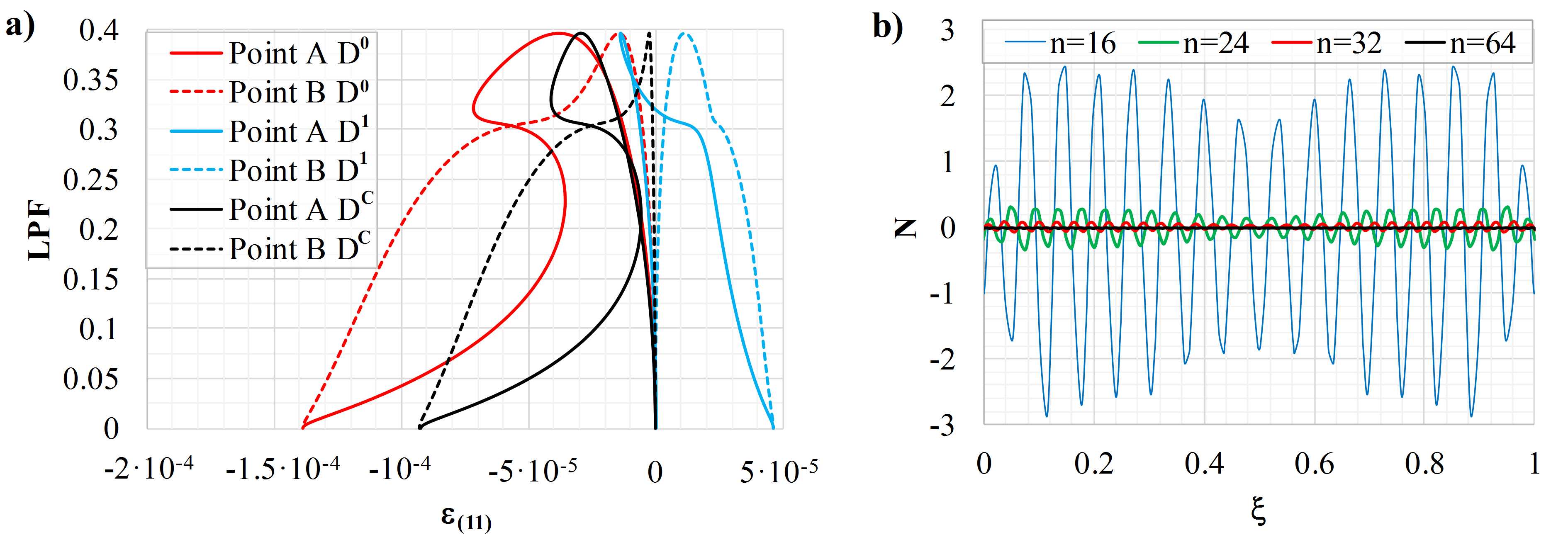}\centering
	\caption{Circular ring subjected to twisting. a) Comparison of the reference axial strain at section A using the different constitutive models and FSR formulation. b) Distribution of normal force for different meshes using the $D^C$ model. }
	\label{fig:Ring4}
\end{figure}

Next, the influence of constitutive relations is assessed. The results for the reference axial strain $\ii{\epsilon}{}{(11)}$ at section A are given in Fig.~\ref{fig:Ring4}a for different constitutive models using 32 quintic elements. 
Interesting equilibrium paths are obtained and the values differ significantly. The $D^1$ model returns extension, while the $D^0$ and $D^C$ return compression. As a check, for the angle of $\varphi=180^{\circ}$, there should be no normal force in the ring. In this configuration, the reference strains of the $D^C$ model are $\epsilon_{(11)}=-0.0000926$ and $\chi_3=0.100014$ which give nearly zero value of the normal force, cf.~\eqqref{eq:section forces}. This validates the presented $D^C$ model and its usage is recommended instead of the $D^0$, $D^1$, and the models suggested in \cite{2022borkovicb}. Furthermore, the normal force distribution along the modeled part of a ring is given in Fig.~\ref{fig:Ring4}b. Clearly, the normal force shows oscillatory behavior that mitigates with $h$-refinement, which suggests presence of membrane locking.

\begin{figure}
	\includegraphics[width=\linewidth]{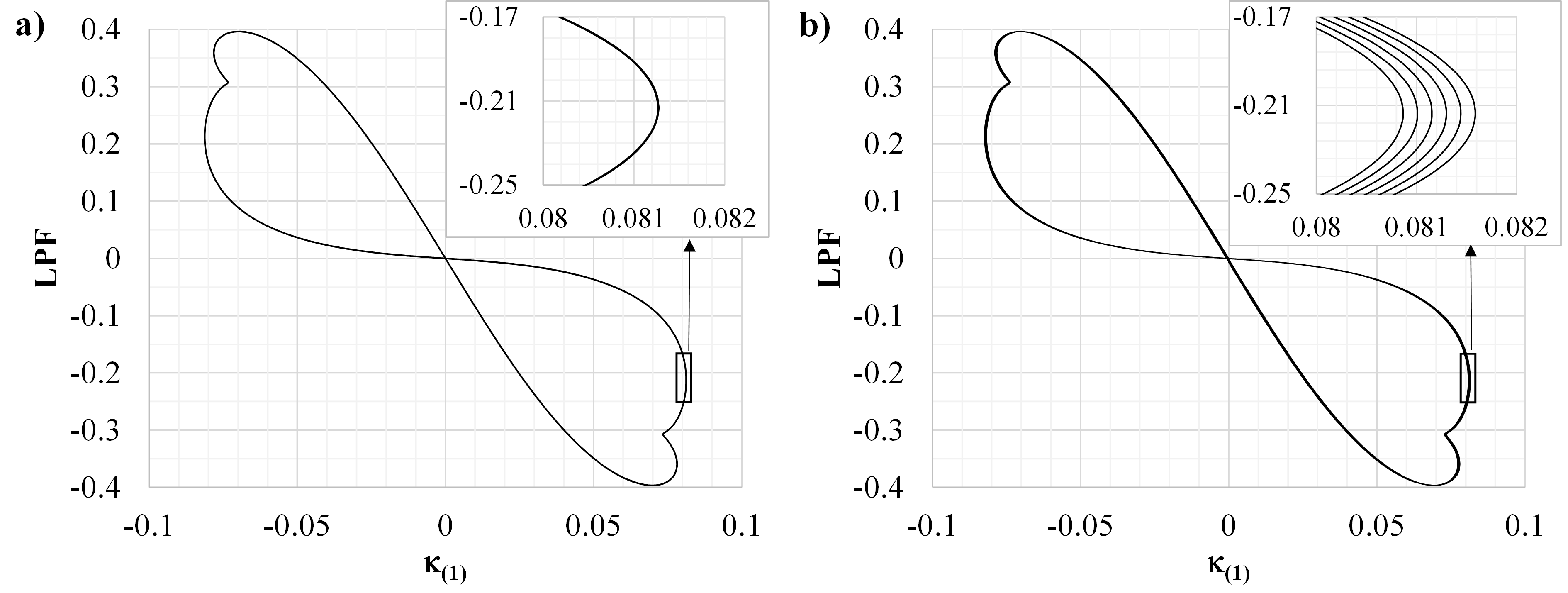}\centering
	\caption{Circular ring subjected to twisting.  Equilibrium path of the torsional curvature at point A during six cycles of twisting: a) FSR; b) SR. }
	\label{fig:Ring6}
\end{figure}

The final benchmark is related to the path-independence. Due to the cyclic response of this ring, path-dependence can be easily detected \cite{2020magisano, 2022borkovicb}. Let us observe the torsional strain at point A while the ring is twisted six times ($\varphi=12\pi$). It is known that the path-dependence mitigates with the increase in the mesh density \cite{2022borkovicb}. Therefore, for this test a sparse mesh with 8 quintic elements is used. The results are calculated with the SR and FSR formulations and compared in Fig.~\ref{fig:Ring6}.
This test confirms the previous observation. The FSR formulation is path-independent while the SR formulation with incremental update of the local vector basis is not.

\subsection{Straight beam bent to helix}

\begin{figure*}[h]
	\includegraphics[width=\linewidth]{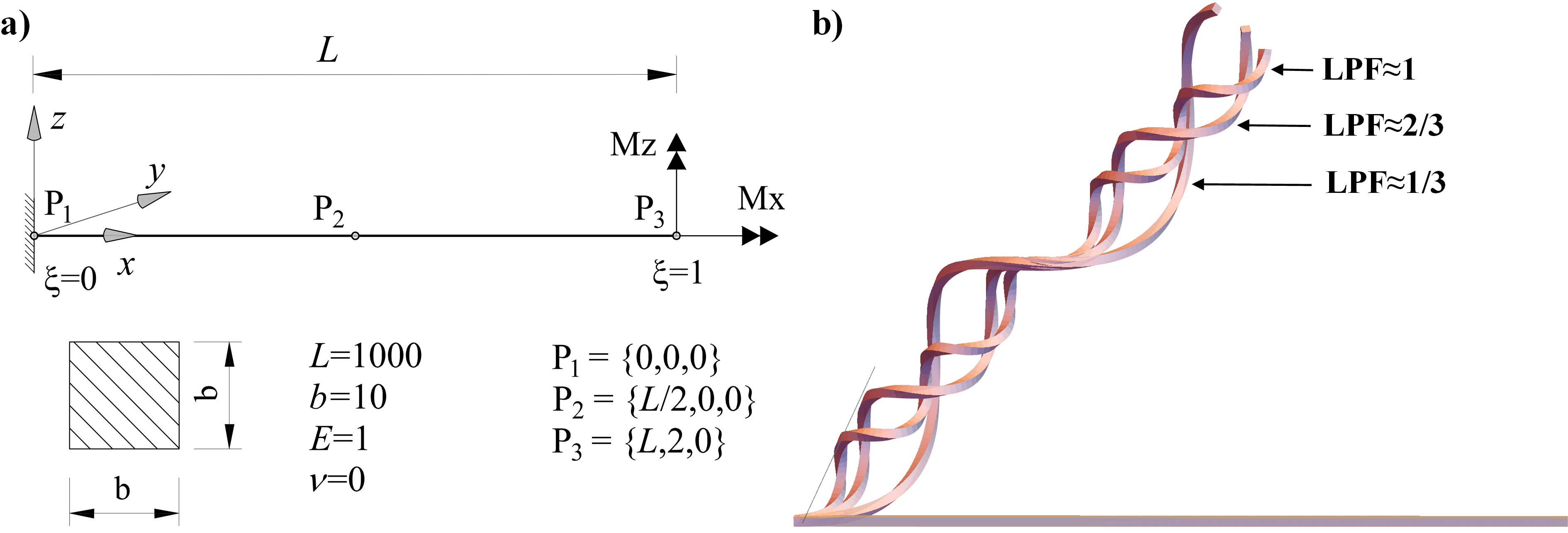}\centering
	\caption{Straight beam bent to helix. a) Geometry and load. b) Deformed configurations.}
	\label{fig:Helix1}
\end{figure*}

\begin{figure}
	\includegraphics[width=\linewidth]{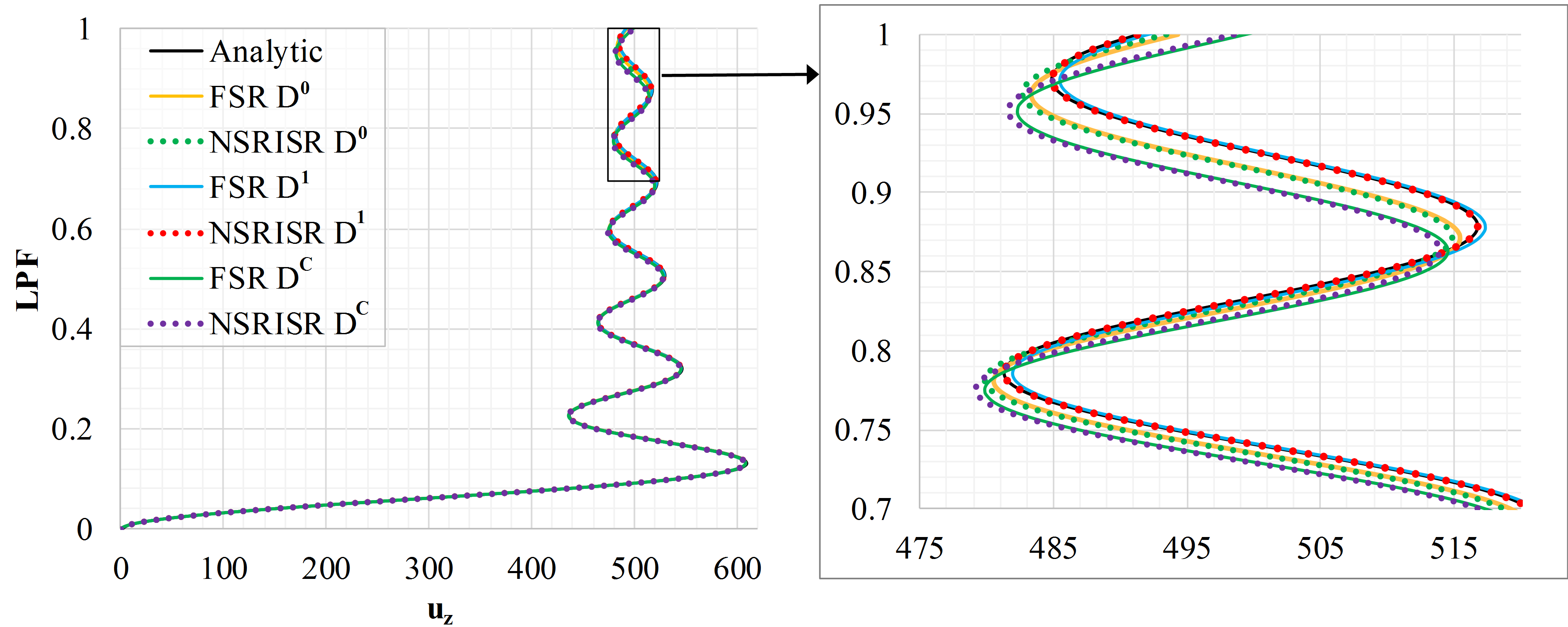}\centering
	\caption{Straight beam bent to helix. Comparison of the $z$-component of the tip displacement for different constitutive models and two formulations. Zoomed part of the equilibrium path is shown on the right.}
	\label{fig:Helix2}
\end{figure}

In this example, we have considered the response of an initially straight cantilever beam loaded with two end moments, as indicated in Fig.~\ref{fig:Helix1}a. 
Since the FSR formulation requires a well-defined FS frame, the beam axis is defined with a quadratic spline, and a small initial curvature is imposed by moving the third control point by 2 along the $y$-direction. In this way, the initial curvature of the beam analyzed with the FSR approach is practically constant with the value of $K_3\approx4 \cdot 10^{-6}$. The NSRISR model is used for comparison, but without the initial curvature. The beam is discretized with 30 quintic elements and subjected to the tip moments $M_x=M_z=20$. The three characteristic configurations plotted in Fig.~\ref{fig:Helix1}b reveal the complex response of this beam that deforms into a near-perfect helix. Fig.~\ref{fig:Helix2} illustrates the $z$-component of the tip displacement for different constitutive models and formulations.. 
Additionally, the analytical solution for small-curvature beams is employed for the comparison \cite{2014meier, 2022borkovicb}. 

Regarding the different constitutive models, all return same equilibrium paths for small LPF. As the LPF increases, the curviness also increases and the differences of displacement become evident. As anticipated, the $D^1$ model is aligned with the analytical solution that assumes inextensibilty of the beam axis. The results of the fully uncoupled $D^0$ model are between the strong- and small-curvature models.

The comparison of the different formulations shows that the FSR approach returns almost identical results as the NSRIS. The differences are most prominent at displacement limit points and they are attributed to the initial curviness applied for the FSR model, which does not exist for the NSRISR model. This example shows that the problem of a non-existent FS frame for straight configurations can be alleviated by imposing the small curvature without significantly affecting the response.

\begin{figure}
	\includegraphics[width=\linewidth]{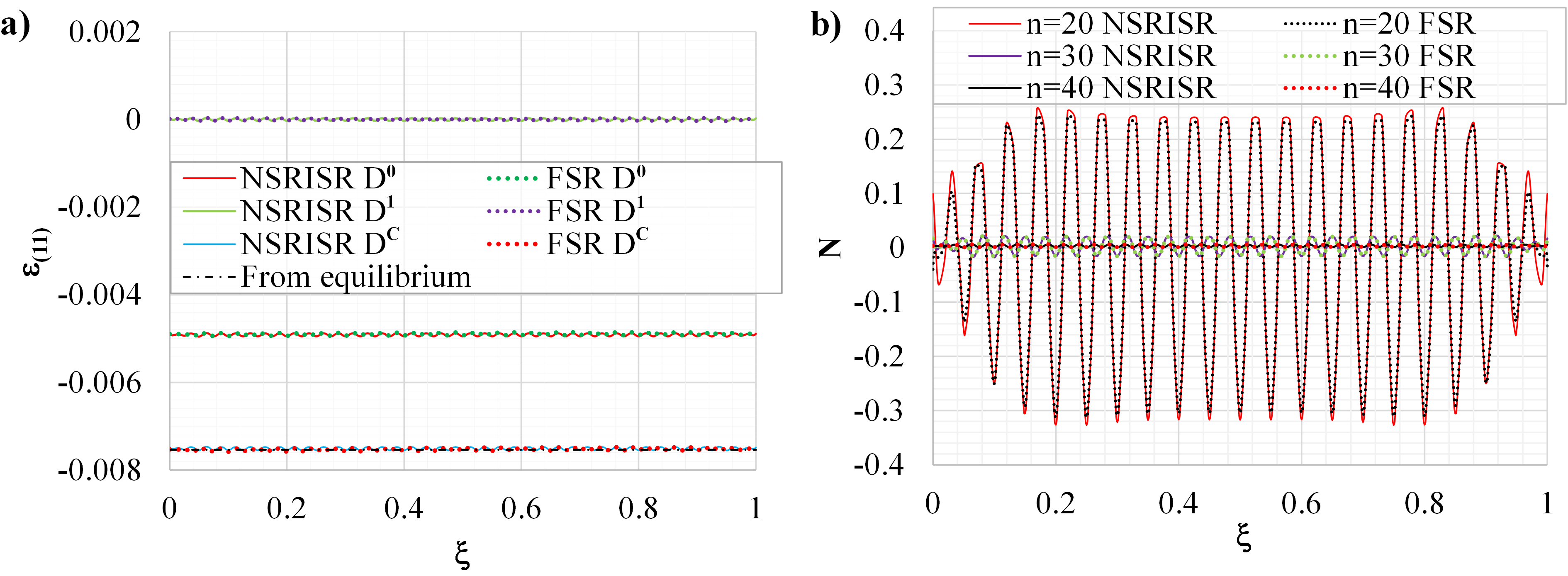}\centering
	\caption{Straight beam bent to helix. a) Distributions of the reference axial strain of the beam axis for LPF=1 using the NSRISR and FSR formulations and different constitutive relations. b) Distribution of the normal force $N$ for LPF=1 and $D^C$ model using the NSRISR and FSR formulations for three mesh densities of quintic elements. }
	\label{fig:Helix3}
\end{figure}

An interesting aspect of this example is that the beam is in a state of pure bending. The fact that the axial strain exists while there is no normal force is discussed in \cite{2022borkovicb} and ameliorated results are presented here. Similar to the example of pure in-plane bending of a beam, cf.~Subsection \ref{pureplane}, axial strain can be obtained from the equilibrium conditions with respect to the stress resultant and stress couples, \eqqref{eq:section forces}:
\begin{equation}
	\label{eq:sfspace}
	\begin{aligned}
	N&=\frac{E}{g} \sqrt{\frac{g^*}{g}}\left(A \ii{\epsilon}{}{11} + 1.5 I_{\zeta \zeta} \ii{\chi}{}{2} \ii{\kappa}{}{2}  +1.5 I_{\eta \eta} \ii{\chi}{}{3} \ii{\kappa}{}{3} \right), \\ 
	M_2&=\frac{E I_{\zeta \zeta}}{g} \sqrt{\frac{g^*}{g}} \left(\chi_2 \ii{\epsilon}{}{11} + \kappa_2 \right),\\
	M_3&=\frac{E I_{\eta \eta}}{g} \sqrt{\frac{g^*}{g}} \left(\chi_3 \ii{\epsilon}{}{11} + \kappa_3 \right).
\end{aligned}
\end{equation}
By assuming that the axial strain is constant along the beam, we can solve these equations at some fixed section. For example at the clamped end, we have $N(0)=M_2(0)=0$ and $M_3(0)=20$, and the resulting axial strain is $\ii{\epsilon}{}{(11)}=-0.00754$. This result is compared in Fig.~\ref{fig:Helix3}a with the numerical results obtained by the NSRISR and FSR formulations using 40 quintic elements. Both formulations return the same results since the deformed configurations for LPF=1 match, as shown in Fig.~\ref{fig:Helix2}. It is important that the values calculated with the $D^C$ model correspond to the solution of equilibrium equations \eqref{eq:sfspace}. Again, the $D^1$ model returns zero axial strain.

The issue of oscillatory normal force along the length of the beam is emphasized in Fig.~\ref{fig:Helix3}b where the results of both formulations are compared for three mesh densities. Evidently, the NSRISR and FSR return relatively similar results. The oscillations of the stress resultant reduce with $h$-refinement, as in the previous example. Evidently, the $D^C$ model correctly predicts zero normal force in this example but requires a dense mesh.

\section{Conclusions}

A geometrically exact isogeometric formulation of the spatial Bernoulli-Euler (BE) beam based on the Frenet-Serret (FS) frame is presented. This new formulation, designated as the Frenet-Serret Rotation (FSR), employs an additive decomposition of the total twist angle into the twist of the FS frame and an independent twist angle. The weak form is rigorously derived and linearized, including the contributions from the constraints and external loads. Nonlinear terms of strain with respect to the material axes are considered and the effect of strong curvature is captured through axial-bending coupling. Two simplified constitutive models for the calculation of internal forces are derived and compared with the exact one.

The FSR formulation is well-suited for the analysis of beams that undergo large rigid-body motions, because it is objective by definition. Since the FSR requires $C^2$ continuity, the NURBS-based IGA is an ideal framework for its implementation. By the consistent treatment of the virtual power and the finite element implementation, the formulation returns results that are indistinguishable from standard methods. The main shortcoming of the FSR approach is that it fails for configurations that do not have an uniquely defined FS frame. If such a configuration is encountered, the implementing of a switch to some other formulation would be a straightforward task. A switch of this kind was not required, however, in any of the presented standard numerical examples. Moreover, it is shown that a straight configuration can be approximated with a slightly curved geometry without significantly affecting the beam's response.
All in all, the FSR formulation presents a significant contribution to the theory of beams, while having a potential to efficiently and accurately simulate specific mechanical systems composed of deformable slender bodies that undergo large rigid-body motions.

A nonlinear rotation-free model of the spatial BE beam follows as a special case of the FSR formulation by omitting the independent twist angle from the DOFs. Importantly, this reduced model still contains the FS part of torsion, which makes it more generally applicable than existing rotation-free formulations of spatial beams. In particular, it is capable of giving approximate solutions for beams that are predominantly bent with respect to the binormal of the beam axis. 
The structural response of such beams can be well-approximated by neglecting the independent twist angle, as demonstrated by the behavior of the ring subjected to symmetrical twisting.

The definition of the BE beam's metric requires several assumptions in order to decouple axial and torsional effects. The aim to capture the effect of the axial-bending coupling is fulfilled by the consistent derivation of the axial strain at an arbitrary point. Through the strict considerations of the current and initial beam metric, a computational model suitable for the nonlinear analysis of strongly curved BE beams is obtained. The influence that large curviness has on axial-bending coupling is evident for standard academic examples.

Further work aims to extend the developed formulation to the dynamic analysis of curved slender bodies with large rigid-body motion.

\section*{Acknowledgments}

We acknowledge the support of the Austrian Science Fund (FWF): M 2806-N.

\section*{Appendix A. Geometric stiffness matrix}
\setcounter{equation}{0}
\renewcommand\theequation{A\arabic{equation}}

Since the linear increments of the material base vectors are:
\begin{equation}
\label{eq: ap1 variation of base vector g_alpha}
\Delta \ivdef{g}{}{m} = \Delta \iv{u}{}{,m} = \iv{v}{}{,m} \Delta t,
\end{equation}
the linearized increments of the virtual reference strain rates are:
\begin{equation}
\label{eq: ap2 variations of reference strains}
\begin{aligned}
\Delta \delta \ii{d}{}{11} &= \Delta \ivdef{g}{}{1} \cdot \delta \iv{v}{}{,1}  = \iv{v}{}{,1} \cdot \delta  \iv{v}{}{,1} \Delta t, \\
\Delta \delta \imd{\kappa}{}{1} &= \Delta \left( \ivdef{T}{}{1}\cdot \delta  \iv{v}{}{,1} - \ivdef{T}{}{2}\cdot \delta  \iv{v}{}{,11} + \ivdef{T}{}{3}\cdot \delta  \iv{v}{}{,111} +\delta \imd{\theta}{}{,1} \right) \\
&= \Delta \ivdef{T}{}{1}\cdot \delta \iv{v}{}{,1} -\Delta \ivdef{T}{}{2}\cdot \delta \iv{v}{}{,11}+\Delta \ivdef{T}{}{3}\cdot \delta \iv{v}{}{,111}, \\
\Delta \delta \imd{\kappa}{}{2} &= \Delta \left[ \ivdef{n}{}{} \cdot \left( \delta \iv{v}{}{,11} - \idef{\Gamma}{1}{11} \delta \iv{v}{}{,1}\right) \sin \idef{\theta}{}{} + \icdef{K}{}{3} \delta \imd{\theta}{}{} \right] = \Delta \ivdef{n}{}{} \cdot \left( \delta \iv{v}{}{,11} - \idef{\Gamma}{1}{11} \delta \iv{v}{}{,1}\right) \sin \idef{\theta }{}{}\\
&\quad + \ivdef{n}{}{} \cdot \left( - \Delta \idef{\Gamma}{1}{11} \delta \iv{v}{}{,1}\right) \sin \idef{\theta}{}{} + \ivdef{n}{}{} \cdot \left( \delta \iv{v}{}{,11} - \idef{\Gamma}{1}{11} \delta \iv{v}{}{,1}\right) \Delta \sin \idef{\theta}{}{} + \Delta \icdef{K}{}{3} \delta \imd{\theta}{}{}, \\
\Delta \delta \imd{\kappa}{}{3} &= \Delta \left[ \ivdef{n}{}{} \cdot \left( \delta \iv{v}{}{,11} - \idef{\Gamma}{1}{11} \delta \iv{v}{}{,1}\right) \cos \idef{\theta}{}{} - \icdef{K}{}{2} \delta \imd{\theta}{}{} \right] =
 \Delta \ivdef{n}{}{} \cdot \left( \delta \iv{v}{}{,11} - \idef{\Gamma}{1}{11} \delta \iv{v}{}{,1}\right) \cos \idef{\theta}{}{} \\ &\quad + \ivdef{n}{}{} \cdot \left( - \Delta \idef{\Gamma}{1}{11} \delta \iv{v}{}{,1}\right) \sin \idef{\theta}{}{} + \ivdef{n}{}{} \cdot \left( \delta \iv{v}{}{,11} - \idef{\Gamma}{1}{11} \delta \iv{v}{}{,1}\right) \Delta \cos \idef{\theta}{}{} - \Delta \icdef{K}{}{2} \delta \imd{\theta}{}{}.
\end{aligned}
\end{equation}
The task is to express these linearized strains as a function of DOFs. In the following, we will adopt $\Delta t = 1$, for brevity. 

The increment of the Christoffel symbol is:
\begin{equation}
	\label{eq: ap6 variations of reference strains}
	\begin{aligned}
		\Delta \idef {\Gamma}{1}{11} &= \Delta \left(\frac{\ivdef{g}{}{1,1} \cdot \ivdef{g}{}{1}}{g^*} \right)= \frac{1}{g^\sharp} \left[ \left( \ic{K}{\sharp}{} \ve{n}^\sharp -\ii{\Gamma}{\sharp 1}{11} \iv{g}{\sharp}{1} \right)\cdot \iv{v}{}{,1} + \iv{g}{\sharp}{1} \cdot \iv{v}{}{,11} \right].
	\end{aligned}
\end{equation}
The increments of the gradients of velocities equal the increments of basis vectors and they follow from \eqqref{eq:461}:
\begin{equation}
\label{eq: variation v,2 v,3 1}
\begin{aligned}
\Delta \iv{v}{}{,2} &= -\frac{1}{\ipre{g}{}{}} \left( \ivpre{g}{}{2} \cdot \iv{v}{}{,1} \right) \ivpre{g}{}{1} + \ivpre{g}{}{3}  \imd{\omega}{}{}, \\
\Delta \iv{v}{}{,3} &= -\frac{1}{\ipre{g}{}{}} \left( \ivpre{g}{}{3} \cdot \iv{v}{}{,1} \right) \ivpre{g}{}{1} - \ivpre{g}{}{2}  \imd{\omega}{}{}.
\end{aligned}
\end{equation}
Using the additive decomposition of total angular velocity, \eqqref{eq: rs21}, these expressions can be written as a function of the independent twist velocity $\imd{\theta}{}{}$:
\begin{equation}
\label{eq: variation v,2 v,3 2}
	\begin{aligned}
		\Delta \iv{v}{}{,2} &= \left( -\frac{1}{\ipre{g}{}{}} \ivpre{g}{}{1} \otimes \ivpre{g}{}{2} -\frac{\ipre{\Gamma}{1}{11}}{\icpre{K}{}{}} \ivpre{g}{}{3} \otimes \vepre{b} \right) \iv{v}{}{,1} + \frac{1}{\icpre{K}{}{}} \ivpre{g}{}{3} \otimes \vepre{b} \:  \iv{v}{}{,11} + \ivpre{g}{}{3} \: \imd{\theta}{}{}, \\
		\Delta \iv{v}{}{,3} &= \left( -\frac{1}{\ipre{g}{}{}} \ivpre{g}{}{1} \otimes \ivpre{g}{}{3} +\frac{\ipre{\Gamma}{1}{11}}{\icpre{K}{}{}} \ivpre{g}{}{2} \otimes \vepre{b} \right) \iv{v}{}{,1} - \frac{1}{\icpre{K}{}{}} \ivpre{g}{}{2} \otimes \vepre{b} \:  \iv{v}{}{,11} - \ivpre{g}{}{2} \: \imd{\theta}{}{}.
	\end{aligned}
\end{equation}
Now, the increments of curvature components follow from \eqqref{eq: 25}:
\begin{equation}
	\label{eq: variations K2 K3}
	\begin{aligned}
		\Delta  \icdef{K}{}{2} &= \Delta \left( - \ivdef{g}{}{1,1} \cdot \ivdef{g}{}{3}\right) = -\iv{v}{}{,11} \cdot \ivpre{g}{}{3} - \ivpre{g}{}{1,1} \cdot \Delta \iv{v}{}{,\bar{3}} = \ivpre{n}{}{} \cdot \left(\iv{v}{}{,11} - \ipre{\Gamma}{1}{11}  \iv{v}{}{,1}\right) \sin \theta^\sharp + \icpre{K}{}{3} \: \imd{\theta}{}{},  \\
		\Delta  \icdef{K}{}{3} &= \Delta \left( \ivdef{g}{}{1,1} \cdot \ivdef{g}{}{2}\right) = \iv{v}{}{,11} \cdot \ivpre{g}{}{2} + \ivpre{g}{}{1,1} \cdot \Delta \iv{v}{}{,\bar{2}} = \ivpre{n}{}{} \cdot \left( \iv{v}{}{,11} - \ipre{\Gamma}{1}{11}  \iv{v}{}{,1}\right) \cos \theta^\sharp - \icpre{K}{}{2} \: \imd{\theta}{}{},
	\end{aligned}
\end{equation}
and they equal the increments of curvature changes. Furthermore, let us define the following increments:
\begin{equation}
	\label{eq: g iK}
	\begin{aligned}
		\Delta \frac{1}{g^*} &= - \frac{2}{\ipre{g}{2}{}} \: \ivpre{g}{}{1} \cdot \iv{v}{}{,1},\\
		\Delta \frac{1}{\icdef{K}{}{}} &= -\frac{1}{\icpre{K}{2}{}} \vepre{n} \cdot \left(  \iv{v}{}{,11} - \ipre{\Gamma}{1}{11} \iv{v}{}{,1}\right).
	\end{aligned}
\end{equation}
Now, the linearized increments of the normal and binormal follow from Eqs.~\eqref{eq:17}, \eqref{eq: ap6 variations of reference strains} and \eqref{eq: g iK}, \cite{2018radenkovicb}:
\begin{equation}
	\label{eq: variations n b}
	\begin{aligned}
		\Delta  \vdef{n} &= - \left( \frac{\ipre{\Gamma}{1}{11}}{\icpre{K}{}{}} \vepre{b} \otimes \vepre{b} + \frac{1}{\ipre{g}{}{}} \ivpre{g}{}{1} \otimes \vepre{n} \right) \iv{v}{}{,1} +\frac{1}{\icpre{K}{}{}} \vepre{b} \otimes \vepre{b} \:  \iv{v}{}{,11},  \\
		\Delta  \vdef{b} &= \left( \frac{\ipre{\Gamma}{1}{11}}{\icpre{K}{}{}} \vepre{n} \otimes \vepre{b} - \frac{1}{\ipre{g}{}{}} \ivpre{g}{}{1} \otimes \vepre{b} \right)  \iv{v}{}{,1} -\frac{1}{\icpre{K}{}{}} \vepre{n} \otimes \vepre{b} \:  \ivpre{v}{}{,11},
	\end{aligned}
\end{equation}
while the increments of the sine and cosine of the angle $\theta$ are:
\begin{equation}
	\label{eq: variations sin cos}
	\begin{aligned}
		\Delta  \sin \theta^* = \imd{\theta}{}{} \cos \theta^\sharp \quad \textrm{and} \quad  \Delta  \cos \theta^* = -  \imd{\theta}{}{} \sin \theta^\sharp.
	\end{aligned}
\end{equation}
These expressions allow us to write the increments of virtual curvature rates as:
\begin{equation}
	\label{eq: variation kapa 2}
	\begin{aligned}
		\Delta \delta \imd{\kappa}{}{2} &=  \iv{v}{}{,1} \cdot \left[ \frac{\ipre{\Gamma}{1}{11} \icpre{K}{}{2}}{\ipre{g}{}{} \icpre{K}{}{}} \left( \frac{\ipre{\Gamma}{1}{11} \ipre{g}{}{}}{\icpre{K}{}{} } \vepre{b} \otimes \vepre{b} -\frac{\icpre{K}{}{}}{\ipre{\Gamma}{1}{11}} \vepre{n} \otimes \vepre{n} +\ivpre{g}{}{1} \otimes \vepre{n} + \vepre{n} \otimes \ivpre{g}{}{1} \right) \right] \delta \iv{v}{}{,1} \\
		& - \iv{v}{}{,1} \cdot  \left( \frac{\ipre{\Gamma}{1}{11} \icpre{K}{}{2}}{\icpre{K}{2}{} } \vepre{b} \otimes \vepre{b} + \frac{\icpre{K}{}{2}}{\ipre{g}{}{} \icpre{K}{}{}} \vepre{n} \otimes \ivpre{g}{}{1} \right)  \delta \iv{v}{}{,11} - \iv{v}{}{,1} \cdot  \left( \frac{\ipre{\Gamma}{1}{11} \icpre{K}{}{3}}{\icpre{K}{}{} } \vepre{n}  \right)  \delta \imd{\theta}{}{}  \\
		& - \iv{v}{}{,11} \cdot \left( \frac{\ipre{\Gamma}{1}{11} \icpre{K}{}{2}}{\icpre{K}{2}{} } \vepre{b} \otimes \vepre{b} + \frac{\icpre{K}{}{2}}{\ipre{g}{}{} \icpre{K}{}{}} \ivpre{g}{}{1} \otimes \vepre{n} \right)\delta \iv{v}{}{,1} +\iv{v}{}{,11} \cdot \left( \frac{\icpre{K}{}{2}}{\icpre{K}{2}{} } \vepre{b} \otimes \vepre{b} \right)\delta \iv{v}{}{,11} \\
		&+ \iv{v}{}{,11} \cdot  \left( \frac{\icpre{K}{}{3}}{\icpre{K}{}{} } \vepre{n}  \right)  \delta \imd{\theta}{}{} - \imd{\theta}{}{} \frac{\ipre{\Gamma}{1}{11} \icpre{K}{}{3}}{\icpre{K}{}{}}  \vepre{n} \cdot \delta \iv{v}{}{,1} + \imd{\theta}{}{} \frac{\icpre{K}{}{3}}{\icpre{K}{}{}}  \vepre{n} \cdot \delta \iv{v}{}{,11} - \imd{\theta}{}{} \icpre{K}{}{2} \: \delta \imd{\theta}{}{},
	\end{aligned}
\end{equation}
and

\begin{equation}
	\label{eq: variation kapa 3}
	\begin{aligned}
		\Delta \delta  \imd{\kappa}{}{3} &=  \iv{v}{}{,1} \cdot \left[ \frac{\ipre{\Gamma}{1}{11} \icpre{K}{}{3}}{\ipre{g}{}{} \icpre{K}{}{}} \left( \frac{\ipre{\Gamma}{1}{11} \ipre{g}{}{}}{\icpre{K}{}{} } \vepre{b} \otimes \vepre{b} -\frac{\icpre{K}{}{}}{\ipre{\Gamma}{1}{11}} \vepre{n} \otimes \vepre{n} +\ivpre{g}{}{1} \otimes \vepre{n} + \vepre{n} \otimes \ivpre{g}{}{1} \right) \right] \delta \iv{v}{}{,1} \\
		& - \iv{v}{}{,1} \cdot  \left( \frac{\ipre{\Gamma}{1}{11} \icpre{K}{}{3}}{\icpre{K}{2}{} } \vepre{b} \otimes \vepre{b} + \frac{\icpre{K}{}{3}}{\ipre{g}{}{} \icpre{K}{}{}} \vepre{n} \otimes \ivpre{g}{}{1} \right)  \delta \iv{v}{}{,11} + \iv{v}{}{,1} \cdot  \left( \frac{\ipre{\Gamma}{1}{11} \icpre{K}{}{2}}{\icpre{K}{}{} } \vepre{n}  \right)  \delta \imd{\theta}{}{}  \\
		& - \iv{v}{}{,11} \cdot \left( \frac{\ipre{\Gamma}{1}{11} \icpre{K}{}{3}}{\icpre{K}{2}{} } \vepre{b} \otimes \vepre{b} + \frac{\icpre{K}{}{3}}{\ipre{g}{}{} \icpre{K}{}{}} \ivpre{g}{}{1} \otimes \vepre{n} \right)\delta \iv{v}{}{,1} +\iv{v}{}{,11} \cdot \left( \frac{\icpre{K}{}{3}}{\icpre{K}{2}{} } \vepre{b} \otimes \vepre{b} \right)\delta \iv{v}{}{,11} \\
		&- \iv{v}{}{,11} \cdot  \left( \frac{\icpre{K}{}{2}}{\icpre{K}{}{} } \vepre{n}  \right)  \delta \imd{\theta}{}{} + \imd{\theta}{}{} \frac{\ipre{\Gamma}{1}{11} \icpre{K}{}{2}}{\icpre{K}{}{}}  \vepre{n} \cdot \delta \iv{v}{}{,1} - \imd{\theta}{}{} \frac{\icpre{K}{}{2}}{\icpre{K}{}{}}  \vepre{n} \cdot \delta \iv{v}{}{,11} - \imd{\theta}{}{} \icpre{K}{}{3} \: \delta \imd{\theta}{}{}.
	\end{aligned}
\end{equation}

The linearization of the virtual rate of torsion is straightforward, but significantly more involved. We will rearrange the expression in \eqqref{eq: new rs} as:
\begin{equation}
	\label{eq: k1 novo}
	\imd{\kappa}{}{1}= \frac{1}{\icdef{K}{2}{}} \left(\idef{\Gamma}{1}{11}  \iv{v}{}{,1} -\iv{v}{}{,11}\right) \cdot \left( G_2^* \vdef{b} + G_3^* \vdef{n} \right) + \frac{1}{\icdef{K}{}{}} \vdef{b} \cdot \left(\iv{v}{}{,111} - \frac{G_1^*}{g^*} \iv{v}{}{,1}\right) + \imd{\theta}{}{,1},
\end{equation}
where the following components of $\iv{g}{}{1,11}$ vector are introduced:
\begin{equation}
	\label{eq: G1 and G2}
		\begin{aligned}
	G_1 &= \iv{g}{}{1,11} \cdot \iv{g}{}{1}= g\left[ \ii{\Gamma}{1}{11,1} + \left(\ii{\Gamma}{1}{11}\right)^2 -K \ic{K}{}{} \right], \\
	G_2 &= \iv{g}{}{1,11} \cdot \iv{n}{}{} = \ii{\Gamma}{1}{11} \ic{K}{}{} + \ic{K}{}{,1}, \\ 
	G_3 &= \iv{g}{}{1,11} \cdot \iv{b}{}{} = \ic{K}{}{} \ic{\tau}{}{}.
\end{aligned}
\end{equation}
Now, the linearized increment of the virtual rate of torsional curvature can be expressed as:
\begin{equation}
	\begin{aligned}
	\label{eq: k1 novo var}
	\Delta \delta \imd{\kappa}{}{1} &= \Delta \left(\frac{1}{\icdef{K}{2}{}}\right) \left(\ipre{\Gamma}{1}{11} \delta \iv{v}{}{,1} -\delta \iv{v}{}{,11}\right) \cdot \left( G_2^\sharp \vepre{b} + G_3^\sharp \vepre{n} \right) + \frac{1}{\icpre{K}{2}{}} \left(\Delta \idef{\Gamma}{1}{11} \delta \iv{v}{}{,1} \right) \cdot \left( G_2^\sharp \vepre{b} + G_3^\sharp \vepre{n} \right)  \\
	& \quad  + \frac{1}{\icpre{K}{2}{}} \left(\ipre{\Gamma}{1}{11} \delta \iv{v}{}{,1} -\delta \iv{v}{}{,11}\right) \cdot \left( \Delta G_2^* \vepre{b} + G_2^\sharp \Delta \ivdef{b}{}{} + \Delta G_3^* \vepre{n} + G_3^\sharp \Delta \ivdef{n}{}{} \right)\\
	& \quad +  \left(\Delta\frac{1}{\icdef{K}{}{}}\vepre{b} +\frac{1}{\icpre{K}{}{}} \Delta \vdef{b} \right) \cdot \left(\delta \iv{v}{}{,111} - \frac{G_1^\sharp}{\ipre{g}{}{}} \delta \iv{v}{}{,1}\right) - \frac{1}{\icpre{K}{}{}}\vepre{b}  \cdot \left( \Delta G_1^* \frac{1}{\ipre{g}{}{}} + G_1^\sharp \Delta \frac{1}{\idef{g}{}{}} \right) \delta \iv{v}{}{,1}.
\end{aligned}
\end{equation}
Moreover, we need the following linearized increments:
\begin{equation}
	\label{eq: var G1 and G2}
	\begin{aligned}
		\Delta G_1^* &=  \left(\frac{G_1^\sharp}{g^\sharp} \ivpre{g}{}{1} + G_2^\sharp \vepre{n} + G_3^\sharp \vepre{b} \right)\cdot \iv{v}{}{,1} +\ivpre{g}{}{1} \cdot \iv{v}{}{,111}, \\
		\Delta G_2^* &= \frac{G_3^\sharp}{\icpre{K}{}{}} \vepre{b} \cdot \left( \iv{v}{}{,11} - \ipre{\Gamma}{1}{11}  \iv{v}{}{,1}\right) -G_1^\sharp \vepre{n} \cdot  \iv{v}{}{,1} + \vepre{n} \cdot \iv{v}{}{,111}, \\ 
		\Delta G_3^* &= - \frac{G_2^\sharp}{\icpre{K}{}{}} \vepre{b} \cdot \left( \iv{v}{}{,11} - \ipre{\Gamma}{1}{11}  \iv{v}{}{,1}\right) -G_1^\sharp \vepre{b} \cdot \iv{v}{}{,1} + \vepre{b} \cdot  \iv{v}{}{,111}.
	\end{aligned}
\end{equation}
By inserting Eqs.~\eqref{eq: ap6 variations of reference strains}, \eqref{eq: variations n b}, and \eqref{eq: var G1 and G2}  into \eqqref{eq: k1 novo var}, we obtain:

\begin{equation}
	\label{eq: variation kapa1}
	\begin{aligned}
				\Delta \delta \imd{\kappa}{}{1} &= \iv{v}{}{,1} \cdot \left\{\ieqpre{\tau}{}{} \left[2  \left(\frac{\ipre{\Gamma}{1}{11}}{\icpre{K}{}{}}\right)^2 + \frac{1}{\ipre{g}{}{}} \right] \left(\vepre{n}\otimes\vepre{n}-\vepre{b}\otimes\vepre{b}\right) - \frac{\icpre{\tau}{}{} \ipre{\Gamma}{1}{11} }{g^\sharp \icpre{K}{}{}} \left(\ivpre{g}{}{1} \otimes \vepre{n} + \vepre{n} \otimes \ivpre{g}{}{1} \right) \right.\\
			& \quad \quad \quad \quad  + 2\frac{\ipre{\Gamma}{1}{11}}{\icpre{K}{2}{}} \left(\frac{\ipre{\Gamma}{1}{11} G_2^\sharp}{\icpre{K}{}{}} - \frac{G_1^\sharp}{\ipre{g}{}{}}\right) \left(\vepre{n} \otimes \vepre{b} + \vepre{b} \otimes \vepre{n} \right)  \\
		&\left.\quad \quad \quad \quad + \frac{1}{g^\sharp \icpre{K}{}{}} \left( \frac{G_1^\sharp}{\ipre{g}{}{}} - \frac{\ipre{\Gamma}{1}{11} G_2^\sharp}{\icpre{K}{}{}}\right)\left(\ivpre{g}{}{1} \otimes \vepre{b} + \vepre{b} \otimes \ivpre{g}{}{1} \right) \right\} \delta \iv{v}{}{,1} \\
		&\quad + \iv{v}{}{,1} \cdot \left[ 2\frac{\icpre{\tau}{}{} \ipre{\Gamma}{1}{11}}{\icpre{K}{2}{}} \left(\vepre{b}\otimes\vepre{b} - \vepre{n}\otimes\vepre{n}\right) + \frac{1}{\icpre{K}{2}{}}\left(\frac{G_1^\sharp}{g^\sharp} -2\frac{\ipre{\Gamma}{1}{11} G_2^\sharp}{\icpre{K}{}{}} \right) \left(\vepre{b}\otimes\vepre{n} + \vepre{n} \otimes \vepre{b} \right)  \right.\\
		&\left. \quad \quad \quad \quad +\frac{G_2^\sharp}{g^\sharp \icpre{K}{2}{}} \vepre{b}\otimes\ivpre{g}{}{1} +\frac{\ipre{\tau}{}{}}{g^\sharp \icpre{K}{}{}} \vepre{n}\otimes\ivpre{g}{}{1}\right] \delta \iv{v}{}{,11} \\
		&\quad + \iv{v}{}{,1} \cdot \left[ \frac{\ipre{\Gamma}{1}{11}}{\icpre{K}{2}{}}  \left(\vepre{b}\otimes\vepre{n} + \vepre{n} \otimes \vepre{b} \right)  -\frac{1}{g^\sharp \icpre{K}{}{}} \vepre{b}\otimes\ivpre{g}{}{1}  \right] \delta \iv{v}{}{,111}\\
		&\quad +\iv{v}{}{,11} \cdot \left[  2\frac{\icpre{\tau}{}{} \ipre{\Gamma}{1}{11}}{\icpre{K}{2}{}} \left(\vepre{b}\otimes\vepre{b} - \vepre{n}\otimes\vepre{n}\right) + \frac{1}{\icpre{K}{2}{}}\left(\frac{G_1^\sharp}{g^\sharp} -2\frac{\ipre{\Gamma}{1}{11} G_2^\sharp}{\icpre{K}{}{}} \right) \left(\vepre{b}\otimes\vepre{n} + \vepre{n} \otimes \vepre{b} \right)  \right.\\
		&\left.\quad \quad \quad \quad +\frac{G_2^\sharp}{g^\sharp \icpre{K}{2}{}} \ivpre{g}{}{1} \otimes \vepre{b} +\frac{\ipre{\tau}{}{}}{g^\sharp \icpre{K}{}{}} \ivpre{g}{}{1}\otimes\vepre{n}\right] \delta \iv{v}{}{,1} \\
		&\quad +\iv{v}{}{,11} \cdot \left[2\frac{\ieqpre{\tau}{}{}}{\icpre{K}{2}{}} \left(\vepre{n}\otimes\vepre{n}-\vepre{b}\otimes\vepre{b}\right) + 2\frac{G_2^\sharp}{\icpre{K}{3}{}} \left(\vepre{n} \otimes \vepre{b} + \vepre{b} \otimes \vepre{n} \right) \right] \delta \iv{v}{}{,11}\\
		& \quad +\iv{v}{}{,11} \cdot \left[-\frac{1}{\icpre{K}{2}{}} \left(\vepre{n} \otimes \vepre{b} + \vepre{b} \otimes \vepre{n} \right) \right] \delta \iv{v}{}{,111}  \\
		& \quad +\iv{v}{}{,111} \cdot \left[ \frac{\ipre{\Gamma}{1}{11}}{\icpre{K}{2}{}}  \left(\vepre{b}\otimes\vepre{n} + \vepre{n} \otimes \vepre{b} \right)  -\frac{1}{g^\sharp \icpre{K}{}{}} \ivpre{g}{}{1}\otimes\vepre{b}  \right] \delta \iv{v}{}{,1}\\
		& \quad + \iv{v}{}{,111} \cdot \left[-\frac{1}{\icpre{K}{2}{}} \left(\vepre{n} \otimes \vepre{b} + \vepre{b} \otimes \vepre{n} \right) \right] \delta \iv{v}{}{,11}.
	\end{aligned}
\end{equation}
Let us introduce the following designations:
\begin{equation}
\label{eq: ap12 variations of reference strains}
\begin{aligned}
\iv{G}{}{11} &= N \textbf{I}_{3\times3}+  M_1 \left\{\ieq{\tau}{}{} \left[2  \left(\frac{\ii{\Gamma}{1}{11}}{\ic{K}{}{}}\right)^2 + \frac{1}{g} \right] \left(\ve{n}\otimes\ve{n}-\ve{b}\otimes\ve{b}\right) - \frac{\ic{\tau}{}{} \ii{\Gamma}{1}{11} }{g \ic{K}{}{}} \left(\iv{g}{}{1} \otimes \ve{n} + \ve{n} \otimes \iv{g}{}{1} \right) \right. \\
& \left. \quad + 2\frac{\ii{\Gamma}{1}{11}}{\ic{K}{2}{}} \left(\frac{\ii{\Gamma}{1}{11} G_2}{\ic{K}{}{}} - \frac{G_1}{g}\right) \left(\ve{n} \otimes \ve{b} + \ve{b} \otimes \ve{n} \right) + \frac{1}{g \ic{K}{}{}} \left( \frac{G_1}{g} - \frac{\ii{\Gamma}{1}{11} G_2}{\ic{K}{}{}}\right)\left(\iv{g}{}{1} \otimes \ve{b} + \ve{b} \otimes \iv{g}{}{1} \right)   \right\} \\
&\quad+ M_2  \left[ \frac{\ii{\Gamma}{1}{11} \ic{K}{}{2}}{g \ic{K}{}{}} \left( \frac{\ii{\Gamma}{1}{11} g}{\ic{K}{}{} } \ve{b} \otimes \ve{b} -\frac{\ic{K}{}{}}{\ii{\Gamma}{1}{11}} \ve{n} \otimes \ve{n} +\iv{g}{}{1} \otimes \ve{n} + \ve{n} \otimes \iv{g}{}{1} \right) \right]  \\
&\quad+ M_3  \left[ \frac{\ii{\Gamma}{1}{11} \ic{K}{}{3}}{g \ic{K}{}{}} \left( \frac{\ii{\Gamma}{1}{11} g}{\ic{K}{}{} } \ve{b} \otimes \ve{b} -\frac{\ic{K}{}{}}{\ii{\Gamma}{1}{11}} \ve{n} \otimes \ve{n} +\iv{g}{}{1} \otimes \ve{n} + \ve{n} \otimes \iv{g}{}{1} \right) \right], \\
\iv{G}{}{12} &= M_1 \left[ 2\frac{\ic{\tau}{}{} \ii{\Gamma}{1}{11}}{\ic{K}{2}{}} \left(\ve{b}\otimes\ve{b} - \ve{n}\otimes\ve{n}\right) + \frac{1}{\ic{K}{2}{}}\left(\frac{G_1}{g} -2\frac{\ii{\Gamma}{1}{11} G_2}{\ic{K}{}{}} \right) \left(\ve{b}\otimes\ve{n} + \ve{n} \otimes \ve{b} \right)  +\frac{G_2}{g \ic{K}{2}{}} \ve{b}\otimes\iv{g}{}{1}  \right.\\
&\left.\quad +\frac{\ii{\tau}{}{}}{g \ic{K}{}{}} \ve{n}\otimes\iv{g}{}{1}\right] - M_2 \left( \frac{\ii{\Gamma}{1}{11} \ic{K}{}{2}}{\ic{K}{2}{} } \ve{b} \otimes \ve{b} + \frac{\ic{K}{}{2}}{g \ic{K}{}{}} \ve{n} \otimes \iv{g}{}{1} \right) - M_3 \left( \frac{\ii{\Gamma}{1}{11} \ic{K}{}{3}}{\ic{K}{2}{} } \ve{b} \otimes \ve{b} + \frac{\ic{K}{}{3}}{g \ic{K}{}{}} \ve{n} \otimes \iv{g}{}{1} \right), \\
\iv{G}{}{13} &= M_1 \left[ \frac{\ii{\Gamma}{1}{11}}{\ic{K}{2}{}}  \left(\ve{b}\otimes\ve{n} + \ve{n} \otimes \ve{b} \right)  -\frac{1}{g \ic{K}{}{}} \ve{b}\otimes\iv{g}{}{1}  \right], \\
\iv{G}{}{14} &=  - M_2 \left( \frac{\ii{\Gamma}{1}{11} \ic{K}{}{3}}{\ic{K}{}{} } \ve{n}  \right) + M_3 \left( \frac{\ii{\Gamma}{1}{11} \ic{K}{}{2}}{\ic{K}{}{} } \ve{n}  \right), \\
\iv{G}{}{22} &= M_1 \left[2\frac{\ieq{\tau}{}{}}{\ic{K}{2}{}} \left(\ve{n}\otimes\ve{n}-\ve{b}\otimes\ve{b}\right) + 2\frac{G_2}{\ic{K}{3}{}} \left(\ve{n} \otimes \ve{b} + \ve{b} \otimes \ve{n} \right) \right] + M_2 \left( \frac{\ic{K}{}{2}}{\ic{K}{2}{} } \ve{b} \otimes \ve{b} \right) \\
&\quad + M_3 \left( \frac{\ic{K}{}{3}}{\ic{K}{2}{} } \ve{b} \otimes \ve{b} \right), \\
\iv{G}{}{23} &= M_1 \left[-\frac{1}{\ic{K}{2}{}} \left(\ve{n} \otimes \ve{b} + \ve{b} \otimes \ve{n} \right) \right], \\
\iv{G}{}{24} &= \ii{M}{}{2} \left( \frac{\ic{K}{}{3}}{\ic{K}{}{} } \ve{n}  \right)  -M_3  \left( \frac{\ic{K}{}{2}}{\ic{K}{}{} } \ve{n}  \right) , \\
\iv{G}{}{33} &= \iv{0}{}{3 \times 3} , \\
\iv{G}{}{34} &= \iv{0}{}{3 \times 1} , \\
\iv{G}{}{44} &= -M_2 \ic{K}{}{2} -M_3 \ic{K}{}{3},
\end{aligned}
\end{equation}
which constitute the matrix of generalized section forces:
\begin{equation}
\label{eq: apnestoG matrix def}
\ve{G} = 
\begin{bmatrix}
\iv{G}{}{11} & \iv{G}{}{12} & \iv{G}{}{13} & \iv{G}{}{14} \\
 & \iv{G}{}{22} & \iv{G}{}{23} & \iv{G}{}{24}\\
 &  & \iv{G}{}{33} & \iv{G}{}{34} \\
sym. &  &  & \iv{G}{}{44} \\
\end{bmatrix}.
\end{equation}
Additionally, we need the matrix of basis functions $\iv{B}{}{G}$:
\setcounter{MaxMatrixCols}{20}
\begin{equation}
\label{eq: appB}
\begin{aligned}
	\ve{B}_G &= 
\begin{bmatrix}
	\iv{B}{}{G1} & \iv{B}{}{G2} & ... & \iv{B}{}{GI} & ... & \iv{B}{}{GN} 
\end{bmatrix}, \\
		\trans{$\iv{B}{}{GI} $}&=
\begin{bmatrix}
	\trans{$\iv{N}{}{I,1}$}  &
	\trans{$\iv{N}{}{I,11}$}  &
	\trans{$\iv{N}{}{I,111}$}  &
	\trans{$\iv{N}{\theta}{I}$} 
\end{bmatrix},
\end{aligned}
\end{equation}
which is actually the matrix $\ve{B}$ without the $11^{th}$ row, see \eqqref{eq: vector of reference strains matrix form2}. The difference between these two matrices of the basis functions is due to the fact that the rate of the torsional curvature depends on the $\imd{\theta}{}{,1}$, cf. Eqs~\eqref{eq: new rs}, while its increment does not, cf. \eqqref{eq: variation kapa1}. Now, the part of the virtual power due to the known stress and the increment of the virtual strain rate in \eqqref{matrix form of linearized virtual power} can be expressed as:
\begin{equation}
\label{eq: apn1estoG matrix def}
\int_{\xi}^{} \trans{$\iv{f}{\: \sharp}{}$} \Delta \delta \ve{e} \sqrt{g} \dd{\xi}=
\int_{\xi}^{} \trans{$\iv{f}{\: \sharp}{}$} \Delta \ve{H}^* \ve{B}  \sqrt{g} \dd{\xi} \delta \ivmd{q}{}{} = \transmd{q} \int_{\xi}^{} \trans{$\iv{B}{}{G}$} \ivpre{G}{}{} \iv{B}{}{G} \sqrt{g} \dd{\xi} \delta \ivmd{q}{}{} =
\transmd{q} \trans{$\ve{K}^\sharp_G$} \delta \ivmd{q}{}{}.
\end{equation}
Note that the derived geometric stiffness matrix $\iv{K}{}{G}$ is symmetric. This confirms that the energetically conjugated pairs are correctly adopted.

\section*{Appendix B. Linearization of constraint equation}
\setcounter{equation}{0}
\renewcommand\theequation{B\arabic{equation}} 

In order to linearize the constraint condition \eqref{eq: vp constraint}, the increment of the constraint rate is required. This increment equals the increment of the total twist velocity $\imd{\omega}{}{}$:
\begin{equation}
	\label{eq: variation const}
	\begin{aligned}
		\Delta \delta \imd{c}{}{} &=
		\Delta \left(\frac{1}{\icdef{K}{}{}}\right)  \vepre{b} \cdot \left(\delta \iv{v}{}{,11} - \ipre{\Gamma}{1}{11} \delta \iv{v}{}{,1}\right) + \frac{1}{\icpre{K}{}{}} \Delta \vdef{b} \cdot \left(\delta \iv{v}{}{,11} - \ipre{\Gamma}{1}{11} \delta \iv{v}{}{,1}\right) +  \frac{1}{\icpre{K}{}{}} \vepre{b} \cdot \left( -\Delta \idef{\Gamma}{1}{11} \delta \iv{v}{}{,1}\right) \\
		&= \iv{v}{}{,1} \cdot \left[  \frac{\ipre{\Gamma}{1}{11}}{\ipre{g}{}{} \icpre{K}{}{}} \left( \ivpre{g}{}{1} \otimes \vepre{b} + \vepre{b} \otimes \ivpre{g}{}{1} \right) -\left( \frac{\ipre{\Gamma}{1}{11}}{\icpre{K}{}{}} \right)^2 \left(\vepre{n} \otimes \vepre{b} + \vepre{b} \otimes \vepre{n}\right) - \frac{1}{\ipre{g}{}{}} \ivpre{n}{}{} \otimes \vepre{b} \right] \delta \iv{v}{}{,1} \\
		&\quad + \iv{v}{}{,1} \cdot \left[ \frac{\ipre{\Gamma}{1}{11}}{\icpre{K}{2}{}} \left(\vepre{n} \otimes \vepre{b} + \vepre{b} \otimes \vepre{n}\right) - \frac{1}{\ipre{g}{}{} \icpre{K}{}{}}  \ivpre{b}{}{} \otimes \ivpre{g}{}{1}  \right] \delta \iv{v}{}{,11} \\
		&\quad + \iv{v}{}{,11} \cdot \left[ \frac{\ipre{\Gamma}{1}{11}}{\icpre{K}{2}{}} \left(\vepre{n} \otimes \vepre{b} + \vepre{b} \otimes \vepre{n}\right) - \frac{1}{\ipre{g}{}{} \icpre{K}{}{}}  \ivpre{g}{}{1} \otimes \ivpre{b}{}{}  \right] \delta \iv{v}{}{,1} \\
		& \quad - \iv{v}{}{,11} \cdot \left[ \frac{1}{\icpre{K}{2}{}} \left(\vepre{n} \otimes \vepre{b} + \vepre{b} \otimes \vepre{n}\right)  \right] \delta \iv{v}{}{,11}.
	\end{aligned}
\end{equation}
Let us introduce the following matrix:
\begin{equation}
	\label{eq: Glambda}
	\iveq{G}{}{\lambda} = 
	\begin{bmatrix}
		\iveq{G}{}{\lambda 11} & \iveq{G}{}{\lambda12} \\
		\iveq{G}{}{\lambda21} & \iveq{G}{}{\lambda22} 
	\end{bmatrix},
\end{equation}
with elements:
\begin{equation}
	\begin{aligned}
\iveq{G}{}{\lambda 11} &=   \frac{\ii{\Gamma}{1}{11}}{\ii{g}{}{} \ic{K}{}{}} \left( \iv{g}{}{1} \otimes \ve{b} + \ve{b} \otimes \iv{g}{}{1} \right) -\left( \frac{\ii{\Gamma}{1}{11}}{\ic{K}{}{}} \right)^2 \left(\ve{n} \otimes \ve{b} + \ve{b} \otimes \ve{n}\right) - \frac{1}{\ii{g}{}{}} \iv{b}{}{} \otimes \ve{n}, \\
\iveq{G}{}{\lambda 12} &=  \frac{\ii{\Gamma}{1}{11}}{\ic{K}{2}{}} \left(\ve{n} \otimes \ve{b} + \ve{b} \otimes \ve{n}\right) - \frac{1}{\ii{g}{}{} \ic{K}{}{}}  \ve{b} \otimes \iv{g}{}{1} , \\
\iveq{G}{}{\lambda 21} &=   \frac{\ii{\Gamma}{1}{11}}{\ic{K}{2}{}} \left(\ve{n} \otimes \ve{b} + \ve{b} \otimes \ve{n}\right) - \frac{1}{\ii{g}{}{} \ic{K}{}{}}  \iv{g}{}{1} \otimes \ve{b} , \\
\iveq{G}{}{\lambda 22} &=   - \frac{1}{\ic{K}{2}{}} \left(\ve{n} \otimes \ve{b} + \ve{b} \otimes \ve{n}\right) , 
\end{aligned}
\end{equation}
and the matrix of basis functions:
\begin{equation}
\begin{aligned}
	\ve{B}_{\lambda} &= 
	\begin{bmatrix}
		\iv{B}{}{\lambda 1} & \iv{B}{}{\lambda 2} & ... & \iv{B}{}{\lambda I} & ... & \iv{B}{}{\lambda N} 
	\end{bmatrix}, \\
	\trans{$\iv{B}{}{\lambda I} $}&=
	\begin{bmatrix}
		\trans{$\iv{N}{}{I,1}$}  &
		\trans{$\iv{N}{}{I,11}$} 
	\end{bmatrix}.
\end{aligned}
\end{equation}
Now, the increment of the constraint rate is $\Delta \delta \imd{c}{}{}=\delta \trans{$\ivmd{q}{}{\lambda}$} \trans{$\iv{B}{}{\lambda}$} \ivpre{G}{}{\lambda} \iv{B}{}{\lambda} \ivmd{q}{}{\lambda}$, and we can derive \eqqref{eq: constraint matrixx}:
\begin{equation}
	\label{eq: f6bc}
	\ipre{\lambda}{}{} \Delta \delta \imd{c}{}{} = 
	\delta \trans{$\ivmd{q}{}{\lambda}$} \trans{$\iv{B}{}{\lambda}$} \ipre{\lambda}{}{} \ivpre{G}{}{\lambda} \iv{B}{}{\lambda} \ivmd{q}{}{\lambda}= \delta \trans{$\ivmd{q}{}{\lambda}$} \iv{K}{\sharp}{G \lambda} \ivmd{q}{}{\lambda}.
\end{equation}
The constraint condition and its contribution to the tangent matrix must be evaluated at the fixed coordinate $\xi=\xi_c$, where the boundary condition is defined. Obviously, the resulting geometric stiffness matrix $\iv{K}{}{G \lambda}$ is not symmetric due to the one term in $\iveq{G}{}{\lambda 11}$. However, our simulations show that this term has negligible influence on the performance of nonlinear solver.

\section*{Appendix C. Linearization of external virtual power}
\setcounter{equation}{0}
\renewcommand\theequation{C\arabic{equation}}

Let us consider the contribution to the tangent stiffness that comes from the configuration-dependent external load. The focus is on the virtual power due to the concentrated moment, since it was the load used in the numerical analysis. To model concentrated moment which is constant with respect to the global coordinate system, the load vector must be updated at every time step due to the change of local vector basis. The increment of the external virtual power is:
\begin{equation}
\label{eq: f1}
\Delta \delta P_{ext}=-\Delta \left( \ve{m}^* \cdot \delta \ivmd{$\pmb{\omega}$}{}{}\right) = -\Delta \left( \ve{m}^* \cdot \ivdef{g}{}{i} \delta \omega^i \right)= -\ve{m}^* \cdot \Delta \ivdef{g}{}{i} \delta \omega^i-\ve{m}^* \cdot \ivpre{g}{}{i} \Delta \delta \omega^i,
\end{equation}
while the increments of virtual angular velocities follow from Eqs.~\eqref{eq:45} and \eqref{eq: variation const}. The increment of the virtual twist velocity is:
\begin{equation}
\label{eq: f3}
\begin{aligned}	
\Delta \delta \imd{\omega}{1}{} &= \Delta \left[ \frac{1}{\sqrt{g^*} } \left( \delta \imd{\omega}{}{FS} + \delta \imd{\theta}{}{} \right)  \right] = \Delta \frac{1}{\sqrt{g^*} } \left(\delta \imd{\omega}{}{FS} + \delta \imd{\theta}{}{}\right) + \frac{1}{\sqrt{g^\sharp} } \Delta \delta \imd{\omega}{}{FS} \\
&= \frac{1}{\sqrt{g^\sharp}} \left\{ \iv{v}{}{,1} \cdot \left[ \left( \frac{\ipre{\Gamma}{1}{11}}{g^\sharp \icpre{K}{}{}} \right) \left(2\ivpre{g}{}{1} \otimes \vepre{b} + \vepre{b} \otimes \ivpre{g}{}{1} \right) -\left(\frac{\ipre{\Gamma}{1}{11}}{\icpre{K}{}{}}\right)^2 \left( \ivpre{b}{}{} \otimes \vepre{n} + \vepre{n} \otimes \ivpre{b}{}{} \right) \right. \right. \\
&\quad \quad \left. \quad - \frac{1}{g^\sharp} \ivpre{n}{}{} \otimes \vepre{b} \right] \delta \iv{v}{}{,1} - \iv{v}{}{,11} \cdot \left[ \frac{1}{\icpre{K}{2}{}} \left(\vepre{n} \otimes \vepre{b} + \vepre{b} \otimes \vepre{n}\right)  \right] \delta \iv{v}{}{,11} \\
&\quad \quad \quad+ \iv{v}{}{,1} \cdot \left[ \frac{\ipre{\Gamma}{1}{11}}{\icpre{K}{2}{}} \left(\vepre{n} \otimes \vepre{b} + \vepre{b} \otimes \vepre{n}\right) - \frac{1}{g^\sharp \icpre{K}{}{}} \left( \vepre{b} \otimes \ivpre{g}{}{1} + \ivpre{g}{}{1}\otimes \vepre{b}  \right)\right] \delta \iv{v}{}{,11}  \\
&\quad \quad \quad \left.\ + \iv{v}{}{,11} \cdot \left[ \frac{\ipre{\Gamma}{1}{11}}{\icpre{K}{2}{}} \left(\vepre{n} \otimes \vepre{b} + \vepre{b} \otimes \vepre{n}\right) - \frac{1}{g^\sharp \icpre{K}{}{}}  \ivpre{g}{}{1} \otimes \vepre{b}  \right] \delta \iv{v}{}{,1}  - \imd{\theta}{}{} \left(\frac{\ivpre{g}{}{1}}{g^\sharp}\right) \delta \iv{v}{}{,1} \right\}, 
\end{aligned}
\end{equation}
while the increments of the other two components are:
\begin{equation}
	\label{eq: f31}
	\begin{aligned}	
\Delta \delta \ii{\omega}{2}{} &= \Delta \left( - \frac{1}{\sqrt{g^\sharp} } \ \ivpre{g}{}{3} \cdot \delta \iv{v}{}{,1} \right) =  \iv{v}{}{,1} \cdot \left[ \frac{1}{g^{\sharp 3/2}} \left(\ivpre{g}{}{3} \otimes \ivpre{g}{}{1} + \ivpre{g}{}{1} \otimes \ivpre{g}{}{3} \right) - \frac{\ipre{\Gamma}{1}{11}}{\sqrt{g^\sharp} \icpre{K}{}{}}  \vepre{b} \otimes \ivpre{g}{}{2}\right] \delta \iv{v}{}{,1} \\
& \quad + \frac{1}{\sqrt{g^\sharp} \icpre{K}{}{}} \iv{v}{}{,11} \cdot \left(  \vepre{b}\otimes \ivpre{g}{}{2} \right) \delta \iv{v}{}{,1} + \frac{1}{\sqrt{g^\sharp}} \: \imd{\theta}{}{} \: \ivpre{g}{}{2} \cdot \delta \iv{v}{}{,1}, \\
\Delta \delta \ii{\omega}{3}{} &= \Delta \left( \frac{1}{\sqrt{g^\sharp} } \ivpre{g}{}{2} \cdot \delta \iv{v}{}{,1} \right) =  -\iv{v}{}{,1} \cdot \left[ \frac{1}{g^{\sharp 3/2}} \left(\ivpre{g}{}{2} \otimes \ivpre{g}{}{1} + \ivpre{g}{}{1} \otimes \ivpre{g}{}{2} \right) + \frac{\ipre{\Gamma}{1}{11}}{\sqrt{g^\sharp} \icpre{K}{}{}} \vepre{b} \otimes \ivpre{g}{}{3}\right] \delta \iv{v}{}{,1} \\
& \quad + \frac{1}{\sqrt{g^\sharp} \icpre{K}{}{}} \iv{v}{}{,11} \cdot \left(  \vepre{b}\otimes \ivpre{g}{}{3}  \right) \delta \iv{v}{}{,1} + \frac{1}{\sqrt{g^\sharp}} \: \imd{\theta}{}{} \: \ivpre{g}{}{3} \cdot \delta \iv{v}{}{,1}.
\end{aligned}
\end{equation}
The required increments of the basis vectors are given with Eqs.~\eqref{eq: ap1 variation of base vector g_alpha} and \eqref{eq: variation v,2 v,3 2}.
By the insertion of Eqs.~\eqref{eq: f3}, \eqref{eq: f31} and \eqref{eq: variation v,2 v,3 2} into  \eqqref{eq: f1}, we obtain:
\begin{equation}
\label{eq: f4}
\begin{aligned}
-\Delta \delta P_{ext} &=  \iv{v}{}{,1} \cdot \left\{ \left(\ve{m}^* \cdot \vepre{t}\right) \left[  \frac{1}{g^\sharp} \left(  \ivpre{g}{}{2} \otimes \ivpre{g}{}{3} - \ivpre{g}{}{3} \otimes \ivpre{g}{}{2} \right) - \left( \frac{\ipre{\Gamma}{1}{11}}{\icpre{K}{}{}} \right)^2 \left(\vepre{b} \otimes \vepre{n} + \vepre{n} \otimes \vepre{b} \right)   \right. \right.  \\
&\quad \left. +\frac{\ipre{\Gamma}{1}{11}}{g^\sharp \icpre{K}{}{}} \left(2 \ivpre{g}{}{1} \otimes \vepre{b} + \vepre{b} \otimes \ivpre{g}{}{1} \right) -\frac{1}{g^\sharp} \ \vepre{n} \otimes \vepre{b} \right] + \frac{1}{g^{\sharp  3/2}} \left( \ve{m}^* \cdot \ivpre{g}{}{2}\right) \left( \ivpre{g}{}{3} \otimes \ivpre{g}{}{1} + \ivpre{g}{}{1} \otimes \ivpre{g}{}{3} \right) \\
& \left. \quad - \frac{1}{g^{\sharp 3/2}} \left( \ve{m}^* \cdot \ivpre{g}{}{3}\right) \left( \ivpre{g}{}{2} \otimes \ivpre{g}{}{1} + \ivpre{g}{}{1} \otimes \ivpre{g}{}{2} \right) -\frac{\ipre{\Gamma}{1}{11}}{\sqrt{g^\sharp} \icpre{K}{}{}} \  \ve{m}^*\otimes \vepre{b} \right\} \delta \iv{v}{}{,1} \\
& \quad + \iv{v}{}{,11} \cdot \left\{ \left(\ve{m}^* \cdot \vepre{t} \right) \left[ \frac{\ipre{\Gamma}{1}{11}}{\icpre{K}{2}{}} \left( \vepre{b} \otimes \vepre{n} +\vepre{n} \otimes \vepre{b} \right) - \frac{1}{g^\sharp \icpre{K}{}{} } \ \ivpre{g}{}{1} \otimes\vepre{b}\right]\right\} \delta \iv{v}{}{,1} \\
& \quad + \iv{v}{}{,1} \cdot \left\{ \left(\ve{m}^* \cdot \vepre{t}\right) \left[   \frac{\ipre{\Gamma}{1}{11}}{\icpre{K}{2}{}}  \left(\vepre{b} \otimes \vepre{n} + \vepre{n} \otimes \vepre{b} \right)  - \frac{1}{g^\sharp \icpre{K}{}{}} \left(\ivpre{g}{}{1} \otimes \vepre{b} + \vepre{b} \otimes \ivpre{g}{}{1} \right) \right] \right. \\
& \quad \left. +\frac{1}{\sqrt{g^\sharp} \icpre{K}{}{}} \ve{m}^* \otimes \vepre{b}  \right\} \delta \iv{v}{}{,11}  -\iv{v}{}{,11} \cdot \left[ \frac{1}{\icpre{K}{2}{}} \left(\ve{m}^* \cdot \vepre{t} \right)  \left(\vepre{b} \otimes \vepre{n} + \vepre{n} \otimes \vepre{b} \right)\right] \delta \iv{v}{}{,11}\\
& \quad + \frac{1}{\sqrt{g^\sharp}}\: \iv{v}{}{,1} \cdot \left[ \ve{m}^* - \left(\ve{m}^* \cdot \vepre{t} \right) \vepre{t} \right] \delta\md{\theta}.
\end{aligned}
\end{equation}
If we define submatrices: 
\begin{equation}
\label{eq: f5}
\begin{aligned}
\iveq{G}{}{11} &= \left(\ve{m}^* \cdot \ve{t}\right) \left[  \frac{1}{g} \left(  \iv{g}{}{2} \otimes \iv{g}{}{3} - \iv{g}{}{3} \otimes \iv{g}{}{2} \right) + \left( \frac{\ii{\Gamma}{1}{11}}{\ic{K}{}{}} \right)^2 \left(\ve{b} \otimes \ve{n} + \ve{n} \otimes \ve{b} \right)   \right.  \\
&\quad \left. -\frac{\ii{\Gamma}{1}{11}}{g \ic{K}{}{}} \left(\iv{g}{}{1} \otimes \ve{b} + 2 \ve{b} \otimes \iv{g}{}{1} \right) +\frac{1}{g} \ \ve{b} \otimes \ve{n} \right] - \frac{1}{g^{3/2}} \left( \ve{m}^* \cdot \iv{g}{}{2}\right) \left( \iv{g}{}{3} \otimes \iv{g}{}{1} + \iv{g}{}{1} \otimes \iv{g}{}{3} \right) \\
& \quad + \frac{1}{g^{3/2}} \left( \ve{m}^* \cdot \iv{g}{}{3}\right) \left( \iv{g}{}{2} \otimes \iv{g}{}{1} + \iv{g}{}{1} \otimes \iv{g}{}{2} \right) +\frac{\ii{\Gamma}{1}{11}}{\sqrt{g} \ic{K}{}{}} \ \ve{b} \otimes \ve{m}^*,  \\
\iveq{G}{}{12} &=  \left(\ve{m}^* \cdot \ve{t} \right) \left[ \frac{1}{g \ic{K}{}{} } \ \ve{b} \otimes\iv{g}{}{1} - \frac{\ii{\Gamma}{1}{11}}{\ic{K}{2}{}} \left( \ve{b} \otimes \ve{n} +\ve{n} \otimes \ve{b} \right)\right], \\
\iveq{G}{}{21} &= \left(\ve{m}^* \cdot \ve{t}\right) \left[ \frac{1}{g \ic{K}{}{}} \left(\iv{g}{}{1} \otimes \ve{b} + \ve{b} \otimes \iv{g}{}{1} \right) -\frac{\ii{\Gamma}{1}{11}}{\ic{K}{2}{}}  \left(\ve{b} \otimes \ve{n} + \ve{n} \otimes \ve{b} \right)\right] -\frac{1}{\sqrt{g} \ic{K}{}{}} \ve{b} \otimes \ve{m}^*, \\
\iveq{G}{}{22} &= \frac{1}{\ic{K}{2}{}} \left(\ve{m}^* \cdot \ve{t} \right)  \left(\ve{b} \otimes \ve{n} + \ve{n} \otimes \ve{b} \right)\\
\iveq{G}{}{13} &= \iveq{G}{}{31} = \iveq{G}{}{33} = \textbf{0}_{3\times3}, \\
\iveq{G}{}{41} &= \frac{1}{\sqrt{g}} \left[ \left(\ve{m}^* \cdot \ve{t} \right) \trans{t} -\trans{$\ve{m}^*$} \right],  \\
\iveq{G}{}{42} &= \iveq{G}{}{43} = \iveq{G}{\mathsf{T}}{13} = \iveq{G}{\mathsf{T}}{12}= \iveq{G}{\mathsf{T}}{13},= \textbf{0}_{1\times3}, \\
\iveq{G}{}{44} &= 0,
\end{aligned}
\end{equation}
of the matrix:
\begin{equation}
\label{eq: f6}
\iveq{G}{}{} = 
\begin{bmatrix}
\iveq{G}{}{11} & \iveq{G}{}{12} & \iveq{G}{}{13} & \iveq{G}{}{14} \\
\iveq{G}{}{21} & \iveq{G}{}{22} & \iveq{G}{}{23}  & \iveq{G}{}{24} \\
\iveq{G}{}{31} & \iveq{G}{}{32} & \iveq{G}{}{33} & \iveq{G}{}{34} \\
\iveq{G}{}{41} & \iveq{G}{}{42} & \iveq{G}{}{43} & \iveq{G}{}{44} 
\end{bmatrix},
\end{equation}
then the geometric stiffness matrix due to the configuration-dependent load is $\iv{K}{\sharp Q}{G}= \trans{$\iv{B}{}{G}$} \iveq{G}{\sharp}{} \iv{B}{}{G}$. This matrix must be evaluated at the point of the application of load and added to the tangent stiffness matrix in \eqqref{eq: Kt}.


\bibliography{spatial_beam} 
\bibliographystyle{ieeetr}

\end{document}